\documentclass[a4paper,11pt]{article}
\usepackage{latexsym}
\usepackage[english]{babel}
\usepackage{amssymb,amsmath,amsfonts,amsthm}
\usepackage{dsfont,mdframed,mathrsfs,stmaryrd}
\usepackage[active]{srcltx}
\usepackage{graphicx,color,framed}
\usepackage{scrtime}
\usepackage{cite}
\usepackage[normalem]{ulem}
\usepackage{mathtools}
\usepackage{comment,cancel}
\usepackage{units}
\usepackage{parskip}
\usepackage{braket}
\usepackage{hyperref}
\usepackage{float}
\usepackage[margin=1in]{geometry}
\usepackage[toc,page]{appendix}
\usepackage{multirow}
\usepackage{array}

\usepackage{multirow}

\usepackage{soul}

\usepackage{tikz}
\usepackage[font=footnotesize]{caption}

\newcolumntype{H}{>{\setbox0=\hbox\bgroup}c<{\egroup}@{}}

\hypersetup{
linktoc=all,
    citecolor=black,
    filecolor=black,
    linkcolor=black,
    urlcolor=black
}

\makeatletter

\@addtoreset{equation}{section}
\makeatother

\def\be{\begin{equation}}
\def\ee{\end{equation}}
\def\bea{\begin{eqnarray}}
\def\eea{\end{eqnarray}}
\def\nn{\nonumber}

\def\m{\mu}
\def\n{\nu}

\newcommand{\eq}[1]{(\ref{#1})}
\newcommand{\w}[1]{\\[0.#1cm]}

\DeclareMathOperator{\Tr}{Tr}
\DeclareMathOperator{\tr}{tr}

\newcommand{\SU}[1]{\mathrm{SU}(#1)}

\begin{document}

\thispagestyle{empty}

\begin{flushright}\small
MI-HET-819   \\

\end{flushright}

\bigskip\bigskip

\vskip 10mm

\begin{center}

{\Large{\bf New anomaly free supergravities in six dimensions}}

\vskip 4mm

\end{center}

\vskip 6mm

\begin{center}

K. Becker$^\star$, A. Kehagias$^\dagger$, E. Sezgin$^\star$, D. Tennyson$^\star$ and A. Violaris$^\dagger$ \\

\vskip 8mm

$^\star$\,{\it George P. and Cynthia W. Mitchell Institute \\for Fundamental
Physics and Astronomy \\
Texas A\&M University, College Station, TX 77843-4242, USA}\\
\vskip 4mm
$^\dagger$\,{\it Physics Division, National Technical University of Athens, 15780 Zografou Campus, Athens, Greece}\\

\end{center}

\vskip1.5cm

\begin{center} {\bf Abstract } \end{center}

An extended search for anomaly free matter coupled $N=(1,0)$ supergravity in six dimension is carried out by two different methods which we refer to as the graphical and rank methods. In the graphical method the anomaly free models are built from single gauge group models, called nodes, which can only have gravitational anomalies. We search for anomaly free theories with gauge groups $G_1\times...\times G_n$ with $n=1,2,...$ (any number of factors) and $G_1\times...\times G_n \times U(1)_R$ where $n=1,2,3$ and $U(1)_R$ is the $R$-symmetry group. While we primarily consider models with the tensor multiplet number $n_T=1$, we also provide some results for $n_T\ne 1$ with an unconstrained number of charged hypermultiplets. We find a large number of ungauged anomaly free theories. However, in the case of $R$-symmetry gauged models with $n_T=1$, in addition to the three known  anomaly free theories with $G_1\times G_2 \times U(1)_R$ type symmetry, we find only six new remarkably anomaly free models with symmetry groups of the form $G_1\times G_2\times G_3 \times U(1)_R$. In the case of $n_T=1$ and ungauged models, excluding low rank group factors and considering only low lying representations, we find all anomaly free theories. Remarkably, the number of group factors does not exceed four in this class. The proof of completeness in this case relies on a bound which we establish for a parameter characterizing the difference between the number of non-singlet hypermultiplets and the dimension of the gauge group.

\newpage

\tableofcontents

\section{Introduction}

Low energy limits of superstring theories are described by certain supergravity theories but not all supergravities are known to arise from string theory. Higher derivative extensions of supergravities arise in the context of counterterms associated with UV divergences and anomalies. While local supersymmetry alone cannot fix all possible higher derivative extensions, various physical considerations such as high energy unitarity, in addition to the requirement of anomaly freedom, may help fix them order by order in a higher derivative expansion. The question arises as to whether any one of these supergravities can qualify for being a candidate for a UV completion. Another way to pose this question is to ask for whether string theory is the unique UV completion of matter coupled quantum gravity. Proving such a proposition is highly nontrivial. Furthermore, solving outstanding open problems in string theory, such as the cosmological constant and string landscape problems, may require a much further development of string theory. On the other hand starting from matter coupled supergravities that seem to pass all known consistency tests to date may have desirable properties with regard to these outstanding problems (more on this below). Therefore, it is natural to explore the landscape of such theories. To this end, matter coupled $N=(1,0)$ supergravities provide a very rich arena to explore, and it is the principal aim of this paper to conduct an extensive search for such anomaly free models that extend previously obtained results considerably.

Six dimensions is the highest dimension in which supergravity couples to matter multiplets other than the vector multiplets, in the framework of $(1,0)$ supergravity. These multiplets include the tensor and hypermultiplets. Chiral couplings being abundant leads to gravitational, gauge and mixed anomalies which makes them ideal candidates for studying issues arising in the swampland program. A key requirement for the  consistency of these theories is the absence of the aforementioned anomalies. Long ago, inspired by the discovery of the Green-Schwarz anomaly cancellation mechanism, a remarkably anomaly free $R$-symmetry gauged $(1,0)$ supergravity with $E_6\times E_7\times U(1)_R$ symmetry, with hyperfermions in the $(1,912)_0$ representation of the gauge group was found \cite{Randjbar-Daemi:1985tdc}. Sometime later, two more anomaly free models sharing similar features were found, one with the gauge group $G_2\times E_7\times U(1)_R$ and hyperfermions in $(14,56)_0$ \cite{Avramis:2005qt}, and another with the gauge group $F_4\times Sp(9)\times U(1)_R$ and hyperfermions in $(52,18)_0$ \cite{Avramis:2005hc}. These models are remarkable because the R-symmetry gauging makes it extremely rare to have anomaly free models of their type, unless the gauge group consists of $SU(2)$ and $U(1)$ factors \cite{Suzuki:2005vu}. None of these remarkable models have known string theory origin so far. Yet they have a number of interesting properties. For example, they come
with a positive definite potential proportional to the exponential function of the dilaton, even in the absence of the hypermultiplets. Consequently, these models do not admit maximally symmetric $6D$ spacetime vacua, but rather Minkowski$_4\times S^2$ with a monopole configuration on $S^2$ as the most symmetric solution. In their simplest form, such gauged supergravities have attractive features for cosmology as well \cite{Maeda:1985es,Halliwell:1986bs,Gibbons:1986xp,Gibbons:1987ps,Anchordoqui:2019amx}. The presence of the potential also has interesting consequences in understanding the dyonic string excitations that may be  supported by the theory. 

A systematic search for anomaly free $(1,0)$ supergravities was carried out in \cite{Avramis:2005hc} with certain assumptions. In particular, only single tensor multiplet and groups of the form $G_1\times G_2$, with one of both group exceptional, and only low lying representations were considered. The $R$-symmetry gauged models were also considered with further assumptions. 

A systematic analysis of anomaly free $(1,0)$ supergravities in $6D$ was also carried out for the case of single tensor field in \cite{Kumar:2009us,Kumar:2009ae}, and for multi tensor fields in \cite{Kumar:2010ru}. Using the requirement of anomaly cancellation by Green-Schwarz mechanism and positive gauge kinetic terms alone, it was shown in \cite{Kumar:2009ae} that the set of possible semi-simple gauge groups and matter representations is finite. Such a finiteness proof was extended to the case of fewer than nine tensor multiplets, namely $n_T<9$, again considering non-abelian gauge groups only, and from anomaly freedom and positivity of gauge kinetic terms alone in \cite{Kumar:2010ru}, where it was shown that infinite families exist for $n_T \ge 9$. This result was later shown to hold in the presence of $U(1)$ factors in  gauge group as well, provided that the charges of the matter under $U(1)$'s are ignored \cite{Park:2011wv}. Note however that in all these finiteness proofs, no maximum rank number was identified.

In this paper we shall extend the search for anomaly free $(1,0)$ models in a number of directions. An important starting point is  that given $(1,0)$ supergravity coupled to with $n_H$ hypermultiplets, $n_V$ vector multiplets and $n_T$ tensor multiplets, the necessary condition for the the absence of gravitational anomalies is given by \cite{Randjbar-Daemi:1985tdc}
\be
n_H=n_V-29 n_T+273\ .
\label{R4}
\ee
We shall follow two different approaches in search of anomaly free $(1,0)$ models. In the first approach, which we refer to as the ``graphical method", we first find {\it single} non-abelian group solutions that have $n_H-n_V\le \kappa$, where $\kappa$ is a constant  which will be specified in later sections. Each solution is represented by a node. We then find out which of these solutions can be combined such that conditions for the factorization of the anomaly polynomial required by the Green-Schwarz mechanism hold, as we shall explain in detail later. When two solutions are combined we draw an edge between the respective nodes in the graph. The solutions containing more than two factors will then be subgraphs of this graph that are fully connected. The search for anomaly free models than becomes a matter of determining these subgraphs, also known as ``cliques" in graph theory. 
In this approach, we shall primarily consider the case of $n_T=1$, with no abelian group factors, including all the exceptional groups except for $G_2$ and the classical groups $SU(N), SO(N), Sp(N)$ for $N \geq 10$. In addition, for the classical groups, we shall only consider the fundamental, the adjoint, the second rank symmetric and antisymmetric representations. Restricting the possible representations will facilitate the discussion in section 5, since each representation is treated separately, while the lower bound in the rank $N$ and the exclusion of the group $G_2$ restrict the solution space enough to make a complete enumeration of all the models feasible.

In a second approach, which we refer to as the ``rank method", we start with a given value of $n_T$ and $n_V$, for the groups $G_1\times\cdots \times G_n$, or $G_1\times\cdots \times G_n \times G_R$, where $G_R$ is the gauged $R$-symmetry group which may be $U(1)_R$ or $Sp(1)_R$. This fixes the value of $n_H$. Next we partition this value of $n_H$ into representations of the full gauge group in all possible ways, for groups and representations that we will specify further below.
 
We will not include $U(1)$ factors in $G_1\times\cdots \times G_n$ but we will allow any exceptional and classical groups whose lowest dimensional representations ${\rm dim}_R \ge 10$, which mean the groups $A_N$ for $N=9,10$, and $B_N, C_N, D_N$ for $5\le N\le 10$.  We will also allow some cases with $n_T\ne 1$. In the case of $n_T >1$, we will employ the generalization of the Green-Schwarz anomaly cancellation mechanism due to Sagnotti \cite{Sagnotti:1992qw}. A significant aspect of the rank method is that we allow all possible representations, including hyperfermions that are simultaneously charged under different group factors, consistent with the condition \eq{R4}.\footnote{Note that simultaneously charged hyperfermions are also allowed in the graphical method.} Only in the case of the exceptional group $G_2$, and the cases of $n_T>1$, we will restrict the search to models with no hypersinglets. 
  
For groups of low rank and low dimensional representations, the partition of $n_H$ discussed above amounts to solving a Diophantine equation, and it can have solutions as many as hundreds of thousands, even millions. Once the partitions are determined, we calculate the resulting anomaly polynomials which can be represented as constant real symmetric $(n+1)\times (n+1)$ matrices for ungauged models, and $(n+2)\times (n+2)$ matrices for gauged models. As we shall see, the generalized anomaly cancellation mechanism requires that this matrix has rank $r\le n_T+1$ and that if $r=n_T+1$ the  nonzero eigenvalues $\lambda_m$ of the anomaly matrix include an eigenvalue $\lambda_0$ such that $\lambda_0\lambda_m <0$ for $m>0$. As we shall see, this is consistent with $SO(1,n_T)$ symmetry which is the isometry of the coset parametrized by the $n_T$ dilatonic scalar fields. Finally, as the the parameters that enter the anomaly matrix enter the gauge kinetic terms, we shall require that these kinetic terms have the correct sign for a region in the moduli space of the dilatons.

In summary, we find a large number of ungauged models with $n_T=0,1$ which are anomaly free by the Green-Schwarz-Sagnotti mechanism. One key result is the derivation of a bound on the rank of the gauge groups. We also find that the $R$-symmetry gauging has profound effect on the solution space in the sense that the possible anomaly free models with $R$-symmetry gauging reduces drastically. For example, in the case of $Sp(1)_R$ gauged models, of the tens of thousands of partitions we checked, none yield factorization of the anomaly polynomial. 

Among a very large number of possible models in which the condition \eq{R4} is satisfied, and therefore free from gravitation anomalies, we confirm that for groups of the form $G_1\times G_2\times G_R$ only the three models mentioned above, namely those with $E_6\times E_7\times U(1)_R, G_2\times E_7\times U(1)_R$ and $F_4\times Sp(9)\times U(1)_R$ are free from all local anomalies. Here, extending the search to groups of the form $G_1\times G_2\times G_3\times G_R$ we find that, again among a very large number of possible models that satisfy \eq{R4}, we find that only six models  are free from all local anomalies! Their symmetry groups and the hyperfermion representation contents are as follows:
\allowdisplaybreaks
\begin{align}
& E_7\times E_8\times SO(20)\times U(1)_R: & &(56,1,20)_0 + (1,1,512)_0\ ,
\label{d1}\w2
& E_6\times E_6 \times Sp(5)\times U(1)_R: & & (78,1,10_0)+ (1,1,132)_0\ ,
\label{d2}\w2
& E_7\times SU(10) \times SU(10) \times U(1)_R: && (1,10,45)_0+(1,252,1)_0\ ,
\label{d3}\w2
& F_4\times SO(13)\times Sp(7)\times U(1)_R: & & (26,1,14)_0+(1,13,14)_0+(1,1,350)_0 
\nn\\
&&& + (1,64,1)_0\ ,
\label{d4}\w2
& Sp(5)\times Sp(6)\times SO(12)\times U(1)_R:  &&  (10,1,12)_0+ (44,12,1)_0 +(1,1,32)_0
\nn\\
&&& + (1,208,1)_0,
\label{d5}\w2
& Sp(5)\times Sp(6)\times SO(13)\times U(1)_R: &&  (10,1,13)_0 + (10,65,1)_0 + (132,1,1)_0\ ,
\label{d6}
\end{align}

The $2N$-plet of $Sp(N)$, the $56$-plet of $E_7$, the $32$-plet of $SO(12)$, the $64$-plet of $SO(13)$, the $512$-plet of SO(20), the $132$-plet of $Sp(5)$, the $65$-plet and $208$-plet of of $Sp(6)$, and the $350$-plet of $Sp(7)$, are pseudo-real. The number of hypermultiplets $n_h= n_V+244$, and the couplings of the hypermultiplets are governed by the quaternionic K\"ahler coset $Sp(n_H,1)/(Sp(n_H)\times Sp(1)_R)$. The gauge group $G_1\times G_2\times U(1)_R \subset Sp(n_H)\times Sp(1)_R$. 

In section 2 we note the contribution of the chiral fields of matter coupled $(1,0)$ supergravity to the 8-form anomaly polynomial, to set our notations, and to highlight the consequences of the $R$-symmetry gauging. We recall the constraints on the field content arising from the requirement of anomaly cancellation by the Green-Schwarz-Sagnotti mechanism, as well as the conditions for the the gauge field kinetic terms to be ghost-free. In section 3, we describe the bosonic part of the tensor-Yang-Mills sector of the (pseudo)action for $(1,0)$ supergravity in $6D$, and illustrate how its gauge variation gives the anomaly. In this section we emphasize the arbitrariness of the coefficients in front of the Chern-Simons modifications of the 3-form field strength, and how they get fixed by the consideration of the Green-Schwarz-Sagnotti anomaly cancellation mechanism.  We describe the solutions off the stated constraints by the rank method in section 4, and by the graphical method in section 5. We summarize the salient features of the remarkably anomaly free six new $R$-symmetry gauged models in section 6. Further comments and conclusions are given in section 5. The results of the search for consistent theories, i.e. namely free from all local anomalies, and with ghost-free gauge field kinetic terms, are provided in the appendices.

{\it Note Added:} While finishing this paper and preparing it for submission a paper appeared, arXiv:2311.00868, which does have overlap with our results. The main difference is that we also consider six-dimensional gauged supergravities, finding new anomaly free models, besides ungauged models which is the focus of arXiv:2311.00868. In addition, we provide a complete enumeration of the anomaly free theories under the given restrictions. Finally, although we didn't find any models with hypermultiplets
 simultaneously charged under three or more factors of the gauge group, this possibility was not excluded a priori.

\section{The anomaly polynomials and consistency conditions}

 The $(1,0)$ super-Poincar\'e algebra in $6D$ admits the following multiplets
\be
\underbrace{ \left(e_\m^m, \psi_{\m +}^A, B_{\m\n}^{-} \right)}_{\rm graviton}\ ,\qquad \underbrace{\left(B_{\m\n}^{+}, \chi_{-}^A,\varphi  \right)}_{\rm tensor}\ ,\qquad \underbrace{\left(A_\m, \lambda_{+}^A\right)}_{\rm vector}\ ,\qquad \underbrace{\left(4\phi,\psi_{-} \right)}_{\rm hyper}\ .
\ee
The two-form potentials, $B_{\m\n}^{\pm}$, have (anti-)self-dual field strengths. The spinors are symplectic Majorana-Weyl, $A=1,2$ labels  doublet of  $R$ symmetry group, and  chiralities of the fermions are denoted by $\pm$. 

\subsubsection*{$\mathbf{U(1)_{R}}$ gauging}

Let us consider $(1,0), 6D$ supergravity with $U(1)_R$ R-symmetry gauged, and coupled to  $n_T$ tensor, $n_V$ vector and $n_H$ hypermultiplets such that the total gauge symmetry group is 
\be
G=G_1\times\cdots \times G_n\times U(1)_R\ ,
\label{g1}
\ee
where $G_1\times\cdots \times G_n$ is semi-simple. The contributions of the chiral fields to the anomaly polynomial are as follows\footnote{ We are using the conventions of \cite{Avramis:2005hc} with the exception that the fermions and the field strengths of the 2-form potentials have opposite chirality/duality assignments here. The field strengths are Lie algebra valued and anti-Hermitian as in \cite{Avramis:2005hc}}
\bea
P(\psi_{\mu +}) &=& c\Big( \frac{245}{360} \tr R^4 -\frac{43}{288} (\tr R^2)^2
+\frac{19}{6} F^2 \tr R^2 +\frac{10}{3} F^4\Big) \ ,
\label{gravitino}\w2
P(\chi_-) &=& - c n_T \Big( \frac{1}{360} \tr R^4 + \frac{1}{288} (\tr R^2)^2 -\frac16 F^2 \tr R^2 +\frac23 F^4 \Big)\ ,
\label{dilatino}\w2
P(B_{\mu\nu}) &=& c (n_T-1) \Big( -\frac{28}{360} \tr R^4 + \frac{1}{36} (\tr R^2)^2 \Big)\ ,
\label{tensor)}\w2
P(\lambda_+) &=& c \Big\{ n_V  \Big[ \frac{1}{360} \tr R^4 +\frac{1}{288} (\tr R^2)^2\Big]
\nn\w2
&& -\frac16  \tr R^2  \left( n_V F^2 +  \sum_{i=1}^n \Tr F_i^2  \right) + 4 F^2 \sum_{i=1}^n \Tr F_i^2 
\nn\w2
&& +\frac23 \left(n_V F^4 + \sum_{i=1}^n \Tr F_i^4  \right) \Big\}\ ,
\label{gaugino1}\w2
P(\psi_-) &=& c\Big\{ n_H \Big[ -\frac{1}{360} \tr R^4 -\frac{1}{288} (\tr R^2)^2 \Big] 
\nn\w2
&&  + \sum_{i=1}^n \sum_{R} \Big(  \frac16 \tr R^2 \tr_R F_i^2 -\frac23 \tr_R F_i^4\Big)\Big\}\ ,
\label{hyperino}
\eea
where $\Tr$ and $\tr_R$ are traces in the adjoint and representation $R$, respectively, and\footnote{According to \cite{Alvarez-Gaume:1984zlq,Bilal:2008qx} the anomaly polynomial is equal to $( 2\pi\times \mbox{index density})$. Using the index density formula given in eq. (2.67) of \cite{Alvarez-Gaume:1984zlq}, gives the expressions above, which differ by the factor $c$ compared to the expressions used in \cite{Avramis:2005hc}.}
\be
U(1)_R: \qquad n_V = 1+ \sum_{i=1}^n {\rm dim}\, G_i\ , \qquad c= \frac{1}{16\times (2\pi)^3} \ .
\label{defc}
\ee
The hyperfermions are  $R$-symmetry singlets by supersymmetry, and they may carry representations of the factor groups simultaneously. If the representation carried by the hyperfermions is pseudo-real, denoting it by $\psi^a$, they obey the symplectic-Majorana-Weyl condition, $(\psi^a)^\star = \epsilon_{ab} \psi^b$ in a suitable representation of Dirac gamma-matrices, and where $\epsilon_{ab}$ is the symplectic invariant tensor. Consequently, their contribution $P(\psi_-)$ to the anomaly polynomial must be divided by half. If the hyperfermion representation is complex, then we work with Weyl spinors, and consequently the $R$-symmetry is $U(1)_R$. Finally, if the hyperfermion representation is real, then we introduce a doublet index $A'=1,2$ to define the pseudo-reality condition 
\be 
\big(\psi^{IA'} \big)^\star = \epsilon_{A'B'} \psi^{IB'}\ ,
\ee
where $I$ labels the real representations. In referring to such hyperfermions we will simply count the real representation, thus ignoring the index $A'$, and therefore we do not divide by a factor of two. Note that the index $A'$ is associated with a global $Sp(1)$ symmetry, and therefore there are no anomalies associated with it. 

Summing up the above contributions, one finds the well known result that the fatal $\tr R^4$ terms cancel provided that the condition \eq{R4} holds. In analysing the gauge anomalies we need the group theoretical relations
\begin{align}
\tr_R F_i^2 &= a_R^i\, \tr F_i^2\ ,
\nn\w2
\tr_R F_i^4 &= b_R^i \tr F_i^4 +  c_R^i \tr (F_i^2)^2 \ ,
\label{gtr}
\end{align}
where $F_i$ is the field strength associated with the group factor $G_i$, and $\tr$ is the trace in the fundamental representations. These relations define the coefficients $a_R^i, b_R^i$ and $c_R^i$ and they can be calculated from the (modified) Casimir invariants of the Lie algebra \cite{Okubo:1981td}. The $2n^{\text{th}}$ order index of a given representation is defined to be
\begin{equation}
    l_{2n}(R) = \sum_{w\in \Lambda_{R}} (w,w)^{n}
\end{equation}
where $w$ are the weights in the weight lattice $\Lambda_{R}$ of the representation. The inner product $(\cdot,\cdot)$ is the standard Killing metric on the weight space. The coefficient $a_{R}$ is the index $l_{2}(R)$, normalised such that it gives 1 on the fundamental representation. The coefficient $b_{R}$ is not, however, $l_{4}(R)$ but the modified fourth order index $l'_{4}(R)$ which we give below. The coefficients $a_{R}$, $b_{R}$, $c_{R}$ for any representation $R$ of an arbitrary group\footnote{This is with the exception of the $D_{4}$ algebra, where the expression for $l'_{4}(R)$ above vanishes identically. This is despite the fact that $\textrm{SO}(8)$ has non-trivial fourth order Casimirs. This peculiarity arises because $\textrm{SO}(8)$ has 3 independent fourth order Casimirs coming from triality. For details on this case see \cite{Okubo:1981td}. The expression $l'_{4}(R)$ is 0 for $\SU{2}$, $\SU{3}$ and the exceptional groups, as expected since they have no independent fourth order Casimirs and hence we take the coefficient $b_{R}$ to be 0.} of dimension $d$, rank $r$ and with fundamental and adjoint representations $F$ and $A$ respectively, are given by the following formulae.
\begin{align}
    a_{R} &= \frac{l_{2}(R)}{l_{2}(F)} \, , \\
    b_{R} &= \frac{l'_{4}(R)}{l'_{4}(F)} \, , \qquad l'_{4}(R) = l_{4}(R) - \frac{r+2}{r} \frac{d}{d+2} \left( \frac{l_{2}(R)}{\dim R} - \frac{1}{6}\frac{l_{2}(A)}{d} \right) l_{2}(R) \, , \\
    c_{R} &= \frac{3}{2+d} \left( a_{R}^{2}\left(\frac{d}{\dim R} - \frac{1}{6} \frac{l_{2}(A)}{l_{2}(R)}\right) - b_{R}\left( \frac{d}{\dim F} - \frac{1}{6} \frac{l_{2}(A)}{l_{2}(F)}\right) \right)\, .
\end{align}

Using the relation \eq{gtr}, the vanishing of the fatal  $\tr F_i^4$ terms in the total anomaly polynomial requires that
\be
B_i \equiv  \big(b_i^{\rm adj} - \sum_R n_i^R b_i^R \big) =0\ , \quad 
\ee
for each group factor, and where $n_i^R$ is the number of representations $R$ of the group factor $G_i$ carried by the hyperfermions. Using this relation, and denoting by $n_{RS}^{ij}$ the number of hyperfermions that carry the representations $R$ and $S$ of the group factors $G_i$ and $G_j\,  (i<j)$ simultaneously, the sum of all the anomaly polynomials listed above yields
\begin{align}
 I_8 &= \frac{1}{8} c (n_T-9) (\tr R^2)^2 +\frac16 c \left( n_T - n_V +19 \right) F^2 \tr R^2
\nn\w2
& +4 c F^2  \sum_{i=1}^n a_i^{\rm adj} \tr F_i^2 -4 c\sum_{i<j} \sum_{R,S} n^{ij}_{RS} \tr_R F_i^2 \tr_S F_j^2 
\nn\w2
& +\frac23 c (n_V-n_T +5 ) F^4  +\frac23 c \sum_{i=1}^n \left( c_i^{\rm adj} -\sum_R n_R^i c_R^i \right) (\tr F_i^2)^2
\nn\w2
& -\frac16 c \tr R^2 \sum_{i=1}^n \left( a_i^{\rm adj} -\sum_R n_R^i a_R^i \right) \tr F_i^2
\label{me1}
\end{align}
The anomaly polynomial for the ungauged case can be obtained from above by setting $F\to 0$.

\subsubsection*{$\mathbf{Sp(1)_{R}}$ gauging}

Next, we consider the gauging of the $R$ symmetry group $Sp(1)_R$ such that the gauge If we chose the full gauge group is
\be
G=G_1\times\cdots \times G_n\times Sp(1)_R\ .
\label{g2}
\ee
Now we let $F^2 \to \tr F^2$ and $F^4 \to \tr F^4$ in \eq{gravitino} and \eq{dilatino}, where the traces are in the fundamental representation of $Sp(1)_R$, and \eq{hyperino} remains the same. Note that, now we have
\be
Sp(1)_R: \qquad n_V = 3+ \sum_{i=1}^n {\rm dim}\, G_i\ .
\ee
The gauginos for the R-symmetry are in the ${\bf 3} \times {\bf 2} $ representation of $Sp(1)_R$ and they count as three complex fermions, in the formula for $n_V$ above. In calculating their contribution to the gauge anomaly we need to take into account the fact that ${\bf 3} \times {\bf 2}= {\bf 4} + {\bf 2}$ under $Sp(1)_R$. Thus, the total contribution of the gauginos to the anomaly polynomial now takes the form
\bea
P(\lambda_+) &=& c n_V  \Big[ \frac{1}{360} \tr R^4 +\frac{1}{288} (\tr R^2)^2\Big]
\nn\w2
&& -\frac16 c \tr R^2 \left(   (n_V-3) \tr F^2 + \tr F^2+\tr_4 F^2+ \sum_{i=1}^n \Tr F_i^2  \right)  
\label{gaugino2}\w2
&& + 4c \tr F^2 \sum_{i=1}^n \Tr F_i^2
 +\frac23 c\left( (n_V-2) \tr F^4 + \tr_4 F^4+\sum_{i=1}^n \Tr F_i^4   \right)\ ,
\nn
\eea
where $F$ is the $Sp(1)_R$ valued field strength, and $\tr$ and $\tr_4 $ reefer to traces in the fundamental and ${\bf 4}$-plet representations of $Sp(1)_R$. Taking into account the following group theoretic relations in the equation above,
\be
\tr_4 F^4 = 41 (\tr F^2)^2\ ,\quad \tr F^4 = \frac12 (\tr F^2)^2\ ,\quad \tr_4 F^2 = 10 \tr F^2\ ,
\ee
and summing up all the contributions to the anomaly polynomial, we obtain\footnote{Here, the $\tr F^{2} \Tr F_{i}^{2}$ term corrects a typo in equation (2.7) of \cite{Avramis:2005hc}.}
\begin{align}
I_8 &= \frac{1}{8} c (n_T-9) (\tr R^2)^2 +\frac16 c \left( n_T - n_V +11 \right) \tr F^2 \tr R^2
\nn\w2
& +\frac13 c (n_V-n_T +85 ) (\tr F^2)^2  +\frac23 c \sum_{i=1}^n \left( c_i^{\rm adj} -\sum_R n_R^i c_R^i \right) (\tr F_i^2)^2
\nn\w2
& +4c \tr F^2  \sum_{i=1}^n \Tr F_i^2  -4c\sum_{i<j} \sum_{R,S} n^{ij}_{RS} \tr_R F_i^2 \tr_S F_j^2 -\frac16 c\tr R^2 \sum_{i=1}^n \left( a_i^{\rm adj} -\sum_R n_R^i a_R^i \right) \tr F_i^2
\label{me2}
\end{align}

\subsection*{Factorization conditions}

The Green-Schwarz-Sagnotti mechanism requires that the anomaly polynomial factorizes as 
\footnote{
 If the gauge group includes the abelian R-symmetry group $U(1)_R$, the Green-Schwarz anomaly cancellation mechanism has yet another generalization \cite{Berkooz:1996iz} in which the anomaly polynomial may have an additional term of the form $F\wedge X_6$ where $F$ is the $U(1)_R$ field strength, and $X_6$ is a gauge invariant 6-form polynomial built out of the Riemann curvature 2-form and the field strengths associated with the gauge group. Such a contribution to the anomaly can be cancelled by coupling linear multiplets to (1,0) supergravity. A linear multiplet has a 4-form gauge potential and 3 scalars as its bosonic field content. Dualizing the 4-form potential to a scalar field, and as discussed in detail in \cite{Park:2011wv}, it gets eaten up by the abelian vector field which becomes massive. We shall not consider here such models in which the $U(1)_R$ symmetry is broken. 
 }
\be
I_8 = \frac{c}{2} \eta_{\alpha\beta} Y^\alpha Y^\beta\ ,\qquad \alpha=0,1,...,n_T\ ,
\label{I8}
\ee
where $\eta_{\alpha\beta}$ is the invariant tensor of $SO(n_T,1)$, and for the gauge group \eq{g1} 
\be
Y^\alpha = \frac12  a^\alpha \tr R^2 +\sum_{z=1}^{n+1} b_z^\alpha \left(\frac{2}{\lambda_z} \tr F_{z}^2\right)\ ,
\label{X4}
\ee
where
\be
\lambda_{n+1}=1\ ,\qquad \tr_{n+1} F^2 =\begin{cases} F^2 & \mbox{for}\ U(1)_R \\ \tr F^2 &\mbox{for} \ Sp(1)_R \end{cases}
\ee
The normalization factors $\lambda_i$ for $i=1,...,n$ are fixed such that the smallest topological charge of an embedded $SU(2)$ instanton is $1$, and they are given by
\medskip
\begin{align}
&& A_n && B_n && C_n && D_n && E_6 && E_7 && E_8 && F_4 && G_2 
\nn\\
\lambda && 1 && 2 && 1 && 2 && 6 && 12 && 60 && 6 && 2
\end{align}
The sign in \eq{I8} is crucial. It is determined by the considerations discussed in detail in the next section. Comparison of \eq{I8} with \eq{me1} gives
\bea
a\cdot a &=&  (n_T-9)\ ,
\label{c1}\w2
a\cdot b_i &=& -\frac16  \lambda_i A_i\ ,
\label{c2}\w2
b_i \cdot b_i &=& \frac13  \lambda_i^2 C_i\ ,
\label{c3}\w2
b_i \cdot b_j &=& - \lambda_i\lambda_j A_{ij}\ ,
\label{c4}\w2
a\cdot b_{n+1} &=& \begin{cases} -\frac16 c (n_V-n_T-19) & \text{for} \ U(1)_R\\ -\frac16  (n_V-n_T-11) & \text{for}\ Sp(1)_R \end{cases}
\label{c5}\w2
b_{i} \cdot b_{n+1} &=& \lambda_i a_i^{\rm adj}\ ,
\label{c6}\w2
b_{n+1}\cdot b_{n+1} &=& \begin{cases} -\frac13  (n_T-n_V-5) & \text{for}\ U(1)_R\\  -\frac16 (n_T-n_V-85) & \text{for}\ Sp(1)_R\end{cases}
\label{c7}
\eea
where the inner products are with respect to the metric $\eta_{\alpha\beta}$, and 
\begin{align}
A_i &= a_i^{\rm adj} -\sum_R n_R^i a_R^i\ ,
\label{A1}\w2
C_i &= c_i^{\rm adj} -\sum_R n_R^i c_R^i\ ,
\label{C}\w2
A_{ij} &= \sum_{R,S} n^{ij}_{RS} a^{i}_R a^{j}_S\ , \quad i\ne j\ .
\label{A2}
\end{align}
Defining 
\be
x_0 := \tr R^2\ ,\quad x_z := \tr F_{z}^2\ , 
\ee
we have
\be 
Y^\alpha = \sum_{r=0}^{n+1} v^\alpha_r x_r\ ,\qquad r=(0,i,n+1)\ ,
\ee
where the constants $v_r^\alpha$ are defined as
\be
v_0^\alpha = \frac12 a^\alpha\ ,\qquad v^\alpha_z = \frac{2}{\lambda_z}  b^\alpha_z \ .
\ee
Thus, the total anomaly polynomial takes the form
\be
I_8 = I_{rs} x_r x_s \ ,\quad r=(0,i, n+1) ,\quad i=1,...,n\ ,
\label{I8P}
\ee
where $I_{rs}$, which we refer to as the anomaly matrix, is given by
\be
I_{rs} = \eta_{\alpha\beta} v_r^\alpha v_s^{\beta}\ ,
\ee
Since $\alpha=0,1,...,n_T$, it follows that the factorization of the anomaly polynomial requires that the anomaly matrix has $\text{rank(I)} \le n_T+1$. If $\text{rank(I)} =n_T+1$, then one of the eigenvalues must have opposite sign to all the others.

\section{The (pseudo)action and gauge kinetic terms}
\label{sec:pseudo-action}

The couplings of two-derivative $(1,0)$ supergravity in $6D$ to a single tensor multiplet, $n_V$ vector multiplets and $n_H$ hypermultiplets was given completely in\cite{Nishino:1986dc}. The field equations of multi-tensors coupled to $(1,0)$ supergravity in leading order in fermions were obtained in \cite{Romans:1986er}. The vector fields were included in \cite{Ferrara:1996wv}, and the hypermultiplet couplings as well in \cite{Nishino:1997ff}, where the complete supertransformations were also found. Subsequently, the complete field equations without hypermultiplets were found in \cite{Ferrara:1997gh}. Finally, the results of \cite{Nishino:1997ff,Ferrara:1997gh} were completed to include the higher order fermion terms in \cite{Riccioni:2001bg}, where a pseudo-action was given from which all but the (anti)self duality equations for the 3-form field strengths can be derived\footnote{The equation of motion for the vector fields is subtle because those in \cite{Nishino:1997ff} cannot be derived from a Lagrangian, pseudo or not. The root of this phenomenon lies in the distinction between the covariant versus consistent anomalies \cite{Ferrara:1996wv}.} In this case, the (anti)self-duality equations are to be imposed after varying the pseudo-action with respect to the fields. A proper action in presence of multi-tensor multiplets from which all field equations can be derived can be constructed but at the expense of not having manifest Lorentz invariance. 

In the field equations \cite{Nishino:1997ff} or the (pseudo)action\cite{Riccioni:2001bg}, it is known that there are arbitrary constants which show up which hare not determined by the Noether procedure to establish supersymmetry. It is important to make contact between these constants, and the coefficients $v^\alpha_z$ that appeared in the description of anomaly polynomials and the Green-Schwarz mechanism. To this end, and putting aside the four derivative extension of the action for now, we begin by defining the Chern-Simons modified field strength as
\be
H^\alpha_{\mu\nu\rho} = 3\partial_{[\mu} B_{\nu\rho]}^\alpha + \omega^\alpha_{\mu\nu\rho}\ ,
\ee
where $dH^\alpha=X^\alpha $, and 
\bea
\omega^\alpha_{\mu\nu\rho} &=& 6 c^\alpha_z \Tr_z \big( A_{[\mu} \partial_\nu A_{\rho]} + \tfrac23 A_{[\mu}A_\nu A_{\rho]} \big)\ ,
\nn\w2
X^\alpha_{\mu\nu\rho\sigma} &:=& 4\partial_{[\mu} \omega^\alpha_{\nu\rho\sigma]}\ ,
\eea
and $c^\alpha_z$ are {\it arbitrary} constants. Another ingredient that we need is the representative of the coset $SO(1,n_T)/SO(n_T)$ parametrized by the dilatons. Denoted by $L_\alpha{}^A= (L_\alpha{}^0, L_\alpha{}^a)$, and choosing given that it is an element of $SO(n_T,1)$ means that 
\be
\eta_{\alpha\beta} =- L_\alpha{}^0 L_\beta{}^0 + L_\alpha{}^a L_\beta{}^a\ ,
\ee
where $a=1,...,n_T$ and $\eta =\text{diag}\,(-,+,+,...,+)$. In terms of this, we can define the positive definite scalar matrix 
\be
M_{\alpha\beta} = L_\alpha{}^0 L_\beta{}^0 + L_\alpha{}^a L_\beta{}^a\ .
\ee
With the above definitions at hand, and putting aside the hypermultiplets, the bosonic part of the (pseudo)action up to two-derivative terms (thus ignoring the Lorentz Chern-Simons term), and assuming $U(1)$ gauging for concreteness, takes the form 
\bea
e^{-1} {\mathcal L} &=& \frac14 R -\frac{1}{48} M_{\alpha\beta} H_{\mu\nu\rho}^\alpha H^{\mu\nu\rho \beta}-\frac14 \partial_\m L_\alpha \partial^\mu L^\alpha  -\frac14 L_\alpha c^\alpha_z \Tr_z \left(F_{\mu\nu} F^{\mu\nu}\right) 
\nn\w2
&&-\frac12 \left(L_\alpha c^\alpha_{n+1}\right)^{-1}  -\frac{1}{192} \varepsilon^{\m_1...\mu_6} \eta_{\alpha\beta} B_{\mu_1\mu_2}^\alpha X^\beta_{\mu_3...\mu_6}\ ,
\label{ga}
\eea
where $L_\alpha \equiv L_\alpha{}^0$. Requiring perturbative unitarity, that is non-ghostly gauge kinetic terms, imposes the condition
\be
L_\alpha c^\alpha_z  >0\ ,\qquad z=1,...,n+1\ .
\label{gc}
\ee
These conditions may be satisfied in a particular region in the moduli space of the $n_T$ dilatons. Of course, all of the $(n+1)$ conditions must be satisfied simultaneously in that region.

The action of this Lagrangian is neither gauge invariant nor supersymmetric\footnote{The breakdown in supersymmetry is seen at higher order in gauge fermion variations, as shown in \cite{Riccioni:1998th}.}. This is to be expected since the terms that break these invariances are exactly what are needed to restore gauge and supersymmetry in the one-loop corrected effective action.
To display the breakdown in gauge invariance, consider the Yang-Mills field equations that follows from \eq{ga}, with the instruction that the duality condition on $H^\alpha$ is to be used {\it after} varying the action. The bosonic part of this duality condition is given by\footnote{Note that this implies the duality relations: $L_\alpha H_{\mu\nu\rho}^\alpha = -(1/3!) \varepsilon_{\mu\nu\rho\sigma\lambda\tau} L_\alpha H^{\sigma\lambda\tau \alpha}$ and $L_\alpha^a H_{\mu\nu\rho}^\alpha = (1/3!) \varepsilon_{\mu\nu\rho\sigma\lambda\tau} L_\alpha^a H^{\sigma\lambda\tau \alpha}$. }
\be\label{eq:duality_condition}
M_{\alpha\beta} H_{\mu\nu\rho}^\beta = \frac{1}{3!} \varepsilon_{\mu\nu\rho\sigma\lambda\tau}\, \eta_{\alpha\beta} H^{\sigma\lambda\tau\,\beta}\ .
\ee
The resulting Yang-Mills field equation is
\be
D_\mu \left(L_\alpha c^\alpha_z F^{\mu\nu}_z\right) = J^\nu\ ,
\ee
where 
\be
\begin{aligned}
J_z^\mu &= \varepsilon^{\mu\nu\rho\sigma\lambda\tau}\eta_{\alpha\beta}c_{z}^{\beta} \Big( \frac{1}{12}  H^{\alpha}_{\nu\rho\sigma}F_{z\,\lambda\tau} -\frac{1}{24} \omega^\alpha_{\mu\nu\rho} F_{z\,\lambda\tau} -\frac{1}{16} c^\alpha_{z'} A_{z'\,\nu} \Tr_z(F_{\rho\sigma} F_{\lambda\tau}) \Big) \ .
\end{aligned}
\ee
In obtaining this result we have used the duality condition \eqref{eq:duality_condition}, after varying the action. From this, we find 
\be
{\cal A}= \int \epsilon_z D_{\mu}J_z^{\mu} = 
\frac14 \eta_{\alpha\beta}  \int \left(c^\alpha_z \Tr_z F^2 \right)\,\left( \epsilon_{z'}\,c^\beta_{z'}  dA_{z'} \right)\ ,
\ee
where $\epsilon(x)$ is the Yang-Mills gauge parameter, and {\cal A} is the ``consistent" anomaly, in the sense that it obeys the Wess-Zumino consistency condition.
On the other hand, as explained in detail in \cite{Alvarez-Gaume:1984zlq,Bilal:2008qx}, the anomaly as the gauge variation of the one-loop effective action $\Gamma$ is related to the anomaly polynomial as
\be
\delta \Gamma = {\cal A} =  \frac{1}{4! (2\pi)^3} \int \epsilon Q_6^1 
=\frac{2c}{3} \int \epsilon Q_6^1 = I_6^1\ ,
\ee
where $Q_6^1$ is related to $P_8= \Tr F^4$, and $I_6^1$ to $I_8$, by descent. The last two equalities follow from \eq{defc} and \eq{gaugino1}. Comparing this equation with \eq{I8}, it follows that the constants $c^\alpha_z$ that occur in the supergravity action are related to the coefficients $v^\alpha_z$ that arise in the factorization of he anomaly polynomial as 
\be 
c^\alpha_z = \sqrt{2c} \, v^\alpha_z\ .
\ee

\subsection*{The case of $n_T=1$}

In the case of $n_T=1$, instead of using $\eta_{\alpha\beta}$, it is often convenient to work with its similarity transformed form given by
\be
\Omega_{\alpha\beta} = \left(\rho^T \eta \rho\right)_{\alpha\beta} = \begin{pmatrix} 0 & 1 \\ 1 & 0 \end{pmatrix}\ ,\qquad \rho= \frac{1}{\sqrt 2}\begin{pmatrix} 1 & -1 \\ 1 & 1 \end{pmatrix}\ .
\label{Omega}
\ee
In this case, we can represent $L_\alpha$ by
\be
L_\alpha= \frac{1}{\sqrt{2}}\begin{pmatrix} e^{-\phi} \\ -e^{\phi}\end{pmatrix}\ ,\quad \alpha=0,1\ .
\ee
Consequently the condition \eq{gc} for the positivity of the gauge kinetic terms reads
\be
-e^\phi c^0_z + e^{-\phi} c^1_z >0\ \qquad z=1,...,n+1\ .
\ee
We need there to exist some $\phi$ for which the above holds for all $z$. It is a necessary (but not sufficient) condition that at least one of $c^0_z<0$ or $c^1_z>0$ for all $z$. 

Note that, only in the $n_{T} = 1$ case, the overall sign of the factorisation of the anomaly polynomial \eqref{X4} is irrelevant. This is because there is a similarity transformation which takes $\Omega$ to $-\Omega$. Indeed, we find that
\be
-\Omega_{\alpha\beta} = (\sigma^T \Omega \sigma)_{\alpha\beta}\, , \qquad \sigma = \begin{pmatrix}
    1 & 0 \\ 0 & -1
\end{pmatrix}
\ee
To ensure that the anomaly polynomial remains invariant under this change, we need to multiply the coefficients $c^{\alpha}_{z}$ by $\sigma^{T}$. That is, the anomaly polynomial factorises with $X^{\alpha} = \sigma^{\alpha}{}_{\beta}c^{\beta}_{z} \Tr_{z} F^{2}$. Moreover, the ghost conditions should be determined with $L_{\alpha} = (e^{-\phi},e^{\phi})$. This subtlety becomes important when determining the anomaly free and ghost free theories via the `rank method' described in the previous section.

Continuing to consider the case of $n_T=1$, the single self-dual and single anti-self-dual 3-form field strengths can be combined to one which is free from duality conditions. In that case, an action from which all field equations can be obtained by setting $B^0_{\mu\nu}\equiv B_{\mu\nu}$ and $B^1_{\mu\nu}=0$, and relax the duality condition on $H^0\equiv H$. From \eq{ga}, this gives
\bea
e^{-1} {\mathcal L} &=& \frac14 R -\frac{1}{12} e^{2\phi} H_{\mu\nu\rho}^\alpha H^{\mu\nu\rho \beta}-\frac14 \partial_\m L_\alpha \partial^\mu L^\alpha  -\frac14 L_\alpha c^\alpha_z \Tr_z \left(F_{\mu\nu} F^{\mu\nu}\right) 
\nn\w2
&&-\frac12 \left(L_\alpha c^\alpha_{n+1}\right)^{-1}  -\frac{1}{16} \varepsilon^{\m_1...\mu_6} B_{\mu_1\mu_2}X^1_{\mu_3...\mu_6}\ ,
\label{ga1}
\eea
Note that while both $c^0_z$ and $c^1_z$ appear in the Yang-Mills kinetic term, only $v^1_z$ arises in the Chern-Simons term $B\wedge X$ in the action. As before, the action is neither gauge invariant, nor supersymmetric. To write down a gauge invariant and supersymmetric classical Lagrangian, we either set $c^1_z=0$, in which case we obtain the usual formulation of the theory in which $H=dB^0 + \omega^0$, and no Chern-Simons term in the Lagrangian, or we set $c^0_z=0$ in which case we obtain the dual formulation of the theory in which $H^0= dB^0$, and a Chern-Simons term $B^1 \wedge X^0$ appears. Including both terms breaks the classical symmetry of the theory but restores all gauge invariance at the 1-loop level.

\subsection*{The case of $n_T = 0$}

For $n_T = 0$, the inner product is given by $\eta = -1$. Hence, for an anomaly free theory, we require that the anomaly polynomial factorises as $I_8 = - \tfrac{1}{2} X\wedge X$ for some 4-form $X$. While there is no proper action in this case, we can use the pseudo-action to determine the ghost free conditions.

Firstly, we require an $L$ such that $L\cdot L = -1$. The only freedom in this choice is a sign ambiguity which we can fix without loss of generality. We take
\begin{equation}
    L = 1
\end{equation}
The ghost-free condition for $n_T=0$ says that
\begin{equation}
    L\cdot c_{z} = -c_{z} > 0 \, , \quad z = 1,...,n+1
\end{equation}

\subsection*{The case of $n_T=2$}

For $n_T = 2$, we work with the diagonal $\eta_{\alpha\beta}$. We can again use the pseudo-action to determine the ghost-free conditions, for which we require some $L_{\alpha}$ such that $\eta_{\alpha\beta}L^{\alpha}L^{\beta} = -1$. We can parameterise $L_{\alpha}$, which now takes values in the two dimensional hyperboloid $\textrm{SO}(1,2)/\textrm{SO}(2)$, with two scalars $\phi, \theta$ as
\begin{equation}
    L_{\alpha} = \left( \begin{array}{c}
        \cosh\phi \\
        \sinh\phi\,\cos\theta \\
        \sinh\phi\,\sin\theta
    \end{array} \right)\, , \quad \alpha=0,1,2
\end{equation}
Condition \eq{gc} for positive kinetic terms then translates to
\begin{equation}\label{eq:nt=2_ghost_free_condition}
\begin{aligned}
    c_{z}^{0} &< \sqrt{ (c_{z}^{1})^{2} + (c_{z}^{2})^{2} }\tanh\phi \, \sin(\theta + \alpha_{z}) \ ,  
\end{aligned}
\end{equation}
where $\alpha_{z}$ is such that
\begin{equation}
\sqrt{ (v_{z}^{1})^{2} + (v_{z}^{2})^{2} }\sin\alpha_{z} = v^{1}_{z} \ , \qquad \sqrt{ (v_{z}^{1})^{2} + (v_{z}^{2})^{2} }\cos\alpha_{z} = v^{2}_{z}\ .
\end{equation}
Note that condition \eqref{eq:nt=2_ghost_free_condition} gives the following necessary (but not sufficient) condition for a ghost free model.
\begin{equation}
    v_{z}^{0}< \sqrt{ (v_{z}^{1})^{2} + (v_{z}^{2})^{2} } \ , \quad \forall\, z \ .
\end{equation}

\section{Solutions by means of the rank method}

\subsection{The method}

In this method, we start with a given value of $n_T$ and $n_V$, corresponding to one of the following groups 
 \begin{align}
 & G_1\times \cdots \times G_n\ ,\quad n=1,2,3\ .
 \label{gs}\w2
 & G_1\times \cdots \times G_n\times G_R\ ,\quad n=1,2,3\ .
 \label{gs2}
 \end{align}
We shall assume that $G_i, i=1,2,3$ do not contain abelian factors, and are to be picked from the following set of groups:
\be
\text{Rank method:} \quad 
\begin{cases}
G_2, F_4, E_6, E_7, E_8 \\
A_N\ , \quad N=9,10  \\
B_N, C_N, D_N\ ,\quad 5 \le N \le 10 \\
\end{cases}
\label{pm}
\ee
Next, we partition the resulting $n_H$ in all possible ways into  representations of the gauge group. We shall allow all possible representations of $G_1 \times\dots \times G_n$, including those which carry simultaneous representations\footnote{{Hyperfermions charged non-trivially under 3 simple gauge group factors need not be considered here as, for the groups we have restricted to, they would violate the bound $n_H - n_V \leq 273- 29 n_T$. In the graphical method, to be described later, we allow hypermultiplets charged under any number of groups in principal. However, we find no solutions with hypers charged under more than 2 groups.}} except the case of $G_2\times G_2\times G_2$ in which case we shall allow for each $G_2$ the irreps $(\textbf{7},\textbf{14}, \textbf{27}, \textbf{64})$, in particular excluding the singlets. The restrictions on the rank of the classical groups is chosen such that their smallest non-trivial representation has dimension ${\rm dim}_R\ge 10$ and their rank is $\leq 10$. We make these restrictions on the groups and representations to keep the number of partitions manageable.

Once the partitions of $n_H$ are determined, we calculate the resulting anomaly polynomials which can be represented as constant $(n+1)\times (n+1)$ matrices for \eq{gs}, or $(n+2)\times (n+2)$ matrices for \eq{gs2}. As explained in Section 2, the generalized anomaly cancellation mechanism requires that this matrix has rank $r\le n_T+1$ and that if $r=n_T+1$, the  nonzero eigenvalues $\lambda_m$ of the anomaly matrix must include an eigenvalue $\lambda_0$ such that $\lambda_0\lambda_m <0$ for $m>0$, consistent with $SO(1,n_T)$ symmetry of the coset parametrized by the $n_T$ dilatonic scalar fields. 

In the ungauged case, if $n \leq n_{T}$, then the rank condition is automatically satisfied, and similarly if $n \leq n_{T} - 1$ in the gauged case. Factorisation only requires that the eigenvalues have Lorentzian signature. This is a comparatively weaker condition and hence we would expect many solutions in these cases. We will therefore only consider values of $n_{T}$ which lead to non-trivial rank reduction, except for the case $n_T = n =1$. To be more precise, for $n=1$ we consider $n_T = 0,1$; for $n=2$ we consider $n_T = 0,1$ (and 2 for the gauged case); and for $n=3$ we consider $n_T = 1$ (and 0 for the gauged case).

There is an additional condition for the physically acceptable anomaly free models coming from the requirement of perturbative unitarity. Namely, as parameters that enter the anomaly matrix enter the gauge kinetic terms as in \eq{ga}, we also require that these kinetic terms have the correct sign for a region in the moduli space of the dilatons, by imposing the conditions \eq{gc}.

In the $R$-symmetry gauged models we consider $Sp(1)_R$ of $U(1)_R$ as an additional factor. In the rank method, we allow all possible representations for the groups stated above, allowed by the constraint \eq{R4}, except that in the cases involving the exceptional group $G_2$, we shall consider only the lowest lying nontrivial four representations, not allowing singlets. Furthermore,we shall impose the condition\footnote{ In a modern treatment of global anomalies, we say that they are measured by $\mathrm{Hom}(\Omega^{\text{spin}}_{7}(BG_{2}),U(1))$. This leads to an apparent confusion given that $\Omega_{7}^{\text{spin}}(BG_{2}) = 0$ which would seem to indicate that there are no global anomalies provided the local anomalies vanish. A reconciliation of this apparent inconsistency comes from a proper treatment of the integral nature of the Green-Schwarz anomaly cancelling term which implies that \eqref{gac} is still necessary. For more details, see \cite{Lee:2020ewl}.}
\be
1-4\sum_{i} n^i_R c^i_R= 0\ {\rm mod}\ 3\ ,
\label{gac}
\ee
as a necessary condition for freedom from global anomalies due to the presence of the group $G_2$ \cite{Bershadsky:1997sb}. In this paper we are not studying the most general possible global anomalies, and as such this condition though necessary, may not be sufficient one from complete freedom from global anomalies.

In the case of $R$-symmetry ungauged models, if there exists any group factor $G_{1}$ such that among the hypermultiplet representations there is only an adjoint representation of that that group, then eqs. \eq{c2}-\eq{c3} imply that $ a\cdot b_1=0$ and $b_1\cdot b_1=0$. Thus, if $n_T<9$, this implies that $b_1=0$. Therefore, we shall consider this model to be equivalent to one with gauge group $G_2\times \cdots \times G_n$. In the case of $R$-symmetry gauged models, the extra condition \eq{c6} arises and it is inconsistent with the vanishing of $b_1$. The signature of $\eta_{\alpha\beta}$ plays an important role in this conclusion. The relevance of this signature is also emphasized in section \ref{sec:pseudo-action}. In summary, we shall not consider models with a group factor whose adjoint representation alone appears in the hyperfermion content, as they are either trivial or inconsistent.

\subsection{Comments on the solutions}\label{sec:comments_on_solns}

The solutions we find for the ungauged models with $n_T = 1$ are contained in appendices A1-A6, and with $n_T = 0$ in appendix B. In particular, in appendix A1 we leave out the solutions which were already found in \cite{Avramis:2005hc}\footnote{As mentioned in Appendix A.1, the solutions for $Sp(N)$ are corrected, some missing solutions are found, few incorrect solutions as well solutions which harbor ghosts are identified. }  It is useful to record that \cite{Avramis:2005hc} considered the following groups
\be
\text{Ref. \cite{Avramis:2005hc}:}\quad 
\begin{cases}
F_4, E_6, E_7, E_8\\
A_N\ , \quad 4\le N\le 31\\
B_N, D_N\ ,\quad5\le N\le 32\\
C_N,\quad  4\le N\le 34
\end{cases}
\label{ak}
\ee
for the $R$-symmetry ungauged models, and $Sp(1)_R$ or $U(1)_R$ gaugings, in which case the group $G_2$ was also considered. In \cite{Avramis:2005hc}, and in the case of ungauged models only groups of the form $E\times E$ and $E\times C$, where $E$ and $C$ are exceptional and classical groups, and in the case of $R$-symmetry gauged models an additional factor of exceptional group $G_2$ was considered. The following representations were considered in \cite{Avramis:2005hc}:
\begin{align}
E_8: & \qquad 248
\nn\\
E_7: &\qquad  56^\star, 132, 912^\star
\nn\\
E_6: & \qquad  27, 78, 351, 351^\prime, 650
\nn\\
F_4: &\qquad   26, 52, 273, 324
\nn\\
G_2: &\qquad 7, 14, 27, 64
\nn\\
SU(N): &\qquad   N, N^2-1, \frac{N(N-1)}{2}, \frac{N(N+1)}{2}
\nn\\ 
SO(N): &\qquad   N, \frac{N(N-1)}{2},{ 2^{[\frac{N+1}{2}]-1}}^\star
\nn\\
Sp(N): &\qquad   2N^\star, N(2N+1), N(2N-1)-1
\label{akr}
\end{align}
The representations with $^\star$ are pseudo-real, in the case of $SO(N)$ for $N=3,4,5\ {\rm mod}\ 8$. In the case of $E_6$ only even number of ${\bf 27}$-plets of $E_6$ were considered.  It should also be noted that in \cite{Avramis:2005hc}, only single or product of two groups (not counting the $R$-symmetry gauge group), and excluding the product of two classical groups, were considered. 

First, we summarise some of the key properties of the ungauged solutions. Firstly, it is interesting that the vast majority of the solutions are built out of low lying representations which we refer to as `generic' ones. For the classical groups these comprise of the fundamental, the antisymmetric (traceless), the symmetric (traceless), adjoint, and the spinor representations.\footnote{It is symmetric traceless and spinor only for $\textrm{SO}(N)$ groups, and antisymmetric traceless for $\textrm{Sp}(N)$ groups.} In the rank method, we made no restriction on the matter content and so it is interesting that even though there is a large number of partitions of matter consistent with \eq{R4}  containing representation that are not generic (we shall refer to these as `exotic' matter', very few of them turn out to be anomaly free. 
For $n_{T} = 1$, there are only 8 such models, containing either the \textbf{110} of $\textrm{Sp}(5)$, the \textbf{208} of $\textrm{Sp}(6)$, which are 3rd rank antisymmetric irreps, and the \textbf{77}, \textbf{77'} of $\textrm{G}_{2}$. For $n_T = 0$, there are 17 of them, containing the previous representations, as well as the \textbf{273, 189, 182, 76} of $\textrm{G}_{2}$, the \textbf{324, 273} of $\textrm{F}_{4}$, the \textbf{351, 351'} of $\textrm{E}_{6}$, the \textbf{364} of $\textrm{SO}(14)$, which is 3rd rank antisymmetric irrep, and the \textbf{210} of $\textrm{SO}(10)$, which is the 3rd rank totally symmetric irrep. From the swampland perspective, these models with exotic matter seem particularly hard to construct from string theory. As such, they are the natural candidates for further analysis with other swampland consistency conditions. Finally, we note that of the over 2,000 anomaly free and ghost free models we have found, only 32 have no singlets for $n_{T} = 1$, and 12 have no singlets for $n_T = 0$. Those are denoted with an asterisk in the appendices.

Next we discuss the new gauged supergravities we have found. These models are exceedingly rare. Having searched for models of the form $ G_{1} \times ...\times G_{n} \times G_{R} $, for $n=1,2,3$, including all representations and allowing for singlets, we find only 6 new models with $\textrm{U}(1)_{R}$ gauging, and no models at all with $\textrm{Sp}(1)_{R}$ gauging. This brings the total known anomaly free models, where the only $\textrm{U}(1)$ and $\textrm{SU}(2)$ factors come from the R-symmetry gauging, to 9. Details of the 6 new models are given in the next section but we note here that, even though we allowed for singlets, none of the 9 consistent models contain any singlet matter. Four of these models also contain exotic matter, containing either the \textbf{132} of $\textrm{Sp}(5)$, the \textbf{252} of $\textrm{SU}(10)$, which are both the 5th rank totally antisymmetric representation, or the \textbf{350} of $\textrm{Sp}(7)$, which is the 3rd rank totally antisymmetric representation.

Under the assumptions discussed in detail above, for the R-symmetry gauged cases, we find anomaly free models only for $n_T = 1$, not  $n_T = 0$ or $2$\footnote{We expect there are many anomaly free models with $n_T = 2$ of the form $G = G_{1} \times G_{R}$, whose anomaly matrix automatically has rank $r\leq 3$.} Using the rank method, it is easy to show that there can never be an anomaly free gauged model with $n_T = 0$. Indeed, for an anomaly free model, we would need that the matrix of anomaly coefficients has rank 1. In particular, we need the determinant of all $2\times 2$ minors to vanish. From \eqref{me1} and \eqref{me2}, we can see that the anomaly coefficient matrix always has the following minor.
\begin{equation}
    \begin{pmatrix}
        -\tfrac{9}{8} & -\tfrac{1}{12}(n_V - \alpha) \\
        -\tfrac{1}{2}(n_V - \alpha) & \tfrac13\beta(n_V + \gamma)
    \end{pmatrix}
\end{equation}
where $(\alpha,\beta,\gamma) = (19,2,5)$ for $\textrm{U}(1)_{R}$, and $(\alpha,\beta,\gamma) = (11,1,85)$ for $\textrm{Sp}(1)_{R}$ gauging. The vanishing of this determinant puts a quadratic constraint on $n_V$. We find
\begin{equation}
    \textrm{U}(1)_{R}\, : \quad 0=n_{V}^{2} + 70n_{V} + 901 \, , \qquad \qquad \textrm{Sp}(1)_{R}\, : \quad 0=n_{V}^{2} + 32n_V + 4711
\end{equation}
Neither of these equations have positive integer solutions, a necessary requirement for $n_V$, and hence there can be no anomaly free gauged models for $n_T = 0$.

For $n_T = 2$, it is perhaps surprising that we find no anomaly free models of the form $G = G_{1}\times G_{2} \times G_{R}$, as the factorization conditions are already not satisfied. This is the first case in which the rank reduction to $r \leq 3$ is non-trivial for gauged models. Given that the rank reduction condition for $n_T=2$ is less restrictive compared to the case of $n_T=1$, for which solutions do exist, one might naively expect to find even more solutions with $n_T = 2$, but this is not what we have found. Instead, we found that all models with rank reduction from 4 to 3 have $a\cdot b_{i} = 0$ and $b_i\cdot b_i = 0$. However, these anomalies cannot factorise by the discussion above, for the groups we have considered. 

\subsection{Properties of the new remarkably anomaly free models }

We have found remarkably anomaly free six models with gauge groups of the form $G_1\times G_2\times G_3\times U(1)_R$. Their key properties are as follows:

\begin{itemize}
\item  $E_7 \times E_8 \times SO(20) \times U(1)_R$

$(56,1,20)_0 + (1,1,512)_0 $
\begin{align}
& a = (2,-2)
\\
& b(E_8) = (-5,30)\ , 
\\
& b(E_7) = (-1/5,-6/5)\ ,
\\
& b(SO(20)) = (-1,6)\ ,
\\
& b(U(1)_R) = (-2,-48)\ .
\end{align}
The gauge kinetic terms are positive for 
\be
e^{-2\phi} < 1/24\ .
\ee

\item  $E_6 \times E_6 \times Sp(5) \times U(1)_R$ 

$(78,1,10)_0+ (1,1,132)_0$
\begin{align}
& a = (2,-2)
\\
& b(E_6) = (-6,2)\ , 
\\
& b^\prime(E_6) = (3,1)\ ,
\\
& b(Sp(5)) = (-3,1)\ ,
\\
& b(U(1)_R) = (18,2)\ .
\end{align}
The gauge kinetic terms are positive for 
\be
e^{-\phi} >3\ .
\ee

\item  $E_7 \times SU(10) \times SU(10) \times U(1)_R$ 

$(1,10,45)_0+(1,252,1)_0$
\begin{align}
& a = (2,-2)
\\
& b(E_7) = (4,1)\ , 
\\
& b(SU(10)) = (-4,1)\ ,
\\
& b^\prime(SU(10)) = (-4,1)\ ,
\\
& b(U(1)_R) = (-28,2)\ .
\end{align}
The gauge kinetic terms are positive for 
\be
e^{-2\phi} >16\ .
\ee

\item  $F_4 \times SO(13) \times Sp(7) \times U(1)_R$ 

$(26,1,14)_0+(1,13,14)_0+(1,1,350)_0 $
\begin{align}
& a = (2,-2)
\\
& b(F_4) = (-1,1)\ , 
\\
& b(SO(13)) = (1,1)\ ,
\\
& b(Sp(7)) = (-2,1)\ ,
\\
& b(U(1)_R) = (20,2)\ .
\end{align}
The gauge kinetic terms are positive for 
\be
e^{-2\phi} >40\ .
\ee

\item  $Sp(5) \times Sp(6) \times SO(12) \times U(1)_R$ 

$  (10,1,12)_0+ (44,12,1)_0 +(1,1,32)_0$
\begin{align}
& a = (2,-2)
\\
& b(Sp(5)) = (-1,5/2)\ , 
\\
& b(Sp_6) = (-1,3/2)\ ,
\\
& b(SO(12)) = (-1,-3/2)\ ,
\\
& b(U(1)_R) = (-2,-17)\ .
\end{align}
The gauge kinetic terms are positive for 
\be
e^{-2\phi} < 2/17\ .
\ee

\item  $ Sp(5) \times Sp(6) \times SO(13)\times U(1)_R$ 

$(10,1,13)_0 + (10,65,1)_0 + (132,1,1)_0$
\begin{align}
& a = (2,-2)
\\
& b(Sp_5) = (-3,1)\ ,
\\
& b(Sp_6) = (-2,1)\ ,
\\
& b(SO(13)) = (2,1)\ , 
\\
& b(U(1)_R) = (18,2)\ .
\end{align}
The gauge kinetic terms are positive for 
\be
e^{-\phi} > 3\ .
\ee

\end{itemize}

Supergravities with gauged R-symmetry are particularly interesting for many reasons, one being that  they exhibit spontaneous compactification down to 4-dimensional Minkowski space. The internal space is an $S^{2}$ with a single $U(1)_{R}$ monopole threading it \cite{Salam:1984cj}. In \cite{Cvetic:2003xr}, they found that one could derive the Salam-Sezgin model from a reduction of 10-dimensional type 1 supergravity reduced on a non-compact hyperboloid $H^{(2,2)}$, followed by a consistent chiral truncation on an $S^{1}$ reduction. 

The Salam-Sezgin model, on the other hand, consists only of the gravity multiplet and a single vector multiplet for the $U(1)_{R}$ gauge symmetry. In particular, it has no hypermultiplet matter and is anomalous by itself. Understanding how to obtain the 9 anomaly free theories, with their gauge groups and complicated matter content, remains an open problem. The spontaneous compactification to 4-dimensional Minkowski will still be present in these anomaly-free theories (one can simply turn off the non-R-symmetry gauge fields in the solution). To be able to uplift this to a solution of 10 or 11-dimensional supergravity, one will need a non-compact manifold internal space. It is therefore likely that any construction of these theories from string theory will require non-compact reductions. This places them outside the scope of F-theory constructions where much is known about how to obtain non-abelian gauge groups and non-trivial hypermultiplet matter. Furthermore, many of the gauge groups above are not subgroups of $E_{8}\times E_{8}$ or $\textrm{SO}(32)$, making a direct reduction from heterotic supergravity less likely.

\section{Solutions by means of the graphical method}

\subsection{The general procedure}\label{sec:graph_method_procedure}

In the previous sections, we performed an exhaustive search of all possible anomaly free models within a restricted set of groups. Namely, we considered only products of up to three groups within the set \eq{pm} (plus possible R-symmetry factor). One might wonder what it would take to go beyond these assumptions. Namely, can one find an upper bound on the number and rank of simple gauge factors? If we were to proceed via the rank method, we would simply have to increase the gauge groups one-by-one, and increase the number of factors, checking each case individually. That is, we would need to find all partitions which solve the $\tr F^{4}$ and $\tr R^{4}$ constraints and then check to see if the anomaly matrix for each has rank reduction. Then we repeat for the next group. Suppose we find no solutions containing $\textrm{SU}(80)$, does this mean that there will not be any containing $\textrm{SU}(81)$? Or suppose we find no solutions of the form $G_{1} \times  ... \times G_{6}$, does that mean there won't be any for $G_{1} \times ... \times G_{7}$? From the rank method, it is unclear and we would have to check each case by case. Note that it was shown in \cite{Kumar:2010ru} that for $n_T < 9$, the number of ghost free and anomaly free models is finite, but an upper bound was not established.

In this section, we consider if there is an efficient way to answer these questions for the ungauged models with $n_T=1$. To do so, however, we will make a slight restriction on the representation content. We will consider the following
\be
\text{Graphical method:}\quad 
\begin{cases}
F_4, E_6, E_7, E_8\\
A_N\ , \quad N\ge 9  \\
B_N, D_N\ ,\quad N\ge 5  \\
C_N,\quad  N \ge 10  \\
\text{Fund., adj. and (anti)symm. irreps for classical groups}\\
\text{Representations as in \eqref{akr} for exceptional groups} \\
\end{cases}
\label{gm}
\ee
Beside the spinor representation, the matter content considered here is precisely the `generic matter' discussed at the end of section \ref{sec:comments_on_solns}, which comprises all but 8 of the anomaly free ungauged models with $n_T = 1$ found by the rank method. Moreover, for sufficiently large $N$, these representations for the classical groups are the only ones consistent with the $\tr F^{4}$ condition of anomaly cancellation.

We want to find a graphical representation for anomaly free models. Consider an anomaly free model with a single simple gauge group $G_{1}$, and representation content $\sum_R n_R \, \mathbf{R}$ (we leave the singlet representations implicit). We will denote such a model as 
\begin{equation}
    M_{1} = \left\{ G_{1} ,\, \sum_{R} n_{R} \, \mathbf{R} \right\}
\end{equation}
For this to be anomaly free, we need that the $\tr F^{4}$ and $\tr R^{4}$ conditions are satisfied and that the coefficient matrix is factorisable with respect to the inner product $\Omega_{\alpha\beta}$ as in \eqref{Omega}. A necessary and sufficient condition for this is that the anomaly coefficient matrix (see \eq{I8P} for the notation) has negative determinant. 
\begin{equation}
    \det I_{1} < 0 \, , \qquad I_{1} = \begin{pmatrix}
        -1 & -\tfrac{1}{12}A_{1} \\ 
        -\tfrac{1}{12}A_{1} & \tfrac{2}{3}C_{1}
    \end{pmatrix}
\end{equation}
Factorisation implies that we can find $a$ and $v_{1}^{\alpha} = \tfrac{2}{\lambda_{1}}b^{\alpha}_{1}$ satisfying \eq{c1}--\eq{c3}. That is, using the scale invariance to set $a = (2,-2)$, we need
\begin{equation}
    v_{1}^{1} - v_{1}^{0} = - \frac{1}{6} A_{1} \, , \qquad v_{1}^{0}v_{1}^{1} = \frac{2}{3}C_{1}
\end{equation}
Ghost freedom puts a further constraint on the $v_{1}^{\alpha}$ such that there must exist a $\phi$ such that
\begin{equation}\label{eq:ghost-free_condition}
    -v_{1}^{0}e^{\phi} + v_{1}^{1}e^{-\phi} > 0
\end{equation}

We may represent such a solution as a node.\footnote{In graph theory, nodes are also referred to as vertices or points.}
\begin{figure}[H]
\centering
\begin{tikzpicture}
    \filldraw[black] (0,0) circle (4pt) node[anchor=east]{$M_{1}\quad$};
\end{tikzpicture}
\caption{A model with a single gauge group and representation content that (i) satisfies the $\tr F^{4}$ constraint, and (ii) leads to a factorisable anomaly coefficient matrix with ghost-free kinetic terms is denoted by a node. The notation $M$ will also be used to denote the non-trivial representation content for the hyperfermions which, together with the number $n_S$ of the hyperfermion singlets define the model.}
\end{figure}

Now consider an anomaly free model with 2 simple factors. For example, lets consider the following model in which the number of singlets is left implicit.
\begin{equation}\label{eq:2group_example}
    M = \{ \textrm{SU}(10) \times \textrm{SU}(16) ,\, 2(\textbf{45},\textbf{1}) + 14(\textbf{1},\textbf{16}) + (\textbf{1},\textbf{120}) + (\textbf{10},\textbf{16}) \}
\end{equation}
It is easy to check that this satisfies the $\tr F^{4}$ and $\tr R^{4}$ constraints. Moreover, the anomaly coefficient matrix is given by
\begin{equation}
    I = \begin{pmatrix}
        -1 & 1 & 1/2 \\
        1 & 0 & -2 \\
        1/2 & -2 & 2
    \end{pmatrix}\ ,
\end{equation}
which is rank 2 and has eigenvalues with opposite signs, and it is ghost free. We can think of this as two individual models which can be combined into an anomaly free one in a manner that will be described below.

With the number of singlets understood to be implicit, we consider
\begin{alignat}{2}
    M_{1} &= \{ \textrm{SU}(10),\, 2 \, (\textbf{45}) + 16\,(\textbf{10}) \} \ ,& \qquad M_{2} &= \{ \textrm{SU}(16),\, 24\,(\textbf{16}) + (\textbf{120}) \}\ , \nn \\
    I_{1} &= \begin{pmatrix}
        -1 & 1 \\
        1 & 0
    \end{pmatrix} \ , & I_{2} &= \begin{pmatrix}
        -1 & 1/2 \\
        1/2 & 2
    \end{pmatrix} \ .
\end{alignat}
Observe that the $I_{i}$ are given by particular $2\times 2$ minors of $I$ that contain elements from the first row and column associated with the gravitational anomalies. Note that necessarily both $M_{i}$ satisfy the $\tr F_{i}^{4}$ condition, have $\det I_{i} < 0$ and are ghost free. However, we can count the total amount of non-singlet matter in $M_{2}$ to be $24 \times 16 + 120 = 504$. Then we have
\begin{equation}
     504 - \dim(\textrm{SU}(16)) = 249 > 244
\end{equation}
Hence, on its own, $M_{2}$ cannot be anomaly-free as it will have a gravitational anomaly coming from the $\tr R^{4}$ term. All other anomalies of $M_{2}$, both mixed and gauge, can be cancelled however\footnote{$M_{1}$ on the other hand can be made into a completely anomaly free model by including the appropriate number of singlets.}. We will refer to any model where all anomalies can be cancelled apart from possibly the $\tr R^{4}$ anomaly as \emph{quasi-anomaly-free}. We consider totally anomaly free models as special cases of quasi-anomaly-free models.

We can represent the resulting anomaly-free model by the two nodes joined by a link as shown in Fig 2.\footnote{In graph theory, links are also referred to as edges or lines.} 
\begin{figure}[H]
    \centering
    \begin{tikzpicture}
        \filldraw[black] (-2,0) circle (4pt) node[anchor=east]{$M_{1}\quad$};
        \filldraw[black] (2,0) circle (4pt) node[anchor=west]{$\quad M_{2}$};
        \draw[black,thick] (-2,0) -- (2,0);
    \end{tikzpicture}
    \caption{A model with 2 simple gauge group factors with factorisable anomaly coefficient matrix may be denoted by 2 nodes joined by a link.}
\end{figure}
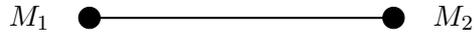
The link denotes that there is a way to combine the representation content of $M_{1}$ and $M_{2}$ into a quasi-anomaly-free theory. Equivalently, there is a way to combine the representations such that the resulting anomaly coefficient matrix is factorisable. Of course, in this example, it is clear that this is possible by construction - we just combine them as in \eqref{eq:2group_example}.

Now lets consider a new quasi-anomaly-free model.
\begin{equation}
    \begin{aligned}
        M_{3} &= \{ \textrm{SU}(16),\, 32 \,(\textbf{16}) \} \\
        I_{3} &= \begin{pmatrix}
            -1 & 0 \\
            0 & 4
        \end{pmatrix}
    \end{aligned}
\end{equation}
Can we join a link between $M_{3}$ and either of the previous models? Let's start with $M_{1}$. We join a link only if the joint model is also quasi-anomaly-free. Equivalently, we draw a link if the associated anomaly coefficient matrix $M_{13}$ is factorisable. We have
\begin{equation}
    I_{13} = \begin{pmatrix}
        -1 & 1 & 0 \\
        1 & 0 & \alpha \\
        0 & \alpha & 4
    \end{pmatrix}
\end{equation}
where $\alpha$ depends on how we form joint representations of $\textrm{SU}(10)\times \textrm{SU}(16)$ from $M_{1}$ and $M_{3}$. Factorisation imposes $\alpha = \pm 2$. However, for $M_{13}$ to be the anomaly coefficient matrix, from \eqref{A2} we see that we need
\begin{equation}\label{eq:combination_condition}
    \alpha = -2\sum_{R,S}n_{R,S}\,a_{R}a_{S}
\end{equation}
The unique solution for this is $\alpha = -2$ with a single joint representation $(\textbf{10},\textbf{16})$. That is, we get have a quasi-anomaly-free model
\begin{equation}
    M_{12} = \{ \textrm{SU}(10) \times \textrm{SU}(16),\, (\textbf{10},\textbf{16}) + 22\, (\textbf{1},\textbf{16}) + 2\,(\textbf{45},\textbf{1}) \}
\end{equation}
Repeating this for $M_{2}$, we have the anomaly coefficient matrix
\begin{equation}
    I_{23} = \begin{pmatrix}
        -1 & \tfrac{1}{2} & 0 \\
        \tfrac{1}{2} & 2 & \alpha \\
        0 & \alpha & 4
    \end{pmatrix}
\end{equation}
Factorisation fixes $\alpha = \pm 3$. However, we still require \eqref{eq:combination_condition} which, for $\textrm{SU}(N)$ groups, implies that $\alpha$ is an even integer. Therefore, there is no way to combine $M_{2}$ and $M_{3}$ into a quasi-anomaly-free model and so we would \emph{not} draw a line between them.\footnote{Note that if we had taken instead $M_2$ to be the model $\{ \textrm{Sp}(10),\, 56\,(\textbf{20})\}$ then we would get the same $I_{2}$ and hence $I_{23}$ would also be as above. This time, we \emph{can} draw a line between the two nodes since the $\textbf{20}$ of $\textrm{Sp}(10)$ is pseudo-real and hence $n_{R,S}$ in \eqref{eq:combination_condition} can take half-integer values.}

Finally, we can repeat the same analysis with a fourth quasi-anomaly-free model $M_{4} = \{ \textrm{SU}(17),\, 16\,(\textbf{17}) + 2\, (\textbf{136}) \} $. By finding all the links between $M_{1},...,M_{4}$, including self-links, we find the following graph.
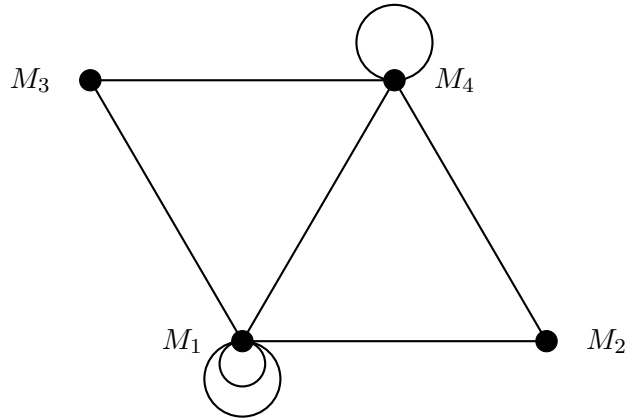
\begin{figure}[H]
    \centering
    \begin{tikzpicture}[roundnode/.style={circle, thick, minimum size=7mm}]
        \filldraw[black] (-2,0) circle (4pt) node[anchor=east]{$M_{1}\quad$};
        \filldraw[black] (2,0) circle (4pt) node[anchor=west]{$\quad M_{2}$};
        \filldraw[black] (0,3.46) circle (4pt) node[anchor=west]{$\quad M_{4}$};
        \filldraw[black] (-4,3.46) circle (4pt) node[anchor=east]{$M_{3}\quad$};
        \draw[black,thick](-2,-0.5) circle (0.5);
        \draw[black,thick](0,3.96) circle (0.5);
        \draw[black,thick](-2,-0.3) circle (0.3);
        \draw[black,thick] (-2,0) -- (2,0);
        \draw[black,thick] (-2,0) -- (-4,3.46);
        \draw[black,thick] (-2,0) -- (0,3.46);
        \draw[black,thick] (-4,3.46) -- (0,3.46);
        \draw[black,thick] (0,3.46) -- (2,0);
    \end{tikzpicture}
    \caption{A graph denoting the different ways to combine models $M_{1}, M_{2}, M_{3}, M_{4}$ into quasi-anomaly-free models with 2 simple gauge group factors.}
    \label{fig:S1...4}
\end{figure}
The double loop around $M_{1}$ denotes the fact that there are 2 ways to combine $M_{1}$ with itself to form a quasi-anomaly-free model.

Can we use this graph to find new models? To answer this, consider a model $M$ with group $G_{1}\times G_{2} \times G_{3}$. It is not difficult to see that $M$ is quasi-anomaly-free if and only if all sub-models $M_{ij}$ with group $G_{i}\times G_{j}$ (for $i\neq j$) are quasi-anomaly-free.\footnote{The `only if' statement is clear. The `if' statement follows from the fact that the factorisation conditions in \eqref{c1}-\eqref{c4} only use data from pairs of groups $G_{i}\times G_{j}$. } Hence, the model $M$ would appear as a totally connected graph with 3 nodes. In fact a quasi-anomaly-free model with $n$ groups would appear as a totally connected graph with $n$ nodes.

To find new models, we simply look for totally connected subgraphs of figure \ref{fig:S1...4}. Ignoring the loops for now, we see that we have 2 totally connected subgraphs of order 3. They are $(M_{1},M_{2},M_{4})$ and $(M_{1},M_{3},M_{4})$. Take the first case. Both the $S_{12}$ link and the $S_{24}$ link contain non-trivial joint representations (they are the $(\textbf{10},\textbf{16})$ and the $(\textbf{16},\textbf{17})$ respectively). To be able to combine these into a single model, we would need at least $27\, \textbf{16}$ in $M_{2}$. This is not the case however, and so this subgraph does not correspond to a new model. The second case, however, does correspond to a new model.
\begin{equation}
\begin{aligned}
    M_{134} &= \{ \textrm{SU}(10)\times \textrm{SU}(16) \times \textrm{SU}(17), \, 2\,(\textbf{45},\textbf{1},\textbf{1}) + 5(\textbf{1},\textbf{16},\textbf{1}) \\
    & \qquad \qquad  + 2\,(\textbf{1},\textbf{1},\textbf{136}) + (\textbf{10},\textbf{16},\textbf{1}) + (\textbf{1},\textbf{16},\textbf{17}) \}
\end{aligned}
\end{equation}
By construction, this model is guaranteed to be quasi-anomaly-free. To check total anomaly freedom, we just need to ask whether the $\tr R^{4}$ condition \eqref{R4} can be satisfied, possibly by adding some neutral hypermultiplets. If we let $\tilde{n}_{H}$ denote the number of hypermultiplets in non-trivial representations and $n_{S}$ denote the number of neutral hypermultiplets, so that
\be
\boxed{{\tilde n}_H := n_H -n_S}\ ,
\ee
then we find
\begin{equation}
    \tilde{n}_{H} - n_V = 874 - 642 = 232 \leq 244
\end{equation}
Hence, if we take $n_S = 12$, then we get a totally anomaly free model.

The final condition we need to check is ghost freedom \eqref{eq:ghost-free_condition}. While each of the individual nodes must be ghost free, the regions in the moduli space in which this occurs may not overlap and hence the total model may not be ghost free. Fortunately, one can check that for the models $M_{1}, M_{3}, M_{4}$, the coefficients $v^{\alpha}_{i}$ in the factorisation are
\begin{equation}\label{eq:M134_v_params}
    v_{1} = (0,2)\, , \qquad v_{3} = (-2,-2) \, , \qquad v_{4} = (0,2) 
\end{equation}
The ghost free condition \eqref{eq:ghost-free_condition} is satisfied for all groups if $e^{-\phi} < 1$.

We have found a new ghost free and anomaly free model with three groups outside the range that we considered. We did not need to exhaustively search all possible combinations of representations for the group $\textrm{SU}(10)\times \textrm{SU}(16)\times \textrm{SU}(17)$. Instead, simply by understanding how we can pair combinations of 2 groups, we arrived at the 3 group solution almost for free.

Can we go beyond this to 4 or more groups? For a 4 group solution, we need a totally connected subgraph with 4 nodes. Ignoring loops, we see that since $M_2$ and $M_3$ aren't linked, we cannot make a 4 group model with $M_{1},...,M_{4}$. However, loops indicate that we can have one model appearing more than once. Let's consider the $M_{13}$ link, and the two self links $M_{11}, M'_{11}$. These correspond to the quasi-anomaly-free models
\begin{align}
    M_{13} &= \{ \textrm{SU}(10)\times \textrm{SU}(16),\, (\textbf{10},\textbf{16}) + 22\,(\textbf{1},\textbf{16}) + 2\,(\textbf{45},\textbf{1}) \} \\
    M_{11} &= \{ \textrm{SU}(10) \times \textrm{SU}(10),\, 16\, (\textbf{10},\textbf{1}) + 2\, (\textbf{45},\textbf{1}) + 16\, (\textbf{1},\textbf{10}) + 2\, (\textbf{1},\textbf{45}) \} \\
    M'_{11} &= \{ \textrm{SU}(10)\times \textrm{SU}(10),\, (\textbf{10},\textbf{10}) + 6\,(\textbf{10},\textbf{1}) + 2\,(\textbf{45},\textbf{1}) + 6\,(\textbf{1},\textbf{10}) + 2\,(\textbf{1},\textbf{45}) \}
\end{align}
Can we combine $M_{13}$ with either $M_{11}$ or $M'_{11}$? Combining $M'_{11}$ would require $M_{1}$ to have at least $26\,(\textbf{10})$ which is not the case. Hence, we can only combine $M_{11}$. We find, by adding more and more $M_{1}$, the following models.
\begin{align}
\begin{split}
    M_{113} &= \{ \textrm{SU}(10)\times \textrm{SU}(10) \times\textrm{SU}(16), \\
    & \qquad (\textbf{10},\textbf{1},\textbf{16}) + (\textbf{1},\textbf{10},\textbf{16})+ 12\,(\textbf{1},\textbf{1},\textbf{16}) + 2\,(\textbf{45},\textbf{1},\textbf{1}) + 2\,(\textbf{1},\textbf{45},\textbf{1}) \} 
\end{split} \\
\begin{split}
    M_{1113} &= \{ \textrm{SU}(10)\times \textrm{SU}(10) \times \textrm{SU}(10) \times\textrm{SU}(16), \\
    & \qquad (\textbf{10},\textbf{1}, \textbf{1},\textbf{16}) + (\textbf{1},\textbf{10},\textbf{1},\textbf{16}) + (\textbf{1},\textbf{1},\textbf{10},\textbf{16}) \\
    & \qquad + 2\,(\textbf{1},\textbf{1},\textbf{1},\textbf{16}) + 2\,(\textbf{45},\textbf{1},\textbf{1},\textbf{1}) + 2\, (\textbf{1},\textbf{45},\textbf{1},\textbf{1}) + 2\,(\textbf{1},\textbf{1},\textbf{45},\textbf{1}) \}
\end{split}
\end{align}
Could we combine another $M_{1}$? This would require $M_{3}$ to have at least $40\,(\textbf{16})$ which is not the case. Hence, we have exhausted all possible additions of $M_{1}$ and $M_{3}$. Again, both $M_{113}$ and $M_{1113}$ are quasi-anomaly-free by construction. We simply need to check the condition \eqref{R4} coming from $\tr R^{4}$, and the ghost free condition \eqref{eq:ghost-free_condition}. It is easy to check that both of these models are anomaly free (provided we add 5 neutral hypers to $M_{113}$ and 14 neutral hypers to $M_{1113}$) and ghost free (for $e^{-\phi}<1$).

We have found a further 2 models simply by considering subgraphs of figure \ref{fig:S1...4}. It is worth noting that the 4-group model $M_{1113}$ would have been very difficult to find via the rank method. Such a model has a $5\times5$ anomaly coefficient matrix and must be rank 2. This is a very stringent constraint which would take a long time to find if we simply searched through all of the possible representation content of an $\textrm{SU}(10)^{3} \times \textrm{SU}(16)$ model. Instead, we simply needed to consider how to pair individual models and then look for subgraphs with 4 nodes (including repeats). Once these were identified, it was an easy check to see that the $\tr R^{4}$ anomaly can be cancelled and that the solution is ghost free.

More generally, suppose we had a set of starting nodes $\mathcal{S}$. Each element of $\mathcal{S}$ would correspond to a quasi-anomaly free and ghost free model with a single gauge group. Finding all possible links between nodes, we obtain a large graph extending figure \ref{fig:S1...4}. From this graph, constructed by finding 1- and 2-group models, we would be able to extract models of 3 or more groups. Such models would be totally connected subgraphs with $n$ nodes for any $n$ (with possible repeats). These totally connected, or complete, subgraphs, also known as `cliques', are well-studied in graph theory, and there are algorithms for efficiently finding them \cite{Coen73,TOMITA200628,CAZALS2008564}. As we have seen, not all cliques will correspond to anomaly free theories, but all anomaly free theories will appear as cliques. With this in mind, the procedure we run goes as follows.
\begin{enumerate}
    \item Find a set $\mathcal{S}$ of models with a single gauge group and matter content, subject to \eqref{gm}, such that the model is quasi-anomaly-free and ghost-free. That is, the model should satisfy:
    \begin{itemize}
        \item The $\tr F^{4}$ anomaly should vanish.

        \item The anomaly coefficient matrix should factorise with respect to $\Omega_{\alpha\beta}$. Equivalently, its determinant should be negative.

        \item The ghost-free condition \eqref{eq:ghost-free_condition} should be satisfied.
    \end{itemize}
    Note that we do \emph{not} need that the $\tr R^{4}$ anomaly vanishes. Draw a node for each model.

    \item Draw a link between each node if they can be combined into a quasi-anomaly-free model with 2 gauge group factors. If there is more than one way to combine two models then multiple links are drawn between them. Links from a node to itself are possible if a model can be combined with itself.

    \item Find all the cliques of any order.

    \item Find all the models associated to each clique (if they exist), including repetitions of models if self-links are present. Check which satisfy the $\tr R^{4}$ constraint, and are ghost free.
\end{enumerate}

At this point, the reader might be concerned that the initial set $\mathcal{S}$ is infinite, and so one will not be able to complete the procedure above in practice. Indeed, we have not currently put an upper bound on the rank of the classical gauge groups. Moreover, since we have dropped the $\tr R^{4}$ condition, we lose a bound on the representation content. This is particularly stark for exceptional groups where any combination of representations satisfies the $\tr F^{4}$ condition.

In the next section, we will show how one can restrict $\mathcal{S}$ to a finite subset. We will see that one can find some constant $\kappa$, which depends on the type of group and is strictly greater than 244, such that any node in $\mathcal{S}$ which does not satisfy the following constraint
\begin{equation}
    \tilde{n}_{H} - n_V \leq \kappa\ ,\qquad \kappa \ge 244\ ,
\end{equation}
can never be combined into a ghost free and anomaly free model. This bound will effectively put a bound on the rank of the classical groups, and on the representation content for all groups. We can therefore run the procedure outlined above, but starting with the following \emph{finite} set of nodes.
\begin{equation}
    \hat{\mathcal{S}} = \{ M \in \mathcal{S}\,|\, \tilde{n}_{H} - n_{V} \leq \kappa \}
\end{equation}

\subsection{Finding the finite set of nodes $\hat{\mathcal{S}}$}\label{sec:proof_of_finite_nodes}

Recall that, for a model with a single simple gauge group to be quasi-anomaly-free and ghost free, there need to exist $v = (v^{0},v^{1})$ such that
\begin{align}
    v^{1}-v^{0} &= -\frac{1}{6}A \, , \\
    v^{1}v^{0} &= \frac{2}{3}C\, , \\
    -v^{0}e^{\phi} + v^{1}e^{-\phi} &> 0 \, ,  \qquad \text{some }\phi \ ,
\end{align}
with $A$ and $C$ from \eq{A1} and \eq{C}, respectively. Following \cite{Kumar:2009ae}, from this we can split the quasi-anomaly-free and ghost free models into 3 kinds. In the following, we write $+,-,0$ to denote whether that component is positive, negative, or 0 respectively.
\begin{itemize}
    \item Type N:\quad  $C > 0$ and $v = (+,+)$ or $(-,-)$.

    \item Type O:\quad   $C = 0$ \text{and} $v = (-,0)$ or $(0,+)$

    \item Type P:\quad  $C < 0$ \text{and} $v = (-,+)$
\end{itemize}
Note that we cannot have other configurations for $v$, else we will break the ghost-free condition. As an example, in figure \ref{fig:S1...4}, we can see from \eqref{eq:M134_v_params} that $M_{1}$ and $M_{4}$ are type O nodes, while $M_{3}$ is a type N node. One can also show that $M_{2}$ also corresponds to a type N node.

When joining 2 nodes, with parameters $v_{1}$ and $v_{2}$, together with a link, we need that
\begin{equation}
    v_{1}^{0}v_{2}^{1} + v_{1}^{1}v_{2}^{0} = -4\sum_{R,S}n_{R,S} a^{1}_{R} a^{2}_{S} \in \mathbb{Z}_{\leq 0}
\end{equation}
In particular, the combination above is strictly negative when there are non-trivial joint representations in the theory (in which case we say the nodes couple non-trivially), and vanishing only if there are no joint representations. Moreover, the above relation must hold for \emph{all} group factors in a model. From this, it was observed in \cite{Kumar:2009ae} that
\begin{itemize}
    \item Type P nodes always couple non-trivially (i.e. with $n_{R,S} \ne 0$) amongst themselves.

    \item Type O nodes always couple non-trivially to type P and type N nodes.

    \item There can be at most 2 type N nodes in a theory. If there are 2, they must be $(+,+)$ and $(-,-)$ and they must couple non-trivially.
\end{itemize}

By writing out the expression for $C$, we see that there are clearly only a finite number of N and O solutions for each simple group $G$. Indeed, we have
\begin{equation}
    C = c_{\text{adj}} - \sum_{R} n_{R} c_{R}
\end{equation}
and since $c_{R} > 0$ for each of the representations we consider here, described in \eq{gm}, there can only be a finite number of combinations of representations such that $C\ge 0$. For each classical group, there are also a finite number of P nodes, since there are only a finite number of solutions of
\begin{equation}
    B := b_{\text{adj}} - \sum_{R}n_{R}b_{R} =0
\end{equation}
For the exceptional groups, however, there can be an infinite number of P nodes.

To find the finite set $\hat{\mathcal{S}}$, we need to put an upper bound on the rank of the classical gauge groups, and limit the number of P nodes for the exceptional groups. To do so, we will carefully examine what happens to the value of $n_H - n_V$ when we couple these nodes to other nodes. We will find that when the rank of the classical group gets too large, or the total representation content of a node gets too large, the value of $n_H - n_V$ will always be greater than 244, no matter how we couple other nodes to it. Hence, these nodes can never be part of a totally anomaly free, ghost free theory.

\subsubsection{Some useful comments on Type N/O/P nodes}\label{sec:comments_on_NOP}

In this section, we derive some useful properties of the type N/O/P nodes for the various groups. Many of these observations are not new, but were found in \cite{Kumar:2009ae}. We reiterate them here for convenience later.

We introduce some notation for the representations of the classical groups. We write $\textbf{f}$, $\textbf{a}$, $\textbf{s}$, and $\textbf{adj}$ for the fundamental, antisymmetric (traceless), symmetric (traceless), and adjoint representations of these groups. For $\textrm{SO}(N)$, the antisymmetric and adjoint coincide so we denote both with $\textbf{a}$. Similarly, the symmetric and adjoint coincide for $\textrm{Sp}(N)$ and so we denote both with $\textbf{s}$. We can find the group theory coefficients $a_{R}, b_{R}, c_{R}$ for each of these representations (listed in \cite[Appendix B]{Avramis:2005hc}) and write the functions $B$ and $C$ for the classical groups as
\begin{align}
    B &= \begin{cases}
        2N - n_{f} - (N-8)n_a - (N+8)n_{s} - 2N n_{\mathrm{adj}} & \quad \textrm{SU}(N) \\
        N-8 - n_{f} - (N-8)n_{a} - (N+8) n_s & \quad  \textrm{SO}(N) \\
        2N+8 - \tfrac{1}{2}n_{f} - (2N-8)n_a - (2n+8) n_s & \quad \textrm{Sp}(N)
    \end{cases} \\[5pt]
    C &= \begin{cases}
        6 - 3 n_{a} - 3 n_s - 6n_{\mathrm{adj}} & \quad \textrm{SU}(N) \\
        3 - 3 n_{a} - 3n_s &\quad \textrm{SO}(N) \\
        3 - 3n_a - 3n_s & \quad \textrm{Sp}(N)
    \end{cases}
\end{align}
We can solve the equation $B=0$ for the 3 cases and group them into type N/O/P. Note that we will exclude any solution which has only a single adjoint representation since, as was discussed in section \ref{sec:comments_on_solns}, these lead to a theory with a vanishing kinetic term for this group. We also exclude the $\textrm{SU}(N)$ solution $(n_{f},n_a,n_s,n_\mathrm{adj}) = (0,1,1,0)$ for the same reason.

The solutions for the groups we are considering are as follows.
\begin{table}[H]
    \centering
    \begin{tabular}{|l|c|c|c|}
    \hline
    Group & Type & Representation content & $\tilde{n}_{H} - n_{V}$ \\
    \hline \hline
        \multirow{5}{*}{$\textrm{SU}(N)$} & \multirow{3}{*}{N} & $2N\,\textbf{f}$ & $N^{2} +1$ \\
        && $(N-8)\,\textbf{f} + \textbf{s}$ & $ \tfrac{1}{2}N^{2} - \tfrac{15}{2}N + 1 $\\
        && $(N+8)\,\textbf{f} + \textbf{a}$ & $ \tfrac{1}{2}N^{2} + \tfrac{15}{2}N + 1 $ \\
        \cline{2-4}
        &  O & $16\,\textbf{f} + 2\,\textbf{a}$ & $15N+1$ \\
        \cline{2-4}
         & P & $ (24-N-(N-8)k)\,\textbf{f} + (3+k)\,\textbf{a} $ & $ -\tfrac{1}{2}(k+1)N^{2} + \tfrac{15}{2}(k+3)N + 1 $  \\
         \hline \hline
         $\textrm{SO}(N)$ &  N & $(N-8) \, \textbf{f}$ & $ \tfrac{1}{2}N^{2} - \tfrac{15}{2}N $ \\
          \hline \hline
        \multirow{3}{*}{$\textrm{Sp}(N)$} &  N & $2(2N+8) \, \textbf{f}$ & $2N^{2} + 15N$ \\
        \cline{2-4}
        &  O & $ 32\, \textbf{f} + \textbf{a} $ & $30N-1$ \\
        \cline{2-4}
        &  P & $2(24-2N)\,\textbf{f} + 2 \,\textbf{a}$ & $-2N^{2} + 45N -2 $  \\ \hline
    \end{tabular}
    \caption{Table showing all solutions of $B=0$ for the classical groups and representations laid out in \eqref{gm}. Here $k\geq 0$ is any integer such that the coefficients are positive. The multiplicity of any representation in the P nodes does not exceed 14 since by assumption $N\geq 10$.}
    \label{tab:B=0_solns}
\end{table}
We can see immediately that all groups have type N solutions at any rank, and $\textrm{SU}(N)$ and $\textrm{Sp}(N)$ have type O solutions at any rank. For the type P solutions, they only exist if the coefficient is positive. Hence, we have type P solutions for $\textrm{SU}(N)$ when $N<24$ and for $\textrm{Sp}(N)$ when $N<12$. Moreover, the multiplicity of any representation in a type P node (with classical groups) does not exceed 14, while the maximum multiplicity for a type O node is 16 of $\textrm{SU}(N)$ and 32 for $\textrm{Sp}(N)$. The type O solutions also all have a positive individual value of $\tilde{n}_{H} - n_{V}$.

For the type N nodes, the individual value of $\tilde{n}_{H} - n_{V}$ grows unbounded with $N$. The minimum value for a type N node is for $\textrm{SO}(10)$ where $\tilde{n}_{H} - n_{V} = -25$. When a type N node is combined with another type N node, their combined value of $\tilde{n}_{H} - n_{V}$ may be less than this. It was shown in \cite{Kumar:2009ae} that there is no way to couple two classical type N nodes together in ghost free way. Hence, the only way to combine them is with the exceptional type N nodes. 

We can also do the same for the exceptional groups. In this case, the function $B$ is identically 0 and so we need only consider the function $C$. We find
\begin{equation}
    C = \begin{cases}
        \tfrac{1}{100} - \tfrac{1}{100}n_{\textbf{248}} & \quad E_{8} \\
        \tfrac{1}{6} - \tfrac{1}{48}n_{\textbf{56}} - \tfrac{1}{6}n_{\textbf{133}} - \tfrac{31}{12}n_{\textbf{912}} & \quad E_{7} \\
        \tfrac{1}{2} - \tfrac{1}{12}n_{\textbf{27}} - \tfrac{1}{2}n_{\textbf{78}} - \tfrac{55}{12}n_{\textbf{351}} - \tfrac{35}{6}n_{\textbf{351'}} - 10n_{\textbf{650}} & \quad E_{6} \\
        \tfrac{5}{12} - \tfrac{1}{12}n_{\textbf{26}} - \tfrac{5}{12}n_{\textbf{52}} - \tfrac{49}{12}n_{\textbf{273}} - \tfrac{23}{4}n_{\textbf{324}} &\quad F_{4}
    \end{cases}
\end{equation}
Since $B\equiv 0$ in these cases, any values of $n_{R}$ are valid. So in principal we can have arbitrarily many type P nodes (although we will see in the next section that one can effectively reduce this to a finite number). We will simply enumerate the type O solutions in each case and list their value of $\tilde{n}_{H} - n_{V}$. As before, we will exclude the trivial solutions which have only a single adjoint representation in their representation content.
\begin{table}[H]
    \centering
    \begin{tabular}{|c|c|c|}
    \hline
    Group & Representation content & $\tilde{n}_{H} - n_{V}$ \\ \hline \hline
      $E_{7}$ & $8\, (\textbf{56})$ & 91 \\
        $E_{6}$ & $6\, (\textbf{27})$ & 84 \\
        $F_{4}$ & $5 \, (\textbf{26})$ & 78 \\
        \hline
    \end{tabular}
    \caption{Type O solutions for the exceptional groups.}
    \label{tab:Exceptional_type_O}
\end{table}
We see that the representation content for any type O node has multiplicities not exceeding 8, and the individual contribution to $\tilde{n}_{H} - n_{V}$ is always positive. We can also see that the type N nodes will be given by the representation content $k\times \text{fundamental}$ for some $k$ less than 8 for $E_{7}$, 6 for $E_{6}$, and 5 for $F_{4}$. The minimal individual value of $\tilde{n}_{H} - n_{V}$ for a type N node with an exceptional group is when there are no non-trivial hypermultiplets, so $\tilde{n}_{H} = 0$.

\subsubsection{Limiting the number of type P nodes for exceptional groups}

In the last section, we saw that, in principal, one can have arbitrarily many type P nodes with exceptional groups because any representation content satisfies $B=0$. Here, we will see that only a finite number of these nodes can ever be coupled with others in such a way that results in a totally anomaly free model. 

\textbf{Case 1 - $M_{1}$ is a type P node with exceptional group}

Let $M_{1}$ denote a type P node with an exceptional gauge group $G_{1} \in \{E_{8},E_{7},E_{6},F_{4}\}$. Let's consider how it can couple to other nodes.

\textbf{Case 1a} - \emph{$M_{1}$ coupled to classical type P nodes}

Suppose we have another node $M_{2}$ which corresponds to a type P node with a classical group. Since both $M_{1}$ and $M_{2}$ are type P, if they couple, they must do so non-trivially, as mentioned earlier. Therefore, there must be a representation of $M_{2}$ with multiplicity at least the dimension of the minimal representation of $G_{1}$. Since $G_{1}$ is exceptional, the minimal multiplicity must be 26. However, as mentioned in the previous section, there are no type P nodes with classical groups where the multiplicity of any representation exceeds 14 (for the groups we are considering). Hence, $M_{1}$ and $M_{2}$ can never couple.

\textbf{Case 1b} - \emph{$M_{1}$ coupled to any type O nodes}

Similarly, suppose $M_{2}$ corresponds to a type O node with any group. If the nodes couple then they must do so non-trivially, in which case the multiplicity of some representation must be at least 26. Hence, the type O node must have an $\textrm{Sp}(N)$ group. Moreover, this can couple to at most 1 type P node with exceptional group, and the exceptional group must be $E_{6}$ or $F_{4}$ (else we exceed the maximum multiplicity of 32).

Suppose we also want to couple a type N node. Then the type O node must also couple non-trivially to that node. This means that the minimal non-trivial representation has to have dimension less than $32-26=6$, which is not the case for us. Hence, we cannot also couple type N nodes. By the same reasoning, if we add more type O nodes then they must couple trivially to the other type O nodes. We are therefore left with the following possibility for the total group $G$ of the composite model.
\begin{equation}
    \mbox{Case 1b:} \qquad G = G_{1}\times G_{\text{O}}\, , \qquad G_{\text{O}}\sim \text{type O}
\end{equation}
where $G_{\text{O}}$ denotes the semisimple group corresponding to all the type O nodes.

We have the following 2 bounds on the total value of $n_{H} - n_{V}$ for the composite model.
\begin{equation}\label{eq:case_2_bounds}
    n_{H} - n_{V} \geq (\tilde{n}_{H}^{1} - n_{V}^{1}) - \dim G_{\text{O}} \, ,\qquad n_{H} - n_{V} \geq -\dim G_{1} + \sum_{i}(\tilde{n}_{H}^{i} - n_{V}^{i})
\end{equation}
Here, $\tilde{n}_{H}^{1}$ corresponds to the total non-trivial representation content of node 1, while $\tilde{n}_{H}^{i}$ label the non-trivial representation content of the type O nodes. Similarly, $n_{V}^{1} = \dim G_{1}$ and $n_{V}^{i}$ correspond to the dimension of the $i^{\text{th}}$ simple factor of $G_{\text{O}}$. Note that, since all the type O nodes couple trivially, the total contribution to $n_{H} - n_{V}$ is given by the sum of the individual values, hence the second expression.

Using the values for $\tilde{n}_{H} - n_{V}$ for the type O nodes with $\textrm{Sp}(N)$ from table \ref{tab:B=0_solns}, we can use the second bound to get a bound on the rank of $G_{\text{O}}$, and hence on $\dim G_{\text{O}}$. For this to be a totally anomaly free theory, we need $n_{H}-n_{V}= 244$ and so we have
\begin{equation}
    244 \geq -78 + \sum_{\text{Type O}}(30N_{i} - 1)
\end{equation}
The largest value we can obtain for $\dim G_{\text{O}}$ such that the above is still satisfied is when $G_{\text{O}} = \textrm{Sp}(10)$. Using the first inequality in \eqref{eq:case_2_bounds}, we then have
\begin{equation}
    \mbox{Case 1b:} \qquad \tilde{n}_{H}^{1} - n_{V}^{1} \leq 244 + 210 = 454
\end{equation}

\textbf{Case 1c} - \emph{$M_{1}$ coupled to any type N and exceptional type P nodes}

The final case to consider is if we couple $M_{1}$ to type P nodes with exceptional groups and type N nodes, of which there can be at most 2. Let us write the total group as
\begin{equation}
\mbox{Case 1c:} \qquad     G = G_{1} \times E_{8}^{\alpha} \times E_{7}^{\beta} \times E_{6}^{\gamma} \times F_{4}^{\delta} \times G_{\text{N}} \, ,\qquad G_{\text{N}} \sim \text{type N}
\end{equation}
As noted, here $G_{\text{N}}$ corresponds to the groups with type N nodes (it can be trivial) and the remaining groups have type P. Since all of the type P nodes must couple non-trivially, we have that the number of hypermultiplets is at least
\begin{equation}
    n_{H} \geq k(248\alpha + \tfrac{1}{2}56\beta + 27\gamma +26\delta)
\end{equation}
where $k$ is the dimension of the minimal non-trivial representation of $G_{1}$.

To determine the contribution from the N nodes we will consider the various possibilities. First, suppose that there are no type N nodes, so $G_{\text{N}}$ is trivial. Then we have that
\begin{align}
\begin{split}
    &\text{Case 1c, \quad $G_{\text{N}}$ trivial:}\\
    &n_{H} - n_{V} \geq k(248\alpha + \tfrac{1}{2}56\beta + 27\gamma +26\delta) - \dim G_{1} - 248\alpha - 133\beta - 78\gamma - 52\delta
\end{split}
\end{align}
Using the fact that $k\geq 26$, $\dim G_{1}\leq 248$, and we require the left hand side of the above equation to be 244, we find that the above equation can only be solved if $\alpha=\beta=\gamma=\delta=0$. That is, we cannot couple $M_{1}$ to any other exceptional P nodes without any N nodes.

We therefore have to couple at least one type N node. Moreover, by the arguments above, it must couple non-trivially to \emph{all} the exceptional type P nodes. Suppose for now that it is the following type N node:
\begin{equation}
    \text{Assume: } \quad G_{\text{N}} = \textrm{SU}(N), \, (N-8)\textbf{N} + \mathbf{\tfrac{1}{2}N(N+1)}
\end{equation}
We can easily run the following argument with the other classical type N nodes, listed in table \ref{tab:B=0_solns}, and we find this is the least restrictive case and hence we will only explicitly check this case here.

Since the type N node couples non-trivially to all type P nodes, we must have that the multiplicity of the fundamental exceeds the sum of the smallest representations of the type P nodes. That is
\begin{align}
    \begin{split}\label{eq:1c_inequality_1}
    &\text{Case 1c, } \quad G_{\text{N}} = SU(N), \, (N-8)\textbf{N} + \mathbf{\tfrac{1}{2}N(N+1)}:\\
    &(248\alpha+\tfrac{1}{2}56\beta+27\gamma+26\delta +k)\leq N-8
    \end{split}
\end{align}
We can then use the following bound on $n_{H} - n_{V}$:
\begin{align}
    \begin{split}
    &\text{Case 1c, } \quad G_{\text{N}} = SU(N), \, (N-8)\textbf{N} + \mathbf{\tfrac{1}{2}N(N+1)}:\\
    &n_{H} - n_{V} \geq (\tilde{n}_{H}-n_{V})_{\text{type N}} - \dim G_{1} - (248\alpha+133\beta+78\gamma+52\delta)
    \end{split}
\end{align}
Using the expression for $\tilde{n}_{H} - n_{V}$ for the type N node given in table \ref{tab:B=0_solns}, we then get
\begin{equation}
    n_{H} - n_{V} \geq \tfrac{1}{2}N^{2} - \tfrac{15}{2}N + 1 - \dim G_{1} - (248\alpha+133\beta+78\gamma+52\delta)
\end{equation}
Using \eqref{eq:1c_inequality_1} in the above equation, and the fact that $k\geq 26$ and $\dim G_{1}\leq 248$ we find that $n_{H} - n_{V}=244$ only if $\alpha=\beta=\gamma=\delta=0$. We can then use the equation above to solve for $N$, assuming $\dim G_{1} \leq 248$. In this case, we find $N\leq 39$.

In fact, we can do better than this. If $G_{1} = E_{8}$, then the type N node must couple non-trivially, implying $N-8 \geq 248$ which contradicts $N\leq 39$. Hence, we can further assume $\dim G_{1}\leq 133$, in which case $N\leq 35$. We can now put a bound on the value of $\tilde{n}_{H}^{1} - n_{V}^{1}$ for $M_{1}$. We have
\begin{align}
    \begin{split}\label{eq:case1c_bound}
        &\text{Case 1c, } \quad G_{\text{N}} = SU(N), \, (N-8)\textbf{N} + \mathbf{\tfrac{1}{2}N(N+1)}:\\
        &\tilde{n}^{1}_{H} - n_{V}^{1} \leq 244 + \dim \textrm{SU}(35) = 1468
    \end{split}
\end{align}
As mentioned, we can run the same argument with the other classical type P nodes and we would find a stricter bound.

We could have 2 classical type N nodes. However, it is not possible to couple them to exceptional type P nodes. See case 3a for a proof of this. Moreover, we cannot have 1 classical type N node and one exceptional type N node for the groups we are considering. See case 3a again for a proof of this.

The final cases to check then are when $G_{N}$ is a product of 1 or 2 exceptional groups. This case is very simple to check because we have
\begin{align}
    \begin{split}\label{eq:1c_constraint_2}
        &\text{Case 1c, } \quad G_{\text{N}} = \text{Exceptional groups}:\\
        & n_{H} - n_{V} \geq k(248\alpha + \tfrac{1}{2}56\beta + 27\gamma +26\delta) - \dim G_{1} - 248\alpha - 133\beta - 78\gamma - 52\delta - \dim G_{\text{N}}
    \end{split}
\end{align}
Once again using $\dim G_{1} \leq 248$, and $\dim G_{\text{N}}\leq 496$, we find that $n_{V} - n_{H} = 244$ only if $\alpha = 0$, $\beta+\gamma+\delta\leq 1$. That is, we can couple exactly 1 other exceptional type P and it must be $F_{4}$, $E_{6}$ or $E_{7}$. Substituting this into the expression above, we find a bound on the value of $\tilde{n}_{H}^{1} - n_{V}^{1}$ for $M_{1}$ to be
\begin{align}
    \begin{split}
        &\text{Case 1c, } \quad G_{\text{N}} = \text{Exceptional groups}:\\
        & n^{1}_{H} - n^{1}_{V} \leq 244+496+133=873
    \end{split}
\end{align}
The least restrictive bound we have found here is \eqref{eq:case1c_bound}. This puts a bound on the possible exceptional type P nodes that will ever be part of a totally anomaly free model. Note that we do not expect this bound to be strict. Instead, we have just shown that anything not satisfying that bound will \emph{never} be part of a totally anomaly free model.

\subsubsection{Bounding the rank of classical groups}

For each classical group, there are only a finite number of solutions to the $\tr F^{4}$ constraint. Hence, to obtain a finite number of nodes, we need only find an upper bound to the rank of the groups such that groups of higher rank can never be part of a fully anomaly free model. Recall from section \ref{sec:comments_on_NOP}, for large enough classical groups there are only type N and type O solutions to the $\tr F^{4}$ constraint. Hence, we need only consider how these solutions couple to other nodes.

\textbf{Case 2 - $M_{1}$ is a type O node with classical group}

Let $M_{1}$ denote a type O node with a classical group. From table \ref{tab:B=0_solns} we see that this can only be $\textrm{SU}(N)$ or $\textrm{Sp}(N)$.

\textbf{Case 2a} - \emph{$M_{1}$ coupled non-trivially to other type O nodes}

Suppose $M_{1}$ combined with other type O nodes. Then they may or may not couple non-trivially. They couple non-trivially if there are at least one $(-,0)$ and one $(0,+)$ node coupled. In this case, all type O nodes couple non-trivially to at least one other. Since the type O nodes have representations with multiplicity less than or equal to 32, this limits the rank of the classical groups to be such that their minimal representation is less than or equal to 32. This limits us to $\textrm{SU}(32)$, $\textrm{SO}(32)$ and $\textrm{Sp}(16)$. From now on, we will assume that no type O nodes couple non-trivially.

\textbf{Case 2b} - \emph{$M_{1}$ coupled trivially to type O nodes}

If the type O nodes are combined with type P or N nodes then they necessarily couple non-trivially. If there are any type O nodes with $\textrm{SU}(N)$ then, since they have representations with multiplicity at most 16, we must have that the sum of the smallest representations of the N/P nodes must be less than or equal to 16. The largest group with this property is $\textrm{Sp}(16)$ (using the fact that the fundamental $\textbf{32}$ is pseudo-real and so contributes with a factor of $\tfrac{1}{2}$). Since we are assuming that none of the type O nodes couple non-trivially to each other, their contributions to $n_{H} - n_{V}$ is precisely the sum of the individual values. Hence we have
\begin{equation}\label{eq:type_O_bound}
    \begin{aligned}
        n_{H} - n_{V} &\geq \sum_{\text{Type O}}(\tilde{n}_{H}^{i} - n^{i}_{V}) - \dim G_{\text{N/P}} \ ,\\
        &\geq \sum_{\text{Type O}}(\tilde{n}_{H}^{i} - n_{V}^{i}) - 528\ ,
    \end{aligned}
\end{equation}
where $G_{\text{N/P}}$ denotes the semisimple group which corresponds to the combined N and P nodes in the composite model. Again, for this to be part of a totally anomaly free model, we need the right hand side of the above equation to be less than or equal to 244. We also saw in section \ref{sec:comments_on_NOP} that, for classical groups $\textrm{SU}(N)$ and $\textrm{Sp}(N)$, their value for $\tilde{n}^{i}_{H} - n_{V}$ is at least $15N-1$. The least restrictive constraint we have on the value of $N$ comes from having exactly 1 type O node, in which case we need
\begin{align}
\mbox{Case 2b, $M_{1}$ with $\textrm{SU}(N)$:} \qquad
 &\tilde{n}_{H}^{1} - n_{V}^{1} \leq 773 \\[3pt] \Rightarrow  \quad & 15N-1\leq 773 \quad \Rightarrow \quad N\leq 51\ .
\end{align} 
If instead there are only type O nodes with $\textrm{Sp}(N)$ groups, then the maximum multiplicity of any node is 32. Hence, the sum of the minimal representations of the N/P nodes cannot exceed 32. The largest groups for which this is possible is $\textrm{SU}(32)$.\footnote{We cannot take $\textrm{Sp}(32)$ in this case because the fundamental of the type O nodes is already pseudo-real and hence we already have a factor of $\tfrac{1}{2}$. The fundamental of $\textrm{Sp}(32)$ would therefore couples and contribute with a factor of $+1$, leading to a multiplicity in the type O node of 64 which cannot happen.} Running the argument as above in this case we find
\begin{align}
\mbox{Case 2b, $M_{1}$ with $\textrm{Sp}(N)$:} \qquad &\tilde{n}_{H}^{1} - n_{V}^{1} \leq 1267 \\[3pt]
\Rightarrow \quad& 30N+1 \leq 1267 \quad \Rightarrow \quad N\leq 42
\end{align}

\textbf{Case 3 - $M_{1}$ a type N node with classical group}

The final case we need to consider is the type N nodes for classical groups. We have already seen that if they are coupled to type O nodes then we cannot have groups larger than $\textrm{SU}(32)$, $\textrm{Sp}(32)$, or $\textrm{SO}(16)$. Hence, we need only consider what happens when we couple them to other type N or P nodes. We may as well assume that the minimal dimension of the non-trivial representations is $N>28$.

\textbf{Case 3a} - \emph{$M_{1}$ coupled to another type N node}

Let us suppose at first that there is another type N node $M_{2}$. If they couple, they must do so non-trivially. Hence, there must be some representation in $M_{2}$ whose multiplicity is at least $N > 28$. We can see from section \ref{sec:comments_on_NOP} that there are no type N nodes with exceptional groups with this property, so $M_{2}$ must correspond to a classical group.\footnote{In fact, we cannot couple an exceptional type N node to $M_{1}$ for the groups we are considering. This would require the exceptional node to have multiplicity at least $N$ for $M_{1}$ with groups $\textrm{SU}(N),\textrm{SO}(N), \textrm{Sp}(N)$. Looking at table \ref{tab:Exceptional_type_O}, we see that the exceptional type N nodes all have multiplicities $<8$ and hence we need $N<8$ which goes outside the initial assumptions we made.} A careful analysis of the ghost conditions shows that the only possible combination is
\begin{align}
        M_{1} &= \{ \textrm{SU}(N),\, 2N\, \textbf{N}\}, && v_{1} = (2,2)  \\
        M_{2} &= \{ \textrm{Sp}(M),\, 2(2M+8)\,\textbf{2M}\}, && v_{2} = (-2,-1)
\end{align}
or vice versa. In this case we find
\begin{equation}
    v^{0}_{1}v^{1}_{2} + v^{1}_{1}v^{0}_{2} = -6
\end{equation}
which implies we have 3 pseudo-real joint representations $(\textbf{N},\textbf{2M})$. We therefore have the following constraints on $N$ and $M$.
\begin{equation}
    \text{Case 3a: }\quad 3M\leq 2N, \quad 3N\leq 4M+16, \quad \Rightarrow \quad N\leq 48, \ M\leq 32.
\end{equation}
Now suppose we couple another node $M_{3}$ to it which by assumption must be type P. Given the values of $v_{1}, v_{2}$, the combinations $v_{1}\cdot v_{3}$ and $v_{2}\cdot v_{3}$ cannot both simultaneously vanish, implying that it will couple non-trivially to at least one of $M_{1}$ and $M_{2}$. This means that it must have a representation of multiplicity at least 15, by our assumption on the rank of the gauge group for $M_{1}$ made at the beginning of this subsection.

The only possibility is if $M_{1}$ and $M_{2}$ are coupled to a P node with an exceptional group. Note that, in this case, we cannot form a non-trivial triple representation between $M_{1}, M_{2}, M_{3}$ else the multiplicity of the joint representation $(\textbf{N},\textbf{2M})$ between $M_{1}, M_{2}$ would need to be $\geq 26$. However, it still must couple non-trivially to at least one of $M_{1}$ or $M_{2}$. Suppose there are joint representations with total multiplicity $k$ between $M_{1}$ and $M_{3}$, and with total multiplicity $k'$ between $M_{2}$ and $M_{3}$. Using the fact that the minimal representation of $F_{4}, E_{6}, E_{7}, E_{8}$ is 26, we find the following new constraints on $N$ and $M$.
\begin{equation}
    3M+26k \leq 2N\, , \quad 3N+26k' \leq 4M+16
\end{equation}
However, these imply $N,M<0$ unless $k,k'=0$. Hence, we cannot couple any type P nodes in this case.

\textbf{Case 3b} - \emph{$M_{1}$ coupled non-trivially only to type P nodes}

We may now assume that we have exactly 1 type N node, given by $M_{1}$. If it couples to any other nodes, they must be type P. Indeed, if we couple type O nodes then they must do so non-trivially which implies the type O nodes must have a multiplicity exceeding 14, a contradiction. If the P nodes couple non-trivially, then we are in the situation explored in case 1a. There we found (at least in the case that $M_{1} = \{\textrm{SU}(N),\, (N-8)\textbf{N} + \mathbf{\tfrac{1}{2}N(N+1)} \}$, although it is easy to extend this to the other cases) that it can couple non-trivially to at most 1 exceptional type P node and it cannot be $E_{8}$. We therefore have
\begin{align}
\mbox{Case 3b:} \qquad
\tilde{n}^{1}_{H} - n_{V}^{1} \leq 244+133 = 377
\label{cond2}
\end{align}
Using table \ref{tab:B=0_solns}, we can bound the maximum rank of the gauge group with this. The least restrictive value of $N$ follows from $\textrm{SO}(N)$ with $(N-8)\textbf{f}$, for the type N node $M_{1}$. This gives $N\leq 35$.

\textbf{Case 3c} - \emph{$M_{1}$ couples trivially to type P nodes}

The only other possibility is if the type P nodes all couple trivially with the type N node $M_{1}$. They then contribute independently to $n_H - n_V$. That is,
\begin{equation}
    n_{H} - n_{V} \geq (\tilde{n}_{H}^{1} - n_{V}^{1}) + (\tilde{n}_{H} - n_{V})_{\text{type P}}
\end{equation}
It is easy to check that any individual type P node has a positive value of $\tilde{n}_{H} - n_{V}$, in which case we find that the model $M_{1}$ is bounded by $\tilde{n}_{H}^{1} - n_{V}^{1} \leq 244$. We therefore assume that we have more than one type P node $M_{2}, M_{3},...$. Since they are all type P they must couple non-trivially. From the previous section we cannot have a combination of exceptional and classical type P nodes. If they are all classical, then the rank of the gauge groups in $M_{3},...$ must be bounded by the largest multiplicity of any representation in $M_{1}$. From section \ref{sec:comments_on_NOP}, we see that the largest possible multiplicity is 14. Hence, the sum of the smallest representations in $M_{3},...$ must not exceed 14. The largest (possible semi-simple) group for which this is possible is $\textrm{Sp}(14)$ (taking into account that the fundamental \textbf{28} is pseudo-real so contributes with a factor if $\tfrac{1}{2}$). This has dimension 406 and so we have
\begin{equation}
    \mbox{Case 3c, Classical P nodes:} \qquad \tilde{n}_{H}^{1} - n_{V}^{1} \leq 244+ 406 = 650 \quad \Rightarrow \quad N \leq 44
\end{equation}
If the P groups are all exceptional, then from the previous section there can be at most 2 of them and their total dimension must be less than $248 + 133 = 381$. We then have
\begin{equation}
    \mbox{Case 3c, Exceptional P nodes:} \qquad \tilde{n}_{H}^{1} - n_{V}^{1} \leq 244 + 381 = 625 \quad \Rightarrow \quad N\leq 43
\end{equation}
In both cases above, the inequality on the right hand side is using the least restrictive constraint on the rank of the classical gauge groups coming from table \ref{tab:B=0_solns}.

\subsubsection{Summary}

We have found a set of bounds on the value of $\tilde{n}_{H}^{1} - n_{V}^{1}$ for certain nodes $M_{1}$. Any nodes which do not satisfy these bounds can never be part of a totally anomaly free model. Hence, from the point of view of the graphical method of finding anomaly free models, we can remove them from the infinite initial set of nodes $\mathcal{S}$.

The least restrictive bound we found for the exceptional nodes was
\begin{equation}
    \tilde{n}_{H}^{1} - n_{V}^{1} \leq 1468
\end{equation}
We then found a bound on the rank of classical groups, the least restrictive case being $N\leq 51$. We could use table \ref{tab:B=0_solns} to determine the maximal value of $\tilde{n}_{H}-n_{V}$ for these ranks. For simplicity, we simply write here\footnote{In fact, our code used precisely this to put a bound on the starting set $\hat{\mathcal{S}}$, i.e. a bound on the rank of classical groups and on the value $\tilde{n}_{H} - n_{V}$ for exceptional groups.}
\begin{equation}
    \hat{\mathcal{S}} = \{ M_{1} \in \mathcal{S}\, |\, \text{Exceptional groups:}\, \tilde{n}_{H}^{1} - n_{V}^{1} \leq 1468, \ \text{Classical groups:}\, N\leq 51 \}
\end{equation}
This will put a bound on both the possible type P exceptional nodes and the value of $N$ for the classical nodes.

We emphasise that with this initial set, we can find \emph{all} anomaly free models with the groups and representations considered in \eqref{gm}. This includes all products of such groups. From our solutions, listed in appendix \ref{app:nt1_solns}, we see that there can be at most a product of 4 simple groups. In this case, they are all products of $\textrm{SU}(N)$ groups. Note that this search does not include groups with low rank, where we expect many more anomaly free theories to appear, even with many products of simple gauge factors. For example, \cite{1986MPLA....1..267B} found an anomaly free model of the form
\be
E_{6}\times E_{6}\times E_{6} \times \textrm{SU}(3):\quad  
    \big[(\textbf{27},\textbf{1},\textbf{1},\textbf{3}) + (\textbf{1},\textbf{27},\textbf{1},\textbf{3}) + (\textbf{1},\textbf{1},\textbf{27},\textbf{3})\big] + ( {\bf 3} \leftrightarrow \bar{\bf 3} )\ .
\ee
The $\textrm{SU}(3)$ factor means that this is outside the range of our search and hence does not appear in our list. Moreover, our search does not allow for abelian factors in the gauge group, and in particular drones. These are $U(1)$ factors under which no hypermultiplet is charged. Including such possibilities would also greatly increase the number of anomaly free models.

\subsection{Limiting the number of loops}

In the previous section, we found that one could reduce the infinite set of quasi-anomaly free and ghost free models with a single gauge group $\mathcal{S}$ to a finite subset $\hat{\mathcal{S}}$ which will contain all of the models present in totally anomaly free theories (within the assumptions given in \eqref{gm}). However, we also saw in section \ref{sec:graph_method_procedure} that the total graph made from these nodes may contain loops from some node $S_{1}$ to itself. These loops correspond to the possibility of having multiple copies of $S_{1}$ within a larger, composite model. To ensure that the procedure we outline for finding new anomaly free models will terminate, we need to show that one can couple a single node to itself a finite number of times.

Firstly, note that if the node $S_{1}$ couples non-trivially to itself then the multiplicities of the representations in $S_{1}$ bound the number of times one can couple it to itself. Indeed, if one couples it $k$ times, then one needs
\begin{equation}
    k \dim \textbf{R} \leq n_{R}
\end{equation}
If instead the node $S_{1}$ couples trivially with itself then it must be type O. In that case \eqref{eq:type_O_bound} limits the number of times one can self-couple $S_{1}$.

\section{Conclusions}

We have extended considerably the previous searches for anomaly free $(1,0)$ supergravities in six dimensions. This is done by considering a larger set of groups and representations. We have considered R-symmetry (un)gauged models but restricted our attention to the number of tensor multiplets $n_T=0,1,2$. We have applied two methods for the search. One of them entails a composition of two quasi-anomaly free models (in the sense that all anomalies vanish except possibly the gravitational anomaly proportional to $\tr R^4$ term in the anomaly polynomial), to obtain an anomaly free model. Another method is based on the partition of a give number of hyperfermions into all possible representations of the full group, and the requirement of the proper factorization of the resulting anomaly polynomials. In both cases, we have also required that the positivity of the gauge kinetic terms in the action in an appropriate region in the dilaton moduli space. 

One of the notable results is that the inclusion of the $R$-symmetry gauge group reduces drastically the number of possible anomaly free models. This was already noted in \cite{Avramis:2005hc} under the conditions in which only limited set of groups of the form $G_1\times G_2\times G_R$, where $G_R$ is either $U(1)_R$ or $SU(2)_R$, and the remainder of the group semi-simple, and only low lying representations for hyperfermions were considered. Here we find that even if we allow all possible representation and extend the group to include a third factor, among several thousands of possible models one can write down, no anomaly free ones arise for $n_T=0,2$, and in the case of $n_T=1$ only six models with groups of the form $G_1\times G_2\times G_3\times U(1)_R$ arise, in addition to the three models known previously with gauge group of the form $G_1\times G_2 \times U(1)_R$. 

As to the $R$-symmetry ungauged models with $n_T=1$, their existence for groups with two or more factors is nontrivial since the Green-Schwarz mechanism requires the anomaly matrices that encode the information on the factorized anomaly polynomial must have less than maximal rank. Yet, it is remarkable that we find a large number of solutions that satisfy this rank reduction. A number of them have already been identified with compactifications of heterotic string on $K3$ with certain $SU(2)$ instanton embeddings in \cite{Avramis:2005hc} where some chains of solutions obtained from them by successive Higgsing were also identified. These have been marked with a $\dagger$ in the appendix. However, it is far from clear if a large number of the other solutions admit a string theory interpretation. It is also noteworthy that while several of this single tensor multiplet coupled models seem to be very similar in many respects, only those which are related to string theory are certain to admit UV completion, while for the rest this is, of course, not ensured, and as such they are candidates to be examined in the framework of the swampland program.  

The are natural further directions to consider in the search for and study of anomaly free supergravities in six dimensions. In addressing the problem of complete classification of the anomaly free models in 6-dimensions, primary challenges have to do with the presence of infinite families of solutions for $n_{T}\geq 9$ \cite{Kumar:2010ru}. If we include abelian factors in the gauge group, the situation becomes even more difficult, with infinite families with different $U(1)$ charges appearing, even for $n_{T} = 0$ \cite{Taylor:2018khc}. Even restricting to semi-simple groups with $n_{T}=1$, the presence of low rank groups with generic matter representations poses a considerable challenge to the enumeration of anomaly free models. This is due to the large number of solutions to the quartic Casimir, and the potential to have non-trivial joint representations under more than 2 groups. Indeed, an enormous number of models with gauge group consisting of only $U(1)$ and $SU(2)$ factors have been found in \cite{Suzuki:2005vu}. Nonetheless, we expect that one can extend the proof in section \ref{sec:proof_of_finite_nodes} to lower rank groups and more general representations. Our proof largely relied on the form of the solutions in table \ref{tab:B=0_solns}, which are the most general solutions to the quartic Casimir, provided the rank of the group is large enough. Considering how these can combine with lower rank groups will give new, but solvable, constraints on the rank of the simple gauge groups in the quasi-anomaly free nodes - making a complete classification of models at $n_T = 1$ possible.

Another interesting open problem is the complete classification of $R$-symmetry gauged anomaly free models. The graphical method, as it has been described, does not seem particularly well suited for finding gauged supergravities. In these cases, the anomaly matrix depends explicitly on the total dimension of the gauge group, including all simple gauge factors. It is not clear, therefore, whether one can decompose a totally anomaly free gauged model into individual quasi-anomaly free gauged models for each simple gauge group factor. One could expand the definition of quasi-anomaly free to include theories with both gravitational and R-symmetry anomalies. The graphical method could then be used as above to construct anomaly free models with larger gauge groups, imposing the vanishing gravitational and R-symmetry anomalies at the end. However, unlike in the ungauged case, the R-symmetry anomaly imposes very strong constraints on the theories, so this method does not seem particularly efficient and does not elude to some deeper structure of gauged models. It would be interesting to see if one could adapt these ideas to produce an efficient algorithm for finding all gauged models. Conversely, a lack of graphical representation of the theories may have interesting consequences for their consistency with string theory as we expand on below.

Next, in the long list of solutions that we have already found, it would be useful to determine which ones may be arising from the compactification of string/M theory, including the F theory constructions, and their successive Higgsing. For example, a general discussion on how one can build 6-dimensional supergravities from F-theory through elliptically fibred Calabi-Yau 3-folds was done in \cite{Kumar:2009ac}. The non-abelian gauge groups arise from certain singular divisors in the base $B$ of the fibration over which D7 branes are wrapped. The matter content arises due to the intersections of these divisors. In \cite{Kumar:2009ac} it was observed that certain matter configurations seem not to correspond to integral divisors and hence are perhaps not consistent with F-theory constructions. For example, a gauge group $\textrm{SU}(N)$ with $(N-8)\textbf{N} + \mathbf{\tfrac{1}{2}N(N+1)}$ does not map to an integral divisor. We note that in our solutions we have many such models. These, along with solutions with other exotic matter are natural starting points for analysis by further consistency conditions coming from the swampland programme. In particular, the three remarkably anomaly free models with $G_1\times G_2\times U(1)_R$ group structure that have been around for a long time now, and the six new remarkably anomaly free models with the $G_1\times G_2\times G_3 \times U(1)_R$ group structure we have found here, deserve a special attention, as they exhibit unusual features that have been impossible to obtain from string/M theory, or the F theory constructions so far. 

It is interesting to note that in our definition of quasi-anomaly-free models we allowed for gravitational anomalies. Such theories can be perfectly consistent as theories provided we go to some decoupling limit where gravity is frozen out. From an F-theory perspective, this would correspond to reducing on some Calabi-Yau with non-compact base. The graphical method then reflects our intuition of building compact geometries (which would produce totally anomaly free models) by gluing together non-compact patches (quasi-anomaly free nodes).\footnote{For an overview of SCFTs constructed from F-theory via non-compact reductions, see e.g. \cite{Heckman:2018jxk}.} Our results seem to put strict constraints on the possible gluings, however. A lack of graphical method for the gauged models may also indicate that these cannot be constructed from F-theory, at least not in a conventional geometric sense.

Finally, it is an open and very interesting problem to determine which one of the growing list of models free from perturbative anomalies may be ruled out by a through study of their potential global anomalies. This is closely related to the question of whether one can give a well-defined global definition of the Green-Schwarz counter term and they are guaranteed to vanish if the bordism group $\Omega^{\text{spin}}_{7}(BG)$ of the classifying space $BG$ of the gauge group $G$ is trivial \cite{Monnier:2018nfs}. If this does not vanish then Dai-Freed anomalies can arise \cite{Garcia-Etxebarria:2018ajm}. Recently, such anomalies were considered in the context of six-dimensional supergravities \cite{Basile:2023zng} and new constraints were found. While most them fall outside of the range we consider here, their constraints do rule out one of the theories listed below, with gauge groups $E_{7}$. Other constraints to consider are also the ability to embed the lattice defined by the anomaly matrix into a unimodular lattice \cite{Seiberg:2011dr}, and various constraints coming from the swampland. Of the latter, a particularly intriguing route is the consistency under the insertion of BPS brane probes \cite{Kim:2019vuc}. The studies on the global anomalies mentioned above deal with ungauged $N=(1,0)$ supergravities in six-dimensions. Much less is known about global anomalies in the case of $R$-symmetry gauged such theories. An attempt has been made to this end in \cite{Pang:2020rir}, and a possible relation to charge lattices arising in a closely related $N=(1,1), 6D$ theory was discussed recently in \cite{Ma:2023tcj}.

\subsection*{Acknowledgements} 

E.S. would like to acknowledge useful discussions with Guillaume Bossard, Axel Kleinschmidt, Hong Lu and Yi Pang. The work of K.B., E.S. and D.T. is supported in part by  NSF grant PHYS-2112859.

\begin{appendix}

\section{Solutions with $n_T=1$}\label{app:nt1_solns}

\subsection{E,C, EE and EC models}\label{app:E_C_EE_EC}

In what follows we have revised the $Sp(N)$ solutions of \cite{Avramis:2005hc}. Furthermore, we have added anomaly free (local and global) models with exceptional group $G_2$ symmetry, that have no hyperfermion singlets, and  have positive gauge kinetic terms. Among the solutions with $G_2$ gauge group, note that only the last one contains a representation that goes beyond the low lying ones assumed in \cite{Avramis:2005hc}. 

The ghost free condition was not taken into account in \cite{Avramis:2005hc}. Indeed, there are models in \cite{Avramis:2005hc} that do not satisfy the ghost free condition. An example is the $SU(N)\times SU(N)$ model in Eq.(3.92) of \cite{Avramis:2005hc} with matter $2(N,N)$. In the list of anomaly free models we provide in this appendix we do take into account the ghost free condition.

In the case of $E_6\times F_4$, only even number of possible \textbf{27}-plets of $E_6$ were considered in \cite{Avramis:2005hc}.  Among the solutions for the ungauged models listed in \cite{Avramis:2005hc}, only $E_7\times SO(11)$ model with hyperfermions in $3 (\textbf{133},\textbf{1}) + 3 (\textbf{1},\textbf{11})$, and the $F_4 \times SU(11)$ model with $5 (\textbf{26},\textbf{1}) + 16 (\textbf{1},\textbf{11}) + 2(\textbf{1},\textbf{55})$  have no singlets.  

Thus, all solutions of \cite{Avramis:2005hc} with corrections stated above and new solutions that we have found for the E, C, EE and EC groups are listed below\footnote{In view of recent paper \cite{Basile:2023zng}, the second $E_7$ model below has global anomalies, but we list it here for completeness.} The ones marked by $^\star$ have no hypersinglets, and those marked by dagger have been identified in \cite{Avramis:2005hc} as coming from heterotic string theory compactification on $K3$ with specific embedding of $SU(2)$ instantons, and/or from Higgsing phenomena.
\begin{enumerate}
\item $E_7$
\begin{enumerate}
\item[(a)$^{\dagger}$] $n\times {\bf 56}\ ,\quad n=0,...,13$ 
\item[(b)] $8\times {\bf 56} + {\bf 133}$ 
\end{enumerate}

\item $E_6$
\begin{enumerate}
\item[(a)$^{\dagger}$] $n\times {\bf 27}\ ,\quad n=0,1,...,11$ 
\item[(b)] $ 3n \times \textbf{27} + (4-n)\times \textbf{78} \ , \quad n=0,...,3 $ 
\item $8\times {\bf 27}+ {\bf 78}$ 
\end{enumerate}

\item  $F_4$
\begin{enumerate}
\item[(a)$^{\dagger}$] $n\times {\bf 26}\ ,\quad n=0,...,11$ 
\item[(b)]  $8\times {\bf 26} + {\bf 52}$ 
\item[(c)] $(11-2n)\times {\bf 26} + n\times {\bf 52} \ , \quad n = 1,...,5 $  
\end{enumerate}

\item $G_2$
\begin{enumerate} 
\item  $(33-2n)\times\textbf{7}+n\times\textbf{14}+1\times\textbf{2}7, \quad n=2,5,8,11,14$ 
\item  $2\times\textbf{7}+2\times\textbf{14}+8\times\textbf{27}$ 
\item  $2\times\textbf{7}+9\times\textbf{14}+2\times\textbf{27}+1\times\textbf{64}$ 
\item  $\textbf{27}+2\times\textbf{77’}+\textbf{77}$ 
\end{enumerate}

\item $SU(N)$
\begin{align}
&N=11: && \textrm{(a) }(22-3n)\times \textbf{11} + n\times \textbf{55} \ , \quad n=0,...,5 
\nn\\
&&& \textrm{(b) }3\times \textbf{11}+\textbf{66}
\nn\\[3pt]
&N=10: && \textrm{(a) } (20-2n)\times  \textbf{10} + n\times \textbf{45}\ , \quad n=0,...,5
\nn\\
&&&\textrm{(b) }2\times \textbf{10} + \textbf{55} 
\nn\\[3pt]
&N=8: &&\textrm{(a) }  8\times {\bf 28} +{\bf 63}
\nn\\[3pt]
&N=7: && \textrm{(a) }  8\times {\bf 7} +8 \times {\bf 21} + {\bf 48}
\nn\\[3pt]
&N=6: &&\textrm{(a) } 16\times {\bf 6} +8 \times {\bf 15} + {\bf 35}
\nn\\
&&& \textrm{(b) }8\times {\bf 6} +10 \times {\bf 15} + 2\times {\bf 35}
\nn\\
&&& \textrm{(c) }8\times {\bf 6} +12 \times {\bf 15} + 2\times {\bf 21}
\nn\\[3pt]
&N=5: &&\textrm{(a) }  24\times {\bf 5} +8 \times {\bf 10} + {\bf 24}
\nn \\
&&&\textrm{(b) } 20\times {\bf 5} +10 \times {\bf 10} + 2\times {\bf 24}
\nn\\
&&& \textrm{(c) }6\times {\bf 5} +12 \times {\bf 10} + 4\times {\bf 24}
\nn\\
&&& \textrm{(d) }6\times {\bf 5} +14 \times {\bf 10} + 2\times {\bf 15} +2\times{\bf 24}
\nn\\
&&& \textrm{(e) } 6\times {\bf 5} +14 \times {\bf 10} + 4\times {\bf 15}
\nn\\
&&&\textrm{(f) } 20\times {\bf 5} +12 \times {\bf 10} + 2\times {\bf 15}
\nn
\end{align}

\item $SO(N)$
\begin{align}
&10\le N\le 30: && \textrm{(a)$^{\dagger}$ }  (N-8)\times {\bf N}  
\nn\\[3pt]
& 10\le N \le 12: && \textrm{(a)$^{\dagger}$ }  8\times {\bf N} +{\bf \frac{N(N-1)}{2}} + 8 \times {\bf 2^{[\frac{N+1}{2}]-1}}  
\nn\\[3pt]
&N=14: &&\textrm{(a) }  (4n+6)\times {\bf 14} + 2n \times {\bf 64}\ ,\quad n=1,2
\nn\\
&&& \textrm{(a) } 8\times \textbf{14} + \textbf{91} + 2\times \textbf{64}
\nn\\[3pt]
&N=13: && \textrm{(a) } (2n+5)\times {\bf 13} + n \times {\bf 64}\ ,\quad n=1,...,4
\nn\\
&&& \textrm{(b) } 8\times \textbf{13} + \textbf{78} + 4\times \textbf{64}
\nn\\[3pt]
&N=12: && \textrm{(a)$^{\dagger}$ }  (n+4)\times {\bf 12} +n \times {\bf 32} \ ,\quad n=1,...,9
\nn\\
&&&\textrm{(b) } 4 \times {\bf 12} +8\times {\bf 32}  + 2\times {\bf 66}
\nn\\[3pt]
&N=11: &&\textrm{(a)$^{\dagger}$ }   (n+3)\times {\bf 11} +n\times {\bf 32} \ , \quad n=1,...,9
\nn\\
&&& (13-4n)\times {\bf 11} +(10-n) \times {\bf 32} + n\times \textbf{55}\ , \quad n=1,...,3
\nn\\[3pt]
&N=10: && \textrm{(a)$^{\dagger}$ }   (n+2)\times {\bf 10} +n\times {\bf 16} \ , \quad n=1,...,10
\nn \\
&&& \textrm{(b) } (12-3n)\times {\bf 10} +(10-n) \times {\bf 16} + n\times {\bf 45}\ , \quad n=1,...,4
\nn
\end{align}

\item $Sp(N)$
\begin{align}
&N=9: && \textrm{(a) } 12\times{\bf 18}+2\times{\bf 152} 
\nn\\[3pt]
&N=8: && \textrm{(a) } (48-16n)\times{\bf 16}+n\times{\bf 119} \ ,\quad\ n=1,2,3
\nn\\[3pt]
&N=7: && \textrm{(a) } (44-12n)\times{\bf 14}+n\times{\bf 90} \ ,\qquad n=0,1,2,3
\nn\\[3pt]
&N=6: && \textrm{(a) } (40-8n)\times{\bf 12}+n\times{\bf 65} \ ,\qquad\ \  n=0,1,...,4
\nn\\[3pt]
&N=5: && \textrm{(a) } (36-4n)\times{\bf 10}+n\times{\bf 44}\ ,\quad \  n=0,1,...,4
\nn\\
&&& \textrm{(b) } 45\times{\bf 10}+1\times{\bf 110} 
\nn
\end{align}

\item {$\textrm{E}_8 \times \textrm{E}_7$} 
\begin{enumerate}
\item[(a)$^{\dagger}$] {$ 20 ( \textbf{1},\textbf{56} )$} 
\item[(b)] $3(\textbf{1},\textbf{56}) + 4(\textbf{1},\textbf{133})$  
\end{enumerate}

\item {$\textrm{E}_8 \times \textrm{E}_6$ }
\begin{enumerate}
\item[(a)$^{\dagger}$] {$ 18 ( \textbf{1},\textbf{27} )$} 
\item {$9(\textbf{1},\textbf{27}) + 4(\textbf{1},\textbf{78})$ } 
\end{enumerate}

\item {$\textrm{E}_8 \times \textrm{F}_4$ }
\begin{enumerate}
\item {$ 17 ( \textbf{1},\textbf{26} )$} 
\item {$12(\textbf{1},\textbf{26}) + 4(\textbf{1},\textbf{52})$ } 
\end{enumerate}

\item {$\textrm{E}_7 \times \textrm{E}_7$ }
\begin{enumerate}
\item[(a)$^{\dagger}$] $ {n ( \textbf{56},\textbf{1} ) + (16-n) ( \textbf{1},\textbf{56} )\ ,\quad n=0,...,8} $ 
\item[(b)] $9(\textbf{56},\textbf{1}) + 4(\textbf{1},\textbf{56}) + n({\bf 133})\ ,\quad n=0,1$  
\end{enumerate}

\item {$\textrm{E}_7 \times \textrm{E}_6$ }
\begin{enumerate}
\item[(a)$^{\dagger}$] $ {n ( \textbf{56},\textbf{1} ) + (14-n) ( \textbf{1},\textbf{27} )\ ,\quad n=0,...,14 } $
\item[(b)] $9(\textbf{56},\textbf{1}) + 2(\textbf{1},\textbf{27})$  
\item[(c)] $9(\textbf{56},\textbf{1}) + 2(\textbf{1},\textbf{27}) + (\textbf{133},\textbf{1})$  
\item[(d)] $2(\textbf{1},\textbf{27}) + 3(\textbf{133},\textbf{1})$ 
\item[(e)] $(n+2)(\textbf{56},\textbf{1}) + 2n(\textbf{1},\textbf{27}) + (5-n)(\textbf{1},\textbf{78})\ ,\quad n=0,1,2 $  
\item[(f)] {$7 ( \textbf{1},\textbf{27} ) + 4 ( \textbf{56},\textbf{1} )$} 
\item[(g)] {$9 ( \textbf{1},\textbf{27} ) +  ( \textbf{1},\textbf{78} ) + 4 ( \textbf{56},\textbf{1} )$} 
\end{enumerate}

\item {$\textrm{E}_7 \times \textrm{F}_4$ }
\begin{enumerate}
\item $ n ( \textbf{56},\textbf{1} ) + (13-n) ( \textbf{1},\textbf{26} )\ ,\quad n=0,...,13 $ 
\item {$ n(\textbf{56},\textbf{1}) + (n+2)(\textbf{1},\textbf{26}) + (7-n)(\textbf{1},\textbf{52})\ , \quad n=0,1,2,...,6 $}
\item $4(\textbf{56},\textbf{1}) + 6(\textbf{1},\textbf{52})$  
\item $4(\textbf{56},\textbf{1}) + 6(\textbf{1},\textbf{26})$  
\item $4(\textbf{56},\textbf{1}) + (\textbf{1},\textbf{52}) + 9(\textbf{1},\textbf{26})$  
\item $9(\textbf{56},\textbf{1}) + (\textbf{1},\textbf{26})$  
\item $9(\textbf{56},\textbf{1}) + (\textbf{1},\textbf{26})  + (\textbf{133},\textbf{1})$  
\item $(\textbf{1},\textbf{26})  + 3(\textbf{133},\textbf{1})$  
\end{enumerate}

\item {$\textrm{E}_6 \times \textrm{E}_6$ }
\begin{enumerate}
\item $ n ( \textbf{27},\textbf{1} ) + (12-n) ( \textbf{1},\textbf{27} )\ ,\quad n=0,...,6  $
\item $ n(\textbf{27},\textbf{1}) + 2n(\textbf{1},\textbf{27}) + (5-n)(\textbf{1},\textbf{78}) \ ,\quad n=0,...,3  $
\item {$7 ( \textbf{27},\textbf{1} ) + 2 ( \textbf{1},\textbf{27} )$} 
\item {$9 ( \textbf{27},\textbf{1} ) +  ( \textbf{78},\textbf{1} ) + 2 ( \textbf{1},\textbf{27} )$} 
\end{enumerate}

\item  $\textrm{E}_6 \times \textrm{F}_4$
\begin{enumerate}
\item $(11-k)(\textbf{27},\textbf{1}) + k(\textbf{1},\textbf{26})$ \qquad $k=0,1,...,10,11$ 

\item $k(\textbf{27},\textbf{1}) + (4+k)(\textbf{1},\textbf{26}) + (5-k)(\textbf{1},\textbf{52})$ \qquad $k=0,1,2,3,4,5$ 

\item $2k(\textbf{27},\textbf{1}) + (5-k)(\textbf{78},\textbf{1}) + (k-1)(\textbf{1},\textbf{26})$ \qquad $k = 1,2,3,4,5^*$ 
\item {$2 ( \textbf{27},\textbf{1} ) + 6 ( \textbf{1},\textbf{26} )$} 
\item {$2 ( \textbf{27},\textbf{1} ) + 6 ( \textbf{1},\textbf{52} )$} 
\item {$2 ( \textbf{27},\textbf{1} ) + 9 ( \textbf{1},\textbf{26} ) +  ( \textbf{1},\textbf{52} )$} 
\item {$7 ( \textbf{27},\textbf{1} ) +  ( \textbf{1},\textbf{26} )$} 
\item {$9 ( \textbf{27},\textbf{1} ) +  ( \textbf{78},\textbf{1} ) +  ( \textbf{1},\textbf{26} )$} 
\end{enumerate} 

\item { $\textrm{F}_4 \times \textrm{F}_4$ }
\begin{enumerate}
\item $n ( \textbf{26},\textbf{1} ) +  (10-n)( \textbf{1},\textbf{26} )\ ,\quad n=0,...,5 $
\item $n ( \textbf{26},\textbf{1} ) + (n+5) ( \textbf{1},\textbf{26} )+(4-n)( \textbf{1},\textbf{52} ) \ ,\quad n=0,...,4 $
\item $ ( \textbf{26},\textbf{1} ) +  6( \textbf{1},\textbf{52} ) $
\item $ ( \textbf{26},\textbf{1} ) +  9( \textbf{1},\textbf{26} ) +( \textbf{1},\textbf{52} ) $
\item {$6 ( \textbf{26},\textbf{1} ) +  ( \textbf{1},\textbf{26} )$} 
\end{enumerate}


\item $G_2\times E_8$ 
\begin{enumerate} 
\item $4({\bf  7,1})+3({\bf  14,1})+2({\bf  64,1})+4({\bf  77,1}) $
\end{enumerate} 

\item $G_2\times E_7$ 
\begin{enumerate} 
\item $2({\bf 1,56})+({\bf 7,1})+3({\bf 14,1})+3({\bf 27,1})+2({\bf 64,1})+({\bf 77,1})$ 
\item $5({\bf 1,56})+24({\bf 7,1})+4({\bf 14,1})+({\bf 27,1})$ 
\end{enumerate} 

\item $G_2\times E_6$ 
\begin{enumerate} 
\item $2({\bf  77,1})+({\bf  182,1})$  

\item $2({\bf  7,1})+10({\bf  14,1})+2({\bf  27,1})+2({\bf  64,1})$  
\end{enumerate} 

\item $G_2 \times F_4$ 
\begin{enumerate} 
\item $({\bf 1,26})+({\bf 7,1})+({\bf 7,26})+({\bf 14,1})+3({\bf 27,1})$ 
\item $({\bf 1,26})+3({\bf 7,1})+5({\bf 27,1})+2({\bf 64,1})$ 
\item $({\bf 1,26})+27({\bf 7,1})+({\bf 14,1})+3({\bf 27,1})$ 
\end{enumerate} 

\item $G_2\times G_2$ 
\begin{enumerate} 
\item $ (\textbf{27},\textbf{1}) + (\textbf{7},\textbf{7}) + (\textbf{7},\textbf{14}) + (\textbf{14},\textbf{7}) $ 
\item $ (\textbf{7},\textbf{1}) + 2(\textbf{1},\textbf{27}) + (\textbf{1},\textbf{64}) + (\textbf{7},\textbf{7}) + (\textbf{7},\textbf{14}) $ 
\item $ 2(\textbf{7},\textbf{1}) + 3(\textbf{27},\textbf{1}) + 2(\textbf{64},\textbf{1}) + (\textbf{7},\textbf{7}) $ 
\item $9(\textbf{7},\textbf{1}) + 5(\textbf{14},\textbf{1}) + (\textbf{27},\textbf{1}) +2(\textbf{1},\textbf{7}) + 2(\textbf{7},\textbf{7})$ 
\item $10(\textbf{7},\textbf{1}) + (\textbf{27},\textbf{1}) + 4(\textbf{1},\textbf{7}) + 3(\textbf{7},\textbf{7})$ 
\item $ 12(\textbf{7},\textbf{1}) + 2(\textbf{14},\textbf{1}) + (\textbf{27},\textbf{1}) + 5(\textbf{1},\textbf{7}) + 2(\textbf{7},\textbf{7}) $ 
\item $ 19(\textbf{7} ,\textbf{1}) + 6(\textbf{14},\textbf{1}) + (\textbf{27},\textbf{1}) + 4(\textbf{1},\textbf{7})$ 
\item $26(\textbf{7},\textbf{1}) + (\textbf{14},\textbf{1}) + (\textbf{27},\textbf{1}) + (\textbf{7},\textbf{7})$ 
\end{enumerate}

\item $E_8 \times \textrm{SO}(N)$:
\begin{align}
& N=14: && \textrm{(a) }22 (\mathbf{1},\mathbf{14}) + 4 (\mathbf{1},\mathbf{64}).
\nn\\[3pt]
& N=13: && \textrm{(a) }21 (\mathbf{1},\mathbf{13}) + 8 (\mathbf{1},\mathbf{64}).
\nn\\[3pt]
& N=12: && \textrm{(a) } 20 (\mathbf{1},\mathbf{12}) + 16 (\mathbf{1},\mathbf{32}),
\nn\\
                             &&&\textrm{(b) } 4 (\mathbf{1},\mathbf{66}) + 3 (\mathbf{1},\mathbf{12}) + 15 (\mathbf{1},\mathbf{32}).
\nn\\
                            &&&\textrm{(c) } 3(\textbf{1},\textbf{12}) + (\textbf{1},\textbf{66}) + 23 (\textbf{1},\textbf{32}) + (\textbf{1},\textbf{77})
\nn\\[3pt]
& N=11: && \textrm{(a) } 19 (\mathbf{1},\mathbf{11}) + 16 (\mathbf{1},\mathbf{32}),
\nn\\
                             &&&\textrm{(b) } 4 (\mathbf{1},\mathbf{55}) + 6 (\mathbf{1},\mathbf{11}) + 15 (\mathbf{1},\mathbf{32}).
\nn\\
                            &&&\textrm{(c) } 4(\textbf{1},\textbf{11}) + (\textbf{1},\textbf{55}) + 23(\textbf{1},\textbf{32}) + (\textbf{1},\textbf{65})
\nn\\[3pt]
& N=10: && \textrm{(a) }  18 (\mathbf{1},\mathbf{10}) + 16 (\mathbf{1},\mathbf{16}),
\nn\\
                             &&&\textrm{(b) } 4 (\mathbf{1},\mathbf{45}) + 9 (\mathbf{1},\mathbf{10}) +  15 (\mathbf{1},\mathbf{16}).
\nn\\
                            &&&\textrm{(c) } 5(\textbf{1},\textbf{10}) + (\textbf{1},\textbf{45}) + 23(\textbf{1},\textbf{16}) + (\textbf{1},\textbf{54})
\nn
\end{align}

\item $E_8 \times \textrm{Sp}(N)$:
\begin{align}
& N=5: && \textrm{(a) }5(\textbf{1},\textbf{10}) + 10(\textbf{1},\textbf{44}) + (\textbf{1},\textbf{110}) 
\nn\\[3pt]
& N=4: && \textrm{(a) }16 (\mathbf{1},\mathbf{8} ) + 13 \left( \mathbf{1},\mathbf{27} \right).
\nn
\end{align}

\item $E_7 \times \textrm{SU}(N)$:
\begin{align}
&10\leq N\leq 24: && \textrm{(a) } 4(\textbf{56},\textbf{1}) + (24-N)(\textbf{1},\textbf{N}) + 3(\textbf{1},\mathbf{\tfrac{1}{2}N(N-1)})
\nn\\
    &&& \textrm{(b) } 9 (\textbf{56},\textbf{1}) + (N-8)(\textbf{N},\textbf{1}) + (\textbf{1},\mathbf{\tfrac{1}{2}N(N+1)}
\nn\\[3pt]
&10\leq N\leq 19: && \textrm{(a) } 12 (\textbf{56},\textbf{1}) + (N-8)(\textbf{1},\textbf{N}) + (\textbf{1},\mathbf{\tfrac{1}{2}N(N+1)})
\nn\\[3pt]
&10\leq N\leq 13 : && \textrm{(a) } 9(\textbf{56},\textbf{1}) + (\textbf{133},\textbf{1}) + (N-8)(\textbf{1},\textbf{N}) + (\textbf{1},\mathbf{\tfrac{1}{2}N(N+1)}
\nn\\[3pt]
&N=13: && \textrm{(a) } 5 ( \textbf{56},\textbf{1} ) +  ( \textbf{1},\textbf{13} ) + 5 ( \textbf{1},\textbf{78} )
\nn\\[3pt]
&N=12: &&  \textrm{(a) } (4+n) (\mathbf{56},\mathbf{1}) + 4n ( \textbf{1},\textbf{12} ) + (6-n)(\mathbf{1},\mathbf{66}) \ , && n=0,1
\nn\\[3pt]
&N=11: && \textrm{(a) } (3+n)(\textbf{56},\textbf{1}) + (1+3n)(\textbf{1},\textbf{11}) +(7-n)(\textbf{1},\textbf{55}) \ , && n=0,1,2
\nn\\[3pt]
&N=10: && \textrm{(a) } n(\textbf{56},\textbf{1}) + 2n(\textbf{1},\textbf{10}) + (10-n) (\textbf{1}
,\textbf{45}) \ , && n=0,1,...,8
\nn\\
    &&& \textrm{(b) } 3 ( \textbf{133},\textbf{1} ) + 2 ( \textbf{1},\textbf{10} ) +  ( \textbf{1},\textbf{55} )
\nn
\end{align}

\item $E_7 \times \textrm{SO}(N)$:
\begin{align}
& 10 \leq N \leq 25: && \textrm{(a) } 9 (\mathbf{56},\mathbf{1}) + (N-8) (\mathbf{1},\mathbf{N}).
\nn\\[3pt]
& 10 \leq N \leq 19: && \textrm{(a) } 12 (\mathbf{56},\mathbf{1}) + (N-8) (\mathbf{1},\mathbf{N}).
\nn\\[3pt]
& 10 \leq N \leq 13: && \textrm{(a) } 9 (\mathbf{56},\mathbf{1}) + (\mathbf{133},\mathbf{1}) + (N-8) (\mathbf{1},\mathbf{N}).
\nn\\[3pt]
& N=16: &&  \textrm{(c) } 4 (\mathbf{56},\mathbf{1}) + 16 (\mathbf{1},\mathbf{16}) + (\mathbf{1},\mathbf{128}).
\nn\\[3pt]
& N=15: &&  \textrm{(c) } 4 (\mathbf{56},\mathbf{1}) + 15 (\mathbf{1},\mathbf{15}) + (\mathbf{1},\mathbf{128}).
\nn\\[3pt]
& N=14: && \textrm{(a) }  (12-4n) (\mathbf{56},\mathbf{1}) + (4n+6) (\mathbf{1},\mathbf{14}) + n(\mathbf{1},\mathbf{64}); &&n=1,2,3. 
\nn\\[3pt]
& N=13: && \textrm{(a) }  (12-2n) (\mathbf{56},\mathbf{1}) + (2n+5) (\mathbf{1},\mathbf{13}) + n (\mathbf{1},\mathbf{64}); &&n=1,\ldots,6.
\nn\\[3pt]
& N=12: &&  \textrm{(a) } (12-n) (\mathbf{56},\mathbf{1}) + (n+4) (\mathbf{1},\mathbf{12}) + n (\mathbf{1},\mathbf{32}); &&n=1,\ldots,12. 
\nn\\
    &&& \textrm{(b) }4( \textbf{56},\textbf{1} )+8 ( \textbf{1},\textbf{32} ) + 3( \textbf{1},\textbf{66} )
\nn\\
    &&& \textrm{(c) } 4 ( \textbf{56},\textbf{1} )+ 9 ( \textbf{1},\textbf{12} ) + 9 ( \textbf{1},\textbf{32} ) +( \textbf{1},\textbf{66} )
\nn\\
    &&& \textrm{(d) } 4(\textbf{56},\textbf{1}) + 9(\textbf{1},\textbf{12}) + 5(\textbf{1},\textbf{32})
\nn\\[3pt]
& N=11: &&  \textrm{(a) } (12-n) (\mathbf{56},\mathbf{1}) + (n+3) (\mathbf{1},\mathbf{11}) + n (\mathbf{1},\mathbf{32}); &&n=1,\ldots,12, 
\nn\\
    &&&\textrm{(b) } 3 (\mathbf{133},\mathbf{1}) + 3 (\mathbf{1},\mathbf{11}).
\nn\\
    &&&\textrm{(c) }4 ( \textbf{56},\textbf{1} ) + 9 ( \textbf{1},\textbf{11} ) +  ( \textbf{1},\textbf{55} ) + 9 ( \textbf{1},\textbf{32} )
\nn\\
    &&&\textrm{(d) }2 ( \textbf{1},\textbf{11} ) + 8 ( \textbf{1},\textbf{32} ) + 4( \textbf{56},\textbf{1} ) + 3 ( \textbf{1},\textbf{55} )
\nn\\
    &&&\textrm{(e) }8 ( \textbf{1},\textbf{11} ) + 5 ( \textbf{1},\textbf{32} ) + 4( \textbf{56},\textbf{1} ) 
\nn\\
    &&&\textrm{(f) }3 ( \textbf{1},\textbf{11} ) + 9( \textbf{56},\textbf{1} ) + ( \textbf{133},\textbf{1} )
\nn\\[3pt]
& N=10: && \textrm{(a) } (12-n) (\mathbf{56},\mathbf{1}) + (n+2) (\mathbf{1},\mathbf{10}) + n (\mathbf{1},\mathbf{16}); &&n=1,\ldots,12, 
\nn\\
    &&&\textrm{(b) } (n+2) (\mathbf{56},\mathbf{1}) + (5-n) (\mathbf{1},\mathbf{45}) + 2n (\mathbf{1},\mathbf{10})
\nn\\
&&& \qquad + 8 (\mathbf{1},\mathbf{16}); &&n=0,\ldots,4.
\nn\\
    &&&\textrm{(c) } 3 (\mathbf{133},\mathbf{1}) + 2 (\mathbf{1},\mathbf{10}), 
\nn\\
    &&&\textrm{(d) } 4 (\mathbf{56},\mathbf{1}) + (\mathbf{1},\mathbf{45}) + 9 (\mathbf{1},\mathbf{10}) + 9 (\mathbf{1},\mathbf{16}),
\nn\\
    &&&\textrm{(e) } 4( \textbf{56},\textbf{1} ) + ( \textbf{1},\textbf{54} ) + 16 ( \textbf{1},\textbf{16} )
\nn\\
    &&&\textrm{(f) } 4( \textbf{56},\textbf{1} ) + 7( \textbf{1},\textbf{10} ) + 5 ( \textbf{1},\textbf{16} )
\nn
\end{align}

\item $E_7 \times \textrm{Sp}(N)$:
\begin{align}
& 5 \leq N \leq 12: && \textrm{(a) }  4 (\mathbf{56},\mathbf{1}) + (48-4N) (\mathbf{1},\mathbf{2N}) + 2 (\mathbf{1},\mathbf{N(2N-1)-1}). 
\nn\\[3pt]
& N=6: && \textrm{(a) }  5 (\mathbf{56},\mathbf{1}) + 8 (\mathbf{1},\mathbf{12}) + 4 (\mathbf{1},\mathbf{65}) .
\nn\\
    &&& \textrm{(b) } 4(\textbf{56},\textbf{1}) + 5(\textbf{1},\textbf{65})
\nn\\
    &&& \textrm{(c) } { 4( \textbf{1},\textbf{12} ) + 5 ( \textbf{1},\textbf{65} ) + (\textbf{1},\textbf{208})}
\nn\\[3pt]
& N=5: && \textrm{(a) } (9-n) (\mathbf{56},\mathbf{1}) + (36-4n) (\mathbf{1},\mathbf{10}) + n (\mathbf{1},\mathbf{44}); && n=0,\ldots,9.
\nn\\
    &&& \textrm{(b) } { 21 ( \textbf{1},\textbf{10} ) + 6 ( \textbf{1},\textbf{44} ) +  ( \textbf{1},\textbf{110} )}
\nn\\
    &&& \textrm{(c) } { ( \textbf{56},\textbf{1} ) + 25 ( \textbf{1},\textbf{10} ) + 5 ( \textbf{1},\textbf{44} ) +  ( \textbf{1},\textbf{110} )}
\nn\\
    &&& \textrm{(d) } { 2( \textbf{56},\textbf{1} )+ 29 ( \textbf{1},\textbf{10} ) + 4 ( \textbf{1},\textbf{44} ) +  ( \textbf{1},\textbf{110} )}
\nn
\end{align}

\item $E_6 \times \textrm{SU}(N)$:
\begin{align}
&10\leq N\leq 25: && \textrm{(a) } 7(\textbf{27},\textbf{1}) + (N-8)(\textbf{1},\textbf{N}) + (\textbf{1},\mathbf{\tfrac{1}{2}N(N+1)})
\nn\\[3pt]
&10\leq N\leq 24: && \textrm{(a) } 2(\textbf{27},\textbf{1}) + (24-N)(\textbf{1},\textbf{N}) + 3(\textbf{1},\mathbf{\tfrac{1}{2}N(N-1)})
\nn\\[3pt]
&10\leq N\leq 20: && \textrm{(a) } 10(\textbf{27},\textbf{1}) + (N-8)(\textbf{1},\textbf{N}) + (\textbf{1},\mathbf{\tfrac{1}{2}N(N+1)})
\nn\\[3pt]
&10\leq N\leq 15^{*}: && \textrm{(a) } 9(\textbf{27},\textbf{1}) + (\textbf{78},\textbf{1}) + (N-8)(\textbf{1},\textbf{N}) + (\textbf{1},\mathbf{\tfrac{1}{2}N(N+1)}
\nn\\[3pt]
&N=13: && \textrm{(a) } 3 ( \textbf{27},\textbf{1} ) +  ( \textbf{1},\textbf{13} ) + 5 ( \textbf{1},\textbf{78} )
\nn\\[3pt]
& N = 12: && \textrm{(a) } 2 (\mathbf{27},\mathbf{1} ) + 6 (
\mathbf{1},\mathbf{66} ). 
\nn\\
    &&& \textrm{(b) } 3 ( \textbf{27},\textbf{1} ) + 4 ( \textbf{1},\textbf{12} ) + 5 ( \textbf{1},\textbf{66} )
\nn\\[3pt]
& N = 11: && \textrm{(a) } n (\textbf{27},\textbf{1}) + (3n-2)(\textbf{1},\textbf{11}) + (8-n)(\textbf{1},\textbf{55}); && n=1,...,4
\nn\\
    &&& \textrm{(b) } 4 ( \textbf{27},\textbf{1} ) + 3 ( \textbf{78},\textbf{1} ) + 3 ( \textbf{1},\textbf{11} ) +  ( \textbf{1},\textbf{66} )
\nn\\[3pt]
& N = 10: && \textrm{(a) }  n (\mathbf{27},\mathbf{1}) + \left[ 4 + 2n \right] (\mathbf{1},\mathbf{10}) + (8-n) \left( \mathbf{1},\textbf{45} \right); &&n=0,\ldots,8. 
\nn\\
    &&& \textrm{(b) } 4 ( \textbf{27},\textbf{1} ) + 3 ( \textbf{78},\textbf{1} ) + 2 ( \textbf{1},\textbf{10} ) +  ( \textbf{1},\textbf{55} )
\nn
\end{align}

\item $E_6 \times \textrm{SO}(N)$: 
\begin{align}
& 10 \leq N \leq 25: && \textrm{(a) } 7 (\textbf{27},\textbf{1}) + (N-8)(\textbf{1},\textbf{N})
\nn\\[3pt]
& 10 \leq N \leq 20: && \textrm{(a) } 10 (\mathbf{27},\mathbf{1}) + (N-8) (\mathbf{1},\mathbf{N}).
\nn\\[3pt]
& 10\leq N \leq 15: && \textrm{(a) } 9(\textbf{27},\textbf{1}) + (\textbf{78},\textbf{1}) + (N-8)(\textbf{1},\textbf{N})
\nn\\[3pt]
&10\leq N\leq 12 : && \textrm{(a) } n (\mathbf{27},\mathbf{1}) + (N+2-n) (\mathbf{1},\mathbf{N}) + (10-n) (\mathbf{1},\mathbf{2^{\lfloor\tfrac{N-1}{2}\rfloor}}); &&n=0,\ldots,9, 
\nn\\
    &&& \textrm{(b) } 2(\textbf{27},\textbf{1})+ 9(\textbf{1},\textbf{N}) + (\textbf{1},\mathbf{\tfrac{1}{2}N(N-1)}) + 9(\textbf{1},\mathbf{2^{\lfloor\tfrac{N-1}{2}\rfloor}})
\nn\\
    &&& \textrm{(c) } 2(\textbf{27},\textbf{1}) + (N-3)(\textbf{1},\textbf{N}) + 5(\textbf{1},\mathbf{2^{\lfloor\tfrac{N-1}{2}\rfloor}})
\nn\\[3pt]
&N=16: &&  \textrm{(a) } 2 (\mathbf{27},\mathbf{1}) + 16 (\mathbf{1},\mathbf{16}) + (\mathbf{1},\mathbf{128}).
\nn\\[3pt]
& N=15: &&  \textrm{(a) } 2 (\mathbf{27},\mathbf{1}) + 15 (\mathbf{1},\mathbf{15}) + (\mathbf{1},\mathbf{128}).
\nn\\[3pt]
& N=14: && \textrm{(a) }  (6-4n) (\mathbf{27},\mathbf{1}) + (4n+10) (\mathbf{1},\mathbf{14}) + (n+1) (\mathbf{1},\mathbf{64}); &&n=0,1.
\nn\\[3pt]
& N=13: && \textrm{(a) }  (10-2n) (\mathbf{27},\mathbf{1}) + (2n+5) (\mathbf{1},\mathbf{13}) + n(\mathbf{1},\mathbf{64}); &&n=1,\ldots,5, 
\nn\\
    &&& \textrm{(b)\textsuperscript{*} } 2(\textbf{27},\textbf{1}) + 8(\textbf{1},\textbf{64}) + (\textbf{1},\textbf{90})
\nn\\[3pt]
& N=12: &&  \textrm{(a) } 2 (\mathbf{27},\mathbf{1}) + 3 (\mathbf{1},\mathbf{66}) + 8 (\mathbf{1},\mathbf{32}).
\nn\\
    &&& \textrm{(b) } 2(\textbf{27},\textbf{1}) + 16(\textbf{1},\textbf{32}) + (\textbf{1},\textbf{77})    
\nn\\[3pt]
&N=11: && \textrm{(a) } 2(\textbf{27},\textbf{1}) + 16(\textbf{1},\textbf{32}) + (\textbf{1},\textbf{65})
\nn\\
    &&& \textrm{(b) } 2 (\mathbf{27},\mathbf{1}) + 3 (\mathbf{1},\mathbf{55}) + 2 (\mathbf{1},\mathbf{11}) + 8 (\mathbf{1},\mathbf{32}), 
\nn\\
    &&& \textrm{(c) } 3({\bf 27,1})+ 5({\bf 1,11})+8({\bf 1,32})+2({\bf 1,55})
\nn\\
    &&&\textrm{(d) }4 (\mathbf{27},\mathbf{1})+ 3 (\mathbf{78},\mathbf{1}) + 3 (\mathbf{1},\mathbf{11}).
\nn\\[3pt]
& N=10: &&  \textrm{(a) } n(\textbf{27},\textbf{1}) + 2n(\textbf{1},\textbf{10}) + (5-n)(\textbf{1},\textbf{45}) + 8(\textbf{1},\textbf{16}) && n=0,1,...,5
\nn\\
    &&& \textrm{(b) } (3-n) (\mathbf{78},\mathbf{1}) + (2n+4) (\mathbf{27},\mathbf{1}) + (n+2) (\mathbf{1},\mathbf{10}) + n (\mathbf{1},\mathbf{16}); &&n=0,1,2.
\nn\\
    &&& \textrm{(c) } 2(\textbf{27},\textbf{1}) + 16(\textbf{1},\textbf{16}) + (\textbf{1},\textbf{54})
\nn
\end{align}

\item $E_6 \times \textrm{Sp}(N)$: 
\begin{align}
&5 \leq N \leq 12: && \textrm{(a) }  2 (\mathbf{27},\mathbf{1}) + (48-4N) (\mathbf{1},\mathbf{2N}) + 2 (\mathbf{1},\mathbf{N(2N-1)-1}). 
\nn\\[3pt]
& N=6: && \textrm{(a) } (2+n)(\textbf{27},\textbf{1}) + 8n(\textbf{1},\textbf{12}) + (5-n)(\textbf{1},\textbf{65}); && n=0,1,2 
\nn\\[3pt]
&  N=5: &&  \textrm{(a) } n (\mathbf{27},\mathbf{1}) + (8+4n) (\mathbf{1},\mathbf{10}) + (7-n) (\mathbf{1},\mathbf{44}); && n=0,...,7.
\nn\\
    &&& \textrm{(b) } {29({\bf 1,10})+4({\bf 1,44})+({\bf 1,110})}
\nn
\end{align}

\item $F_4 \times \textrm{SU}(N)$:
\begin{align}
& 10\leq N\leq 25 : && \textrm{(a) } 6(\textbf{26},\textbf{1}) + (N-8)(\textbf{1},\textbf{N}) + (\textbf{1},\mathbf{\tfrac{1}{2}N(N+1)}) 
\nn\\[3pt]
& 10\leq N\leq 24: && \textrm{(a) } (\textbf{26},\textbf{1}) + (24-N)(\textbf{1},\textbf{N}) + 3(\textbf{1},\mathbf{\tfrac{1}{2}N(N-1)})
\nn\\[3pt]
& 10\leq N\leq 20 && \textrm{(a) } 9(\textbf{26},\textbf{1}) + (N-8)(\textbf{1},\textbf{N}) + (\textbf{1},\mathbf{\tfrac{1}{2}N(N+1)}) 
\nn\\[3pt]
& 10\leq N\leq 16: && \textrm{(a) } 9(\textbf{26},\textbf{1}) + (\textbf{52},\textbf{1})+ (N-8)(\textbf{1},\textbf{N}) + (\textbf{1},\mathbf{\tfrac{1}{2}N(N+1)}) 
\nn\\[3pt]
& 10 \leq N\leq 12: && \textrm{(a) } 6(\textbf{26},\textbf{1}) + 3(\textbf{52},\textbf{1}) + (N-8)(\textbf{1},\textbf{N}) + (\mathbf{\tfrac{1}{2}N(N+1)})
\nn\\
    &&& \textrm{(b) } 6(\textbf{52},\textbf{1}) + (N-8)(\textbf{1},\textbf{N}) + (\textbf{1},\mathbf{\tfrac{1}{2}N(N+1)})
\nn\\[3pt]
& N=13: && \textrm{(a) } 2 ( \textbf{26},\textbf{1} ) +  ( \textbf{1},\textbf{13} ) + 5 ( \textbf{1},\textbf{78} )
\nn\\[3pt]
& N = 12: && \textrm{(a) } (n+1)(\textbf{26},\textbf{1}) + 4n(\textbf{1},\textbf{12}) + (6-n)(\textbf{1},\textbf{66}); && n=0,1,2
\nn\\[3pt]
& N = 11: && \textrm{(a) } n(\textbf{26},\textbf{1}) + (1+3n)(\textbf{1},\textbf{11}) + (7-n)(\textbf{1},\textbf{66}); && n=0,1,...,5
\nn\\[3pt]
&N = 10: && \textrm{(a) } n (\mathbf{26},\mathbf{1}) + (6+2n) (\mathbf{1},\mathbf{10}) + (7-n) \left( \mathbf{1},\textbf{45} \right); &&n=0,...,7. 
\nn
\end{align}

\item $F_4 \times \textrm{SO}(N)$:
\begin{align}
& 10 \leq N \leq 25: && \textrm{(a) } 6 (\mathbf{26},\mathbf{1}) + (N-8) (\mathbf{1},\mathbf{N}).
\nn\\[3pt]
& 10 \leq N \leq 20: && \textrm{(a) } 9 (\mathbf{26},\mathbf{1}) + (N-8) (\mathbf{1},\mathbf{N}).
\nn\\[3pt]
& 10 \leq N \leq 16: && \textrm{(a) } 9 (\mathbf{26},\mathbf{1}) + (\mathbf{52},\mathbf{1}) + (N-8) (\mathbf{1},\mathbf{N}).
\nn\\[3pt]
&10\leq N\leq 12: && \textrm{(a) } n (\mathbf{26},\mathbf{1}) + (N+1-n) (\mathbf{1},\mathbf{N}) + (9-n) (\mathbf{1},\mathbf{2^{\lfloor\tfrac{N-1}{2}\rfloor}}); && n=0,\ldots,8, 
\nn\\
    &&& \textrm{(b) } 6(\textbf{52},\textbf{1}) + (N-8)(\textbf{1},\textbf{N})
\nn\\
    &&& \textrm{(c) } (\textbf{26},\textbf{1}) + 9(\textbf{1},\textbf{N}) + (\textbf{1},\mathbf{\tfrac{1}{2}N(N-1)}) + 9(\textbf{1},\mathbf{2^{\lfloor\tfrac{N-1}{2}\rfloor}})
\nn\\
    &&& \textrm{(d) } (\textbf{26},\textbf{1}) + (N-3)(\textbf{1},\textbf{N}) + 5(\textbf{1},\mathbf{2^{\lfloor\tfrac{N-1}{2}\rfloor}})
\nn\\[3pt]
& N=16:  && \textrm{(a) }  (\mathbf{26},\mathbf{1}) + 16 (\mathbf{1},\mathbf{16}) + (\mathbf{1},\mathbf{128})
\nn\\[3pt]
& N=15: && \textrm{(a)\textsuperscript{*} } (\textbf{26},\textbf{1}) + (\textbf{1},\textbf{119}) + 2(\textbf{1},\textbf{128})
\nn\\
    &&& \textrm{(b) } (\mathbf{26},\mathbf{1}) + 15 (\mathbf{1},\mathbf{15}) + (\mathbf{1},\mathbf{128})
\nn\\[3pt]
& N=14: && \textrm{(a) } (\textbf{26},\textbf{1}) + (\textbf{1},\textbf{104}) + 4(\textbf{1},\textbf{64})
\nn\\
    &&& \textrm{(b) }  (\mathbf{26},\mathbf{1}) + 14 (\mathbf{1},\mathbf{14}) + 2 (\mathbf{1},\mathbf{64}), 
\nn\\
    &&&\textrm{(c) } 5 (\mathbf{26},\mathbf{1}) + 10 (\mathbf{1},\mathbf{14}) + (\mathbf{1},\mathbf{64})
\nn\\[3pt]
& N=13: && \textrm{(a) } (7-2n) (\mathbf{26},\mathbf{1}) + (2n+7) (\mathbf{1},\mathbf{13}) + (n+1) (\mathbf{1},\mathbf{64}); &&  {n=0,\ldots,3.} 
\nn\\
    &&& \textrm{(b) } (\textbf{26},\textbf{1}) + (\textbf{1},\textbf{90}) + 8(\textbf{1},\textbf{64})
\nn\\[3pt]
& N=12: && \textrm{(a) } \textbf{(26},\textbf{1}) + (\textbf{1},\textbf{77}) + 16(\textbf{1},\textbf{32})
\nn\\
    &&&\textrm{(b) } (\mathbf{26},\mathbf{1}) + 3 (\mathbf{1},\mathbf{66}) + 8 (\mathbf{1},\mathbf{32}),
\nn\\
    &&&\textrm{(c) } 2 (\mathbf{26},\mathbf{1}) + 2 (\mathbf{1},\mathbf{66}) + 4 (\mathbf{1},\mathbf{12}) + 8 (\mathbf{1},\mathbf{32}),
\nn\\
    &&&\textrm{(d)\textsuperscript{*} } 7 (\mathbf{26},\mathbf{1}) +  2 (\mathbf{52},\mathbf{1}) + 5 (\mathbf{1},\mathbf{12}) + 1 (\mathbf{1},\mathbf{32}),
\nn\\
    &&&\textrm{(e) } 6 (\mathbf{26},\mathbf{1}) + 3 (\mathbf{52},\mathbf{1}) +  4 (\mathbf{1},\mathbf{12})
\nn\\
& N=11: && \textrm{(a) } (n+1) (\mathbf{26},\mathbf{1}) + (3-n) (\mathbf{1},\mathbf{55}) + (3n+2) (\mathbf{1},\mathbf{11}) + 8 (\mathbf{1},\mathbf{32}); &&n=0,\ldots,2, 
\nn\\
    &&&\textrm{(b) } (n+6) (\mathbf{26},\mathbf{1}) + (3-n) (\mathbf{52},\mathbf{1}) + (n+3) (\mathbf{1},\mathbf{11}) + n (\mathbf{1},\mathbf{32}); &&n=0,\ldots,2, 
\nn\\
    &&& \textrm{(c) } (\textbf{26},\textbf{1}) + (\textbf{1},\textbf{65}) + 16(\textbf{1},\textbf{32})
\nn\\
    &&&\textrm{(d) } 9 (\mathbf{26},\mathbf{1}) + 6 (\mathbf{1},\mathbf{11}) + 3 (\mathbf{1},\mathbf{32}),
\nn\\[3pt]
& N=10: && \textrm{(a) } (n+1) (\mathbf{26},\mathbf{1}) + (3-n) (\mathbf{1},\mathbf{45}) + (2n+4) (\mathbf{1},\mathbf{10}) + 8 (\mathbf{1},\mathbf{16}), && n=0,...,3
\nn\\
    &&&\textrm{(b) }  (n+6) (\mathbf{26},\mathbf{1})+ (3-n) (\mathbf{52},\mathbf{1}) + (n+2) (\mathbf{1},\mathbf{10}) + n (\mathbf{1},\mathbf{16}); &&n=0,\ldots,3, 
\nn\\
    &&& \textrm{(c) } (\textbf{26},\textbf{1}) + (\textbf{1},\textbf{54}) + 16(\textbf{1},\textbf{16})
\nn\\
    &&&\textrm{(d) } 4 (\mathbf{1},\mathbf{45}) + 2 (\mathbf{1},\mathbf{10}) + 8 (\mathbf{1},\mathbf{16}),
\nn
\end{align}

\item $F_4 \times \textrm{Sp}(N)$:
\begin{align}
&5 \leq N \leq 12: && \textrm{(a) } (\mathbf{26},\mathbf{1}) + (48-4N) (\mathbf{1},\mathbf{2N}) + 2 (\mathbf{1},\mathbf{N(2N-1)-1}).
\nn\\[3pt]
& N=6: && \textrm{(a) }  (n+1) (\mathbf{26},\mathbf{1}) + 8n (\mathbf{1},\mathbf{12}) + (5-n) (\mathbf{1},\mathbf{65});&& n=0,1,2 
\nn\\[3pt]
& N=5: && \textrm{(b) }  n (\mathbf{26},\mathbf{1}) + (12+4n) (\mathbf{1},\mathbf{10})+ (6-n) (\mathbf{1},\mathbf{44}); && n=0,\ldots,6.
\nn
\end{align}

\item $G_2\times \textrm{SU}(N)$   
\begin{align}
&N=10: && \textrm{(a) } 2({\bf 1,55})+2({\bf 1,120})+({\bf 7,1}) 
\nn\\
    &&& \textrm{(b) } 4({\bf 1,10})+({\bf 1,45})+13({\bf 7,1})+2({\bf 7,10})+({\bf 14,1})+({\bf 27,1})
\nn
\end{align}

\item $G_2 \times \textrm{SO}(N)$
\begin{align}
&N=16: && \textrm{(a) } ({\bf 1,16})+2({\bf 7,1})+({\bf 7,16})+4({\bf 27,1})+2({\bf 64,1})
\nn\\
    &&& \textrm{(b) } ({\bf 1,16})+26({\bf 7,1})+({\bf 7,16})+({\bf 14,1})+2({\bf 27,1})
\nn\\[3pt]
&N=14: && \textrm{(a) } 6({\bf 1,14})+10({\bf 7,1})+12({\bf 14,1})+({\bf 27,1})
\nn\\
    &&& \textrm{(b) } 6({\bf 1,14})+28({\bf 7,1})+3({\bf 14,1})+({\bf 27,1})
\nn\\[3pt]
& N=13: && \textrm{(a) }({\bf 1,64})+3({\bf 7,1})+({\bf 7,13})+3({\bf 64,1})
\nn\\
    &&& \textrm{(b) } 5({\bf 1,13})+13({\bf 7,1})+9({\bf 14,1})+2({\bf 27,1})
\nn\\
    &&& \textrm{(c) }({\bf 1,64})+15({\bf 7,1})+({\bf 7,13})+4({\bf 27,1})
\nn\\
    &&& \textrm{(d) } 5({\bf 1,13})+31({\bf 7,1})+2({\bf 27,1})
\nn\\[3pt]
&N=11: && \textrm{(a) } 3({\bf  1,7})+({\bf 7,1})+13({\bf  14,1})+({\bf 27,1})+({\bf 64,1}) 
\nn\\
    &&& \textrm{(b) } 3({\bf  1,7})+13({\bf  7,1})+7({\bf  27,1}) 
\nn\\[3pt]
&N=10: && \textrm{(a) } 3({\bf 1,10})+({\bf 1,16})+({\bf 7,1})+({\bf 7,10})+({\bf 14,1})+2({\bf 27,1})
\nn\\
    &&& \textrm{(b) } 2({\bf 1,10})+({\bf 7,1})+6({\bf 14,1})+3({\bf 64,1})
\nn\\
    &&& \textrm{(c) } 5({\bf 1,16})+18({\bf 7,1})+({\bf 7,10})+({\bf 27,1})
\nn\\
    &&& \textrm{(d) } 2({\bf 1,10})+19({\bf 7,1})+3({\bf 14,1})+4({\bf 27,1})
\nn\\
    &&& \textrm{(e) } 3({\bf 1,10})+({\bf 1,16})+27({\bf 7,1})+({\bf 14,1})+2({\bf 27,1})
\nn
\end{align}
\item$G_2\times Sp(N)$
\begin{align}
&N=7: && \textrm{(a) } 9({\bf 1,14})+5({\bf 7,14})+2({\bf 14,1})+({\bf 27,1})
\nn\\
    &&& \textrm{(b) } 11({\bf 1,14})+({\bf 1,90})+7({\bf 7,1})+3({\bf 7,14})
\nn\\[3pt]
&N=6: && \textrm{(a) } ({\bf 1,12})+4({\bf 1,65})+4({\bf 7,1})+({\bf 7,12})
\nn\\[3pt]
&N=5: && \textrm{(a) } 15({\bf 1,10})+13({\bf 7,1})+3({\bf 7,10})+3({\bf 14,1})
\nn
\end{align}

\end{enumerate}

\subsection{$C\times C$ models}

In the graphical method the groups $SU(N), SO(N), Sp(N)$ for $N\ge 10$ are considered, while in the rank method the groups $SU(10), SU(11), SO(10\le N\le 21)$ and $Sp(5\le N\le 10)$ are considered. It worth noting that there exist two infinite families of anomaly free models, namely (3.92) and (3.94) in \cite{Avramis:2005hc}, first found in \cite{Schwarz:1995zw}, which, however, are not ghost-free \cite{Kumar:2009ae}.

\begin{enumerate}
\item { $ \boxed{\textrm{SU}(N) \times \textrm{SU}(M)}$ 
\begin{enumerate}
\item  $ 10 \leq N \leq 16$, $N \leq M \leq 2N: $
\item[] {$ (2N-M)( \textbf{N},\textbf{1} ) + (16-N) ( \textbf{1},\textbf{M} ) + 2 ( \textbf{1},\mathbf{\frac{M(M-1)}{2}} ) +  ( \textbf{N},\textbf{M} )$  }
\item $ 10 \leq N \leq 16$, $N \leq M \leq N+8 : $
\item[] {$ (N+8 - M) (\textbf{N}, \textbf{1}) + (16 - N) (\textbf{1}, \textbf{M}) + (\mathbf{\frac{N(N-1)}{2}}, \textbf{1} ) + 2 (\textbf{1}, \mathbf{\frac{M(M-1)}{2}}) +  ( \textbf{N},\textbf{M} )$ }

\item $10 \leq N \leq 13$, $N \leq M \leq 24:$
\item[] {$ (N-8) (\textbf{N}, \textbf{1}) + (\mathbf{\frac{N(N+1)}{2}}, \textbf{1} ) + (24-M) (\textbf{1}, \textbf{M}) + 3 (\textbf{1}, \mathbf{\frac{M(M-1)}{2}})$}

\item $ 10 \leq N \leq 15$, $N+1 \leq M \leq 31-N$  or
\item[] $14 \leq N \leq 19\ , M=14$ or \qquad $15 \leq N \leq 17\ , M=15$ 
\item[] or $(N, M)=(16, 16)$ or $(N, M)=(15, 17):$ 
\item[] {$ (24-N)(\textbf{N}, \textbf{1}) + 3(\mathbf{\frac{N(N-1)}{2}}, \textbf{1} ) + (M-8) (\textbf{1}, \textbf{M}) + (\textbf{1}, \mathbf{\frac{M(M+1)}{2}})$ }

\item$10 \leq N \leq 16$, $N \leq M \leq 16:$
\item[] {$ (16-M)( \textbf{N},\textbf{1} ) + 2 ( \mathbf{\frac{N(N-1)}{2}}, \textbf{1} ) + (16-N) ( \textbf{1},\textbf{M} ) + 2 ( \textbf{1},\mathbf{\frac{M(M-1)}{2}} ) +  ( \textbf{N},\textbf{M} )$ }

\item  $10 \leq N \leq 15$, $N+1 \leq M \leq 16:$
\item[] {$ (16-M)( \textbf{N},\textbf{1} ) + 2 ( \mathbf{\frac{N(N-1)}{2}}, \textbf{1} ) + (M+8-N) ( \textbf{1},\textbf{M} ) + ( \textbf{1},\mathbf{\frac{M(M-1)}{2}} ) +  ( \textbf{N},\textbf{M} )$ }

\item $10 \leq N \leq 15$, $N+1 \leq M \leq 15:$
\item[] or $(N, M)=(14, 16)$ or $(N, M)=(15, 16):$
\item[] {$ (16-M)( \textbf{N},\textbf{1} ) + 2 ( \mathbf{\frac{N(N-1)}{2}}, \textbf{1} ) + (2M-N) ( \textbf{1},\textbf{M} ) +  ( \textbf{N},\textbf{M} )$ }

\item  $10 \leq N \leq 12$, $10 \leq M \leq 13:$
\item[] {$ (48-4N)( \textbf{N},\textbf{1} ) + 6 ( \mathbf{\frac{N(N-1)}{2}}, \textbf{1} ) + (M-8) ( \textbf{1},\textbf{M} ) + ( \textbf{1}, \mathbf{\frac{M(M+1)}{2}} )$ }

\item  $10 \leq N \leq 12$, $10 \leq M \leq 12$ or 
 \ \ $(N, M)=(14, 10)$ or $(N, M)= (13, 11)$ 
 \item[] or $(N, M)= (13, 10)$ \ \ or $(N, M)=(10, 13)$ \ \ or $(N, M)=(11, 13):$
\item[] {$ (24-N-M)( \textbf{N},\textbf{1} ) + 3 ( \mathbf{\frac{N(N-1)}{2}}, \textbf{1} ) + (16-N) ( \textbf{1},\textbf{M} ) + 2 ( \textbf{1}, \mathbf{\frac{M(M-1)}{2}} ) + ( \textbf{N},\textbf{M} )$ }

\end{enumerate}}

\bigskip
\item { $\boxed{SU(M) \times SO(N) }$} 
\begin{enumerate}
    \item $10\leq N$, \quad $10\leq M$, such that $n_H - n_V \leq 244$:

    \item[] $(24-N)(\textbf{N},\textbf{1}) + 3(\mathbf{\tfrac{1}{2}N(N-1)},\textbf{1}) + (M-8)(\textbf{1},\textbf{M})$

    \item $10\leq N\leq 12$, \quad $10\leq M\leq 14$, \quad $N+M\leq 24$,\\
    $(N,M)^{*} = (10,14)$

    \item[] $(48-4N)(\textbf{N},\textbf{1}) + 6(\mathbf{\tfrac{1}{2}N(N-1)},\textbf{1}) + (M-8)(\textbf{1},\textbf{M})$

    \item $10\leq N\leq 11$, \quad $10\leq M\leq 16$:

    \item[] $(N-8)(\textbf{N},\textbf{1}) + (\mathbf{\tfrac{1}{2}N(N+1)},\textbf{1}) + M(\textbf{1},\textbf{M}) + k(\textbf{1},\mathbf{2^{\lfloor\tfrac{M-1}{2}\rfloor}})$\\
    $k=1$ for $M=15,16$, $k=2$ for $M=14$, $k=4$ for $M=13$, $k=8$ for $M=10,11,12$

    \item $10\leq N\leq 11$, \quad $N\leq M\leq 14$, \quad $N+M\leq 14$,\\
    $(N,M)^{*} = (10,14)$:

    \item[] $(16-M)(\textbf{N},\textbf{1}) + 2(\mathbf{\tfrac{1}{2}N(N-1)},\textbf{1}) + (M-N)(\textbf{1},\textbf{M}) + k(\textbf{1},\mathbf{2^{\lfloor\tfrac{M-1}{2}\rfloor}}) + (\textbf{N},\textbf{M})$\\
    $k=2$ for $M=14$, $k=4$ for $M=13$, $k=8$ for $M=10,11,12$

    \item $10\leq N\leq 11$, \quad $10\leq M\leq 12$:

    \item[] $(40-3N+n(N-8))(\textbf{N},\textbf{1}) + (5-n)(\mathbf{\tfrac{1}{2}N(N-1)},\textbf{1}) + (M-7+n)(\textbf{1},\textbf{M}) + (n+1)(\textbf{1},\mathbf{2^{\lfloor\tfrac{M-1}{2}\rfloor}})$\\
    $n=0,...,12-M$ for $N=11$,\\
    $n=0,...,2$ for $(N,M)= (10,12)$,\\
    $n=0,...,5$ for $(N,M) = (10,10),(10,11)$

    \item $10\leq N\leq 11$, \quad $10\leq M \leq 12$:
    \begin{enumerate}
        \item $(N-8)(\textbf{N},\textbf{1}) + (\mathbf{\tfrac{1}{2}N(N+1)},\textbf{1}) + (24-2M)(\textbf{1},\textbf{M}) + 3(\textbf{1},\mathbf{\tfrac{1}{2}M(M-1)}) + 8(\textbf{1},\mathbf{2^{\lfloor\tfrac{M-1}{2}\rfloor}})$

        \item $(N-8)(\textbf{N},\textbf{1}) + (\mathbf{\tfrac{1}{2}N(N+1)},\textbf{1}) + (M-3)(\textbf{1},\textbf{M}) + 5(\textbf{1},\mathbf{2^{\lfloor\tfrac{M-1}{2}\rfloor}})$

        \item $(N-8)(\textbf{N},\textbf{1}) + (\mathbf{\tfrac{1}{2}N(N+1)},\textbf{1}) + 9(\textbf{1},\textbf{M}) + (\textbf{1},\mathbf{\tfrac{1}{2}M(M-1)}) + 9(\textbf{1},\mathbf{2^{\lfloor\tfrac{M-1}{2}\rfloor}})$
    \end{enumerate}

    \item $N=11$, \quad $10\leq M\leq 11$,\\
    $(N,M)^{*} = (11,11)$:

    \item[] $(N+8-M)(\textbf{N},\textbf{1}) + (\mathbf{\tfrac{1}{2}N(N-1)},\textbf{1}) + (M-10)(\textbf{1},\textbf{M}) + 9(\textbf{1},\mathbf{2^{\lfloor\tfrac{M-1}{2}\rfloor}}) + (\textbf{N},\textbf{M})$

    \item $N=11$, \quad $M=12$:

    \item[] $(\textbf{N},\textbf{1}) + 3(\mathbf{\tfrac{1}{2}N(N-1)},\textbf{1}) + 7(\textbf{1},\mathbf{2^{\lfloor\tfrac{M-1}{2}\rfloor}}) + (\textbf{N},\textbf{M})$

    \item $N=10$, \quad $10\leq M\leq 13$, or $(N,M) = (11,10)$,\\
    $(N,M)^{*} = (10,13)$:

    \item[] $(N-8)(\textbf{N},\textbf{1}) + (\mathbf{\tfrac{1}{2}N(N+1)},\textbf{1}) + k(\textbf{1},\mathbf{2^{\lfloor\tfrac{M-1}{2}\rfloor}}) + (\textbf{1},\mathbf{\tfrac{1}{2}M(M+1)-1})$\\
    $k=16$ for $M=10,11,12$, $k=8$ for $M=13$

    \item $N=10$, \quad $10\leq M\leq 12$:

    \item[] $(2N-n(N-8)-M)(\textbf{N},\textbf{1}) + n(\mathbf{\tfrac{1}{2}N(N-1)},\textbf{1}) + (M-8-n)(\textbf{1},\textbf{M}) + (10-n)(\textbf{1},\mathbf{2^{\lfloor\tfrac{M-1}{2}\rfloor}}) + (\textbf{N},\textbf{M})$\\
    $n=0,...,M-8$
\end{enumerate}

\bigskip
\item $\boxed{ \textrm{SU}(N)\times \textrm{Sp}(M)} $ 

\begin{enumerate}
    \item $15\leq N\leq 48$, \quad $10\leq M\leq 32$, \quad $3M\leq 2N$, \quad $3N\leq 4M+16$:

    \item[] $(2N-3M)(\textbf{N},\textbf{1}) + (2(2M+8)-3M)(\textbf{1},\textbf{2M}) + 3(\textbf{N},\textbf{2M})$

    \item $18\leq N\leq 31$, \quad $10\leq M\leq N-8$,\\
    or $N=32$, \quad $14\leq M\leq 24$:

    \item[] $(N-8-M)(\textbf{N},\textbf{1}) + (\mathbf{\tfrac{1}{2}N(N+1)},\textbf{1}) + (32-N)(\textbf{1},\textbf{2M}) + (\textbf{1},\mathbf{M(2M-1)-2}) + (\textbf{N},\textbf{2M})$

    \item $12\leq N\leq 16$, \quad $10\leq M\leq N$, \\
    or $10\leq N\leq 11$, \quad $5\leq M\leq N$:

    \item[] $(2N-2M)(\textbf{N},\textbf{1}) + (32-2N)(\textbf{1},\textbf{2M}) + (\textbf{1},\mathbf{M(2M-1)-1}) + 2(\textbf{N},\textbf{2M})$

    \item $12\leq N\leq 16$, \quad $10\leq N\leq \tfrac{1}{2}(N+8)$,\\
    or $10\leq N\leq 11$, \quad $5\leq M\leq \tfrac{1}{2}(N+8)$:

    \item[] $(N+8-2M)(\textbf{N},\textbf{1}) +(\mathbf{\tfrac{1}{2}N(N-1)},\textbf{1}) + (32-2N)(\textbf{1},\textbf{2M}) + (\textbf{1},\mathbf{M(2M-1)-1}) + 2(\textbf{N},\textbf{2M})$

    \item $12\leq N\leq 14$, \quad $10\leq M\leq 12$, \quad $(N,M)\neq (14,11)$,\\
    or $10\leq N\leq 11$, \quad $5\leq M\leq 12$,\\
    $(N,M)^{*} = (14,12)$:

    \item[] $(N-8)(\textbf{N},\textbf{1}) + (\mathbf{\tfrac{1}{2}N(N+1)},\textbf{1}) + 2(24-2M)(\textbf{1},\textbf{2M}) + 2(\textbf{1},\mathbf{M(2M-1)-1})$

    \item $10\leq N\leq 11$, \quad $5\leq M\leq 8$:

    \item[] $(16-2M)(\textbf{N},\textbf{1}) + 2(\mathbf{\tfrac{1}{2}N(N-1)},\textbf{1}) + 2(2M+8-N-n(2M-8))(\textbf{1},\textbf{2M}) + n(\textbf{1},\mathbf{M(2M-1)-1}) + 2(\textbf{N},\textbf{2M}) $\\
    $n\geq 0$ such that all coefficients are positive or 0

    \item $10\leq N\leq 11$, \quad $5\leq M\leq \tfrac{2}{3}N$:

    \item[] $(2N-3M)(\textbf{N},\textbf{1}) + (2(2M+8)-3N)(\textbf{1},\textbf{2M}) + 3(\textbf{N},\textbf{2M})$

    \item $10\leq N\leq 11$, \quad $5\leq M\leq 6$:
    \begin{enumerate}
    \item $(N-8)(\textbf{N},\textbf{1}) + (\mathbf{\tfrac{1}{2}M(M+1)},\textbf{1}) + 2(48-8M)(\textbf{1},\textbf{2M}) + 5(\textbf{1},\mathbf{M(2M-1)-1})$

    \item $(24-N-2M)(\textbf{N},\textbf{1}) + 3(\mathbf{\tfrac{1}{2}N(N-1)},\textbf{1}) + (32-2N)(\textbf{1},\textbf{2M}) + (\textbf{1},\mathbf{M(2M-1)-1}) + 2(\textbf{N},\textbf{2M})$
    \end{enumerate}
    \item $10\leq N\leq 11$, \quad $5\leq M\leq 16-N$:

    \item[] $(2N-M)(\textbf{N},\textbf{1}) + (2(24-2M)-N)(\textbf{1},\textbf{2M}) + 2(\textbf{1},\mathbf{M(2M-1)-1}) + (\textbf{N},\textbf{2M})$
\end{enumerate}

\bigskip
\item $\boxed{\textrm{SO}(N)\times \textrm{SO}(M)}$ 
\begin{enumerate}
    \item $10\leq N\leq 14$, \quad $N\leq M\leq 20$, \quad $N+M\leq 30$ or $(N,M) = (12,19),(11,20)$,\\
    $(N,M)^{*} = (12,19)$:

    \item[] $(N-k_{2}n)(\textbf{N},\textbf{1}) + \tfrac{1}{k_{1}}(8-k_{2}n)(\mathbf{2^{\lfloor\tfrac{N-1}{2}\rfloor}},\textbf{1}) + (M-8+k_{2}n)(\textbf{1},\textbf{M}) + n(\textbf{1},\mathbf{2^{\lfloor\tfrac{M-1}{2}\rfloor}})$\\
    $n=0,...,\lfloor \tfrac{8}{k_{2}} \rfloor$\\
    $k_{1} = 1$ for $N\leq 12$, $k_{1} = 2$ for $N=13$, $k_{1} = 4$ for $N=14$, $k_{1} = 8$ for $N=15$, and $k_{1} = 16$ for $N\geq 17$ \\
    $k_{2}$ as above with $N\to M$

    \item $10\leq N\leq 12$, \quad $N\leq M\leq 21$:

    \item[] $(N-3)(\textbf{N},\textbf{1}) + 5(\mathbf{2^{\lfloor\tfrac{N-1}{2}\rfloor}},\textbf{1}) + (M-8)\textbf{1},\textbf{M}$)

    \item $10\leq N\leq 13$, \quad $M=N+8$:

    \item[] $k(\mathbf{2^{\lfloor\tfrac{N-1}{2}\rfloor}},\textbf{1}) + (\textbf{N},\textbf{M})$\\
    $k=16$ for $N\leq 12$, and $k=8$ for $N=13$

    \item $10\leq N\leq 12$, \quad $N\leq M\leq 16$, \quad $N+M\leq 26$,\\
    $(N,M)\neq (12,14)$:

    \item[] $9(\textbf{N},\textbf{1}) + (\mathbf{\tfrac{1}{2}N(N-1)},\textbf{1}) + 9(\mathbf{2^{\lfloor\tfrac{N-1}{2}\rfloor}},\textbf{1}) + (M-8)\textbf{1},\textbf{M)}$

    \item $10\leq N\leq M\leq 12$:

    \item[] $(N-8+n)(\textbf{N},\textbf{1}) + n(\mathbf{2^{\lfloor\tfrac{N-1}{2}\rfloor}},\textbf{1}) + (24-2M+n(M-8))(\textbf{1},\textbf{M}) + (3-n)(\textbf{1},\mathbf{\tfrac{1}{2}M(M-1)}) + 8(\textbf{1},\mathbf{2^{\lfloor\tfrac{M-1}{2}\rfloor}})$
    $n=0,...,12-M$ for $N=11,12$, and $n=0,...,13-M$ for $N=10$

    \item $N=10$, \quad $10\leq M\leq 14$, or $(N,M) = (11,11)$,\\
    $(N,M)^{*} = (11,11),(10,14)$:

    \item[] $(N-8)(\textbf{N},\textbf{1}) + k(\textbf{1},\mathbf{2^{\lfloor\tfrac{M-1}{2}\rfloor}}) + (\textbf{1},\mathbf{\tfrac{1}{2}M(M+1)})$\\
    $k=16$ for $M\leq 12$, $k=8$ for $M=13$, and $k=4$ for $M=14$

    \item $(N,M) = (10,16),(11,16)$:

    \item[] $(N-8)(\textbf{N},\textbf{1}) + M(\textbf{1},\textbf{M}) + (\textbf{1},\mathbf{2^{\lfloor\tfrac{M-1}{2}\rfloor}})$

    \item $(N,M) = (10,11),(10,12),(11,12)$:
    \begin{enumerate}
        \item $(N-8)(\textbf{N},\textbf{1}) + (M-3)(\textbf{1},\textbf{M}) + 5(\textbf{1},\mathbf{2^{\lfloor\tfrac{M-1}{2}\rfloor}})$

        \item $(N-8)(\textbf{N},\textbf{1}) + 9(\textbf{1},\textbf{M}) + (\textbf{1},\mathbf{\tfrac{1}{2}M(M-1)}) + 9(\textbf{1},\mathbf{2^{\lfloor\tfrac{M-1}{2}\rfloor}})$

        \item $(24-2N + n(N-8))(\textbf{N},\textbf{1}) + (3-n)(\mathbf{\tfrac{1}{2}N(N-1)},\textbf{1}) + 8(\mathbf{2^{\lfloor\tfrac{N-1}{2}\rfloor}},\textbf{1}) + (N-8+n)(\textbf{1},\textbf{M}) + n(\textbf{1},\mathbf{2^{\lfloor\tfrac{M-1}{2}\rfloor}})$\\
        $n=0$ for $(N,M)=(11,12)$, $n=0,1$ for $(N,M)=(10,12)$, and $n=0,...,3$ for $(N,M)=(10,11)$
    \end{enumerate}

    \item $(N,M)=(10,11)$:

    \item[] $16(\textbf{N},\textbf{1}) + (\mathbf{\tfrac{1}{2}N(N+1)},\textbf{1}) + (M-8)(\textbf{1},\textbf{M})$
\end{enumerate}

\bigskip
\item $\boxed{\textrm{SO}(N) \times \textrm{Sp}(M)}$ 

\begin{enumerate}
    \item $18\leq N\leq 31$, \quad $10\leq M\leq N-8$,\\
    or $13\leq N\leq 21$, \quad $5\leq M\leq 10$, \quad $M\leq N-8$,\\
    or $N=32$, \quad $14\leq M\leq 24$:

    \item[] $(N-8-M)(\textbf{N},\textbf{1}) + (32-N)(\textbf{1},\textbf{2M}) + (\textbf{1},\mathbf{M(2M-1)-1}) + (\textbf{N},\textbf{2M})$

    \item  $10\leq N\leq 14$, \quad $5\leq M\leq 12$,\\
    or $15\leq N\leq 19$, \quad $5\leq M\leq 24-N$,\\
    or $(N,M) = (18,7),(19,6),(20,5),(21,5)$:

    \item[]$(N-8)(\textbf{N},\textbf{1}) + (48-4M)(\textbf{1},\textbf{2M}) + 2(\textbf{1},\mathbf{M(2M-1)-1})$

    \item $10\leq N\leq 14$, \quad $5\leq M\leq 8$, \quad $M\leq N-4$:

    \item[] $(N-4-M)(\textbf{N},\textbf{1}) + k(\mathbf{2^{\lfloor\tfrac{N-1}{2}\rfloor}},\textbf{1}) + (2(2M+8-n(2M-8))-N)(\textbf{1},\textbf{2M}) + n(\textbf{1},\mathbf{M(2M-1)-1}) + (\textbf{N},\textbf{2M})$\\
    $n=1,2$ for $M=8$, $n=0,1,2$ for $M=7$, $n=0,...,3$ for $M=5,6$\\
    $k=1$ for $N=14$, $k=2$ for $N=13$, $k=4$ for $N=10,11,12$

    \item $10\leq N\leq 14$, \quad $5\leq M\leq 6$:

    \item[] $(N-8)(\textbf{N},\textbf{1}) + 2(48-8M)(\textbf{1},\textbf{2M}) + 5(\textbf{1},\mathbf{M(2M-1)-1})$

    \item $10\leq M\leq 14$, \quad $5\leq M\leq 6$, \quad $2M\leq N$,\\
    $(N,M)^{*} = (14,6)$

    \item[] $(N-2M)(\textbf{N},\textbf{1}) + k(\mathbf{2^{\lfloor\tfrac{N-1}{2}\rfloor}},\textbf{1}) + (32-2N)(\textbf{1},\textbf{2M}) + (\textbf{1},\mathbf{M(2M-1)-1}) + 2(\textbf{N},\textbf{2M})$\\
    $k=2$ for $N=14$, $k=4$ for $N=13$, $k=8$ for $N=10,11,12$

    \item $10\leq N\leq 12$, \quad $5\leq M\leq 6$:

    \item[] $(N-8+n)(\textbf{N},\textbf{1}) + n(\mathbf{2^{\lfloor\tfrac{N-1}{2}\rfloor}},\textbf{1}) + 2(48-4N+n(2N-8))(\textbf{1},\textbf{2M}) + (5-n)(\textbf{1},\mathbf{M(2M-1)-1})$\\
    $n=1$ for $(N,M)=(11,6),(12,6)$, $n=1,2$ for $(N,M) = (10,6)$, $n=1,...,4$ for $(N,M) = (12,5)$, $n=1,...,5$ for $(N,M)=(10,5),(11,5)$

    \item $10\leq N\leq 13$, \quad $5\leq M\leq 7$, \quad $(N,M) \neq (11,7)$,\\
    $(N,M,n)^{*} = (10,7,-1)$:

    \item[] $(N-M-4-kn)(\textbf{N},\textbf{1}) + n(\mathbf{2^{\lfloor\tfrac{N-1}{2}\rfloor}},\textbf{1}) + (32-N)(\textbf{1},\textbf{2M}) + (\textbf{1},\mathbf{M(2M-1)-1}) + (\textbf{N},\textbf{2M})$\\
    $k=2$ for $N=13$, $k=1$ for $N=10,11,12$\\
    $n=0,1$ for $N=13$ or $(N,M) = (12,7)$, $n=-1,0,1,2$ for $(N,M)=(12,6)$, $n=-2,...,3$ for $(N,M)=(12,5)$, $n=-1,0,1$ for $(N,M) = (11,6)$, $n=-2,...,2$ for $(N,M) = (11,5)$, $n=-1$ for $(N,M) = (10,7)$, $n=-1,0$ for $(N,M) = (10,6)$, $n=-2,...,1$ for $(N,M) = (10,5)$

    \item $13\leq N\leq 14$, \quad $M=5$,\\
    $(N,M)^{*} = (14,5)$

    \item[] $(N-13)(\textbf{N},\textbf{1}) + (29-N)(\textbf{1},\textbf{2M}) + 4(\textbf{1},\mathbf{M(2M-1)-1}) + (\textbf{1},\textbf{110})$

    \item $10\leq N\leq 12$, \quad $M=5$:
    \begin{enumerate}
        \item $n(\textbf{N},\textbf{1}) + (13-N+n)(\mathbf{2^{\lfloor\tfrac{N-1}{2}\rfloor}},\textbf{1}) + (60-5N+4n)(\textbf{1},\textbf{2M}) + (N-6-n)(\textbf{1},\mathbf{M(2M-1)-1}) + (\textbf{N},\textbf{2M})$\\
        $n=0,...,N-6$

        \item $n(\textbf{N},\textbf{1}) + (18-N+n)(\mathbf{2^{\lfloor\tfrac{N-1}{2}\rfloor}},\textbf{1}) + (72-6N+4n)(\textbf{1},\textbf{2M}) + (N-9-n)(\textbf{1},\mathbf{M(2M-1)-1}) + (\textbf{N},\textbf{2M})$\\
        $n=0,...,N-9$
    \end{enumerate}

    \item $(N,M)^{*} = (18,5)$:

    \item[] $6(\textbf{1},\textbf{2M}) + 3(\textbf{1},\mathbf{M(2M-1)-1}) + 2(\textbf{1},\textbf{110}) + 2(\textbf{N},\textbf{2M})$

    \item $(N,M) = (13,5)$:
    \begin{enumerate}
        \item $7(\textbf{N},\textbf{1}) + (\mathbf{2^{\lfloor\tfrac{N-1}{2}\rfloor}},\textbf{1}) + 24(\textbf{1},\textbf{2M}) + 3(\textbf{1},\mathbf{M(2M-1)-1})$

        \item $2(\textbf{N},\textbf{1}) + (\mathbf{2^{\lfloor\tfrac{N-1}{2}\rfloor}},\textbf{1}) + 3(\textbf{1},\textbf{2M}) + 5(\textbf{1},\mathbf{M(2M-1)-1}) + (\textbf{N},\textbf{2M})$
    \end{enumerate}

    \item $(N,M) = (12,5)$:

    \item[] $11(\mathbf{2^{\lfloor\tfrac{N-1}{2}\rfloor}},\textbf{1}) + 3(\textbf{N},\textbf{2M})$

    \item $(N,M)=(10,6),(11,6)$:

    \item[] $(N-M-4)(\textbf{N},\textbf{1}) + 5(\mathbf{2^{\lfloor\tfrac{N-1}{2}\rfloor}},\textbf{1}) + (24-N)(\textbf{1},\textbf{2M}) + 2(\textbf{1},\mathbf{M(2M-1)-1}) + (\textbf{N},\textbf{2M})$
\end{enumerate}

\bigskip
\item $\boxed{\textrm{Sp}(N)\times \textrm{Sp}(M)}$

\begin{enumerate}
    \item $5\leq N\leq 8$, \quad $N\leq M\leq 10$, \quad $M\leq N+4$:

    \item[] $2(2N+8-2M)(\textbf{2N},\textbf{1}) + (32-4N)(\textbf{1},\textbf{2M}) + (\textbf{1},\mathbf{M(2M-1)-1}) + 2(\textbf{2N},\textbf{2M})$

    \item $5\leq N\leq 7$, \quad $N<M\leq 8$:

    \item[] $(32-4M)(\textbf{2N},\textbf{1}) + (\mathbf{N(2N-1)-1},\textbf{1}) + 2(2M+8-2N)(\textbf{1},\textbf{2M}) + 2(\textbf{2N},\textbf{2M}) $

    \item $5\leq N\leq M\leq 8$:

    \item[] $(32-4M)(\textbf{2N},\textbf{1}) + (\mathbf{N(2N-1)-1},\textbf{1}) + (32-4N)(\textbf{1},\textbf{2M}) + (\textbf{1},\mathbf{M(2M-1)-1})$

    \item $5\leq N\leq M\leq 7$, \quad $N+M\leq 12$:
    \begin{enumerate}
        \item $2(24-2N-2M)(\textbf{2N},\textbf{1}) + 2(\mathbf{N(2N-1)-1},\textbf{1}) + (32-4N)(\textbf{1},\textbf{2M}) + (\textbf{1},\mathbf{M(2M-1)-1})$

        \item $(32-4M)(\textbf{2N},\textbf{1}) + (\mathbf{N(2N-1)-1},\textbf{1}) + 2(24-2N-2M)(\textbf{1},\textbf{2M}) + 2(\textbf{1},\mathbf{M(2M-1)-1})$
    \end{enumerate}
\end{enumerate}

\subsection{$E \times E\times E$ models}

\item { $\textrm{F}_4 \times \textrm{F}_4 \times \textrm{F}_4$ 

\begin{enumerate}
\item {$5 ( \textbf{26},\textbf{1},\textbf{1} ) + 5 ( \textbf{1},\textbf{26},\textbf{1} ) + 5 ( \textbf{1},\textbf{1},\textbf{26} )$}
\end{enumerate}}

\item { $\textrm{E}_6 \times \textrm{F}_4 \times \textrm{F}_4$ 
\begin{enumerate}
\item {$6 ( \textbf{27},\textbf{1},\textbf{1} ) + 5 ( \textbf{1},\textbf{26},\textbf{1} ) + 5 ( \textbf{1},\textbf{1},\textbf{26} )$}
\end{enumerate}}

\subsection{$E \times E\times C$ models}

\item $\textrm{F}_{4}\times \textrm{F}_{4}\times \textrm{SO}(N)$ 
{\begin{enumerate}
\item $5(\textbf{26},\textbf{1},\textbf{1}) + 5(\textbf{1},\textbf{26},\textbf{1}) + (N-4)(\textbf{1},\textbf{1},\textbf{N}) + 4(\textbf{1},\textbf{1},\mathbf{2^{\lfloor \frac{N-1}{2}\rfloor}})$\ , \quad $N=10,11$
\end{enumerate}}

\item $\textrm{F}_{4}\times \textrm{E}_{6}\times \textrm{SO}(10)$ 
{\begin{enumerate}
\item$5(\textbf{26},\textbf{1},\textbf{1}) + 6(\textbf{1},\textbf{27},\textbf{1}) + 6(\textbf{1},\textbf{1},\textbf{10}) + 4(\textbf{1},\textbf{1},\textbf{16})  $
\end{enumerate}}

\subsection{$E \times C\times C$ models}

\item $\textrm{E}_{6} \times \textrm{SU}(N)\times \textrm{SU}(M)$
\begin{enumerate}
    \item $10\leq N\leq M\leq 14$, \quad $N+M\leq 24$:
    \item[] $2(\textbf{27},\textbf{1},\textbf{1}) + (24-N-M)(\textbf{1},\textbf{N},\textbf{1}) + 3(\textbf{1},\mathbf{\tfrac{1}{2}N(N-1)},\textbf{1}) + (24-N-M)(\textbf{1},\textbf{1},\textbf{M}) + 3(\textbf{1},\textbf{1},\mathbf{\tfrac{1}{2}M(M-1)}) + (\textbf{1},\textbf{N},\textbf{M})$
\end{enumerate}

\item $\textrm{E}_{7} \times \textrm{SU}(N)\times \textrm{SU}(M)$
\begin{enumerate}
    \item $10\leq N\leq M\leq 14$, \quad $N+M\leq 24$:
    \item[] $4(\textbf{56},\textbf{1},\textbf{1}) + (24-N-M)(\textbf{1},\textbf{N},\textbf{1}) + 3(\textbf{1},\mathbf{\tfrac{1}{2}N(N-1)},\textbf{1}) + (24-N-M)(\textbf{1},\textbf{1},\textbf{M}) + 3(\textbf{1},\textbf{1},\mathbf{\tfrac{1}{2}M(M-1)}) + (\textbf{1},\textbf{N},\textbf{M})$
\end{enumerate}

\item $\textrm{F}_{4} \times \textrm{SU}(N)\times \textrm{SU}(M)$
\begin{enumerate}
    \item $10\leq N\leq M\leq 14$, \quad $N+M\leq 24$:
    \item[] $(\textbf{26},\textbf{1},\textbf{1}) + (24-N-M)(\textbf{1},\textbf{N},\textbf{1}) + 3(\textbf{1},\mathbf{\tfrac{1}{2}N(N-1)},\textbf{1}) + (24-N-M)(\textbf{1},\textbf{1},\textbf{M}) + 3(\textbf{1},\textbf{1},\mathbf{\tfrac{1}{2}M(M-1)}) + (\textbf{1},\textbf{N},\textbf{M})$
\end{enumerate}

\item {$\textrm{F}_{4} \times \textrm{SU}(N) \times \textrm{SO}(M)$}
\begin{enumerate}
\item $N=10,\, 11,\,\, M= 10,...,14$:
\item[] $ (\textbf{26},\textbf{1},\textbf{1}) + (24-N-M)(\textbf{1},\textbf{N},\textbf{1}) + 3(\textbf{1},\mathbf{\tfrac{1}{2}N(N-1)},\textbf{1}) + (M-N)(\textbf{1},\textbf{1},\textbf{M}) + k(\textbf{1},\textbf{1},\mathbf{2^{\lfloor \tfrac{M-1}{2}\rfloor}}) + (\textbf{1},\textbf{N},\textbf{M}) $ \\
$k=8$ for $M=10,11,12$, $k=4$ for $M=13$, $k=2$ for $M=14$
\end{enumerate}

\item {$\textrm{E}_{7} \times \textrm{SU}(N) \times \textrm{SO}(M)$ 
\begin{enumerate}
\item $N=10^{*},\, 11,\,\, M= 10,...,13^{*}$:
\item[] $ 4(\textbf{56},\textbf{1},\textbf{1}) + (24-N-M)(\textbf{1},\textbf{N},\textbf{1}) + 3(\textbf{1},\mathbf{\tfrac{1}{2}N(N-1)},\textbf{1}) + (M-N)(\textbf{1},\textbf{1},\textbf{M}) + k(\textbf{1},\textbf{1},\mathbf{2^{\lfloor \tfrac{M-1}{2}\rfloor}}) + (\textbf{1},\textbf{N},\textbf{M}) $ \\
$k=8$ for $M=10,11,12$, and $k=4$ for $M=13$
\end{enumerate}}

\item {$\textrm{E}_{6} \times \textrm{SU}(N) \times \textrm{SO}(M)$
\begin{enumerate}
\item $N=10,\, 11,\,\, M= 10,...,13$:
\item[] $ 2(\textbf{27},\textbf{1},\textbf{1}) + (24-N-M)(\textbf{1},\textbf{N},\textbf{1}) + 3(\textbf{1},\mathbf{\tfrac{1}{2}N(N-1)},\textbf{1}) + (M-N)(\textbf{1},\textbf{1},\textbf{M}) + k(\textbf{1},\textbf{1},\mathbf{2^{\lfloor \tfrac{M-1}{2}\rfloor}}) + (\textbf{1},\textbf{N},\textbf{M}) $ \\
$k=8$ for $M=10,11,12$, and $k=4$ for $M=13$
\end{enumerate}}

\item {$\textrm{F}_{4} \times \textrm{SU}(N) \times \textrm{Sp}(M)$
\begin{enumerate}
\item $N=10,\,11 ,\,\, M=5,\,6,\,7$:
\item[] $ (\textbf{26},\textbf{1},\textbf{1}) + (24-N-2M)(\textbf{1},\textbf{N},\textbf{1}) + 3(\textbf{1},\mathbf{\tfrac{1}{2}N(N-1)},\textbf{1}) + 2(24-N-2M)(\textbf{1},\textbf{1},\textbf{2M}) + 2(\textbf{1},\textbf{1},\mathbf{M(2M-1)-1}) + 2(\textbf{1},\textbf{N},\textbf{2M}) $
\end{enumerate} }   

\item $ \textrm{E}_{7}  \times \textrm{SU}(N)\times \textrm{Sp}(M) $ 
\begin{enumerate}
\item $ N=10,\,11 ,\,\, M= 5,\,6,\,7 $:
\item[] $ 4(\textbf{56},\textbf{1},\textbf{1}) + (24-N-2M)(\textbf{1},\textbf{N},\textbf{1}) + 3(\textbf{1},\mathbf{\tfrac{1}{2}N(N-1)},\textbf{1}) + 2(24-N-2M)(\textbf{1},\textbf{1},\textbf2{M}) + 2(\textbf{1},\textbf{1},\mathbf{M(2M-1)-1}) + 2(\textbf{1},\textbf{N},\textbf{2M}) $
\end{enumerate}

\item $\textrm{E}_{6}\times \textrm{SU}(N) \times \textrm{Sp}(M)$ 
\begin{enumerate}
\item $ N=10,\,11 ,\,\, M= 5,\,6,\,7 $:
\item[] $ 2(\textbf{27},\textbf{1},\textbf{1}) + (24-N-2M)(\textbf{1},\textbf{N},\textbf{1}) + 3(\textbf{1},\mathbf{\tfrac{1}{2}N(N-1)},\textbf{1}) + 2(24-N-2M)(\textbf{1},\textbf{1},\textbf{2M}) + 2(\textbf{1},\textbf{1},\mathbf{M(2M-1)-1}) + 2(\textbf{1},\textbf{N},\textbf{2M}) $
\end{enumerate}

\item $\textrm{F}_{4}\times \textrm{SO}(N) \times \textrm{SO}(M)$
\begin{enumerate}
\item $N=M=10^{*},...,14$:
\item[] $(\textbf{26},\textbf{1},\textbf{1}) + k(\textbf{1},\mathbf{2^{\lfloor \tfrac{N-1}{2}\rfloor}},\textbf{1}) + k(\textbf{1},\textbf{1},\mathbf{2^{\lfloor \tfrac{M-1}{2}\rfloor}}) + (\textbf{1},\textbf{N},\textbf{M})$ \\
$k=8$ for $N=M=10,11,12$, $k=4$ for $N=M=13$, and $k=2$ for $N=M=14$
\item $N=10,\, M=10,11$:
\item[] $5(\textbf{26},\textbf{1},\textbf{1}) + 6(\textbf{1},\textbf{10},\textbf{1}) + 4(\textbf{1},\textbf{16},\textbf{1}) + (M-4)(\textbf{1},\textbf{1},\textbf{M}) + 4(\textbf{1},\textbf{1},\mathbf{2^{\lfloor \tfrac{M-1}{2}\rfloor}})$
\end{enumerate}

\item $ \textrm{E}_{6}\times \textrm{SO}(N) \times \textrm{SO}(M) $
\begin{enumerate}
\item $N=M=10,\,11,\,12^{*}$:

\item[] $2(\textbf{27},\textbf{1},\textbf{1}) + 8(\textbf{1},\mathbf{2^{\lfloor \tfrac{N-1}{2}\rfloor}},\textbf{1}) + 8(\textbf{1},\textbf{1},\mathbf{2^{\lfloor \tfrac{M-1}{2}\rfloor}}) + (\textbf{1},\textbf{N},\textbf{M})$

\item $N= M=10$:
\item[] $6(\textbf{27},\textbf{1},\textbf{1}) + 6(\textbf{1},\textbf{10},\textbf{1}) + 4(\textbf{1},\textbf{16},\textbf{1}) + 6(\textbf{1},\textbf{1},\textbf{10}) + 4(\textbf{1},\textbf{1},\mathbf{16})$
\end{enumerate}

\item $\textrm{F}_{4}\times \textrm{SO}(N) \times \textrm{Sp}(M)$
\begin{enumerate}
\item $N=10,\, 11,\, 12,\,\, M=5,\,6,\,...,\,9$:
\item[] $ (\textbf{26},\textbf{1},\textbf{1}) + \tfrac{1}{2}(2N-2M-6)(\textbf{1},\textbf{N},\textbf{1}) + 5(\textbf{1},\mathbf{ 2^{\lfloor\tfrac{N-1}{2}\rfloor } },\textbf{1}) + (48-N-4M)(\textbf{1},\textbf{1},\textbf{2M}) + 2(\textbf{1},\textbf{1},\mathbf{ \tfrac{1}{2}N(N-1)-1 }) + (\textbf{1},\textbf{N},\textbf{2M}) $

\item $N=10,...,14,\,\, M=5$ or $N=12, M=6$:

\item[] $ (\textbf{26},\textbf{1},\textbf{1}) + (N-2M)(\textbf{1},\textbf{N},\textbf{1}) + k(\textbf{1},\mathbf{ 2^{\lfloor\tfrac{N-1}{2}\rfloor} },\textbf{1}) + (8-2(N-2M))(\textbf{1},\textbf{1},\textbf{10}) + 2(\textbf{1},\textbf{1},\textbf{44}) + 2(\textbf{1},\textbf{N},\textbf{10}) $\\
$k=8$ or $N=10,11,12$, $k=4$ for $N=13$, and $k=8$ for $N=14$
\item[(c)\textsuperscript{*}] $N=10, M=5$:

\item[] $ (\textbf{26},\textbf{1},\textbf{1}) + 7(\textbf{1},\textbf{10},\textbf{1}) + 5(\textbf{1},\textbf{1},\textbf{44}) + (\textbf{1},\textbf{16},\textbf{10}) $
\end{enumerate}

\item $\textrm{E}_{7}\times \textrm{SO}(N) \times \textrm{Sp}(M)$
 \begin{enumerate}
\item $N=10,\, 11,\, 12,\,\, M=5,\,6,\,...,\,9$:
\item[] $ 4(\textbf{56},\textbf{1},\textbf{1}) + \tfrac{1}{2}(2N-2M-6)(\textbf{1},\textbf{N},\textbf{1}) + 5(\textbf{1},\mathbf{ 2^{\lfloor\tfrac{N-1}{2}\rfloor } },\textbf{1}) + (48-N-4M)(\textbf{1},\textbf{1},\textbf{2M}) + 2(\textbf{1},\textbf{1},\mathbf{ \tfrac{1}{2}N(N-1)-1 }) + (\textbf{1},\textbf{N},\textbf{2M}) $

\item $N=10,...,13,\,\, M=5$ or $N=12, M=6$:
\item[] $ 4(\textbf{56},\textbf{1},\textbf{1}) + (N-M)(\textbf{1},\textbf{N},\textbf{1}) + k(\textbf{1},\mathbf{ 2^{\lfloor\tfrac{N-1}{2}\rfloor} },\textbf{1}) + (8-2(N-M))(\textbf{1},\textbf{1},\textbf{10}) + 2(\textbf{1},\textbf{1},\textbf{44}) + 2(\textbf{1},\textbf{N},\textbf{10}) $\\
$k=8$ for $N=10,11,12$, and $k=4$ for $N=13$
\end{enumerate}

\item $\textrm{E}_{6}\times \textrm{SO}(N) \times \textrm{Sp}(M)$
 \begin{enumerate}
\item $N=10,\, 11,\, 12,\,\, M=5,\,6,\,...,\,9$:
\item[] $ 2(\textbf{27},\textbf{1},\textbf{1}) + \tfrac{1}{2}(2N-2M-6)(\textbf{1},\textbf{N},\textbf{1}) + 5(\textbf{1},\mathbf{ 2^{\lfloor\tfrac{N-1}{2}\rfloor } },\textbf{1}) + (48-N-4M)(\textbf{1},\textbf{1},\textbf{2M}) + 2(\textbf{1},\textbf{1},\mathbf{ \tfrac{1}{2}N(N-1)-1 }) + (\textbf{1},\textbf{N},\textbf{2M}) $

\item $N=10,...,13,\,\, M=5$ or $N=12, M=6$:

\item[] $ 2(\textbf{27},\textbf{1},\textbf{1}) + (N-2M)(\textbf{1},\textbf{N},\textbf{1}) + k(\textbf{1},\mathbf{ 2^{\lfloor\tfrac{N-1}{2}\rfloor} },\textbf{1}) + (8-2(N-2M))(\textbf{1},\textbf{1},\textbf{10}) + 2(\textbf{1},\textbf{1},\textbf{44}) + 2(\textbf{1},\textbf{N},\textbf{10}) $\\
$k=8$ or $N=10,11,12$, and $k=4$ for $N=13$
\end{enumerate}

\item $\textrm{F}_{4} \times \textrm{Sp}(N) \times \textrm{Sp}(M)$
\begin{enumerate}
\item $N,M = 5,6,7$:

\item[] $(\textbf{26},\textbf{1},\textbf{1}) + 2(24-N-2M)(\textbf{1},\textbf{N},\textbf{1}) + 2(\textbf{1},\mathbf{ \tfrac{1}{2}N(N-1)-1 },\textbf{1}) + 2(24-N-2M)(\textbf{1},\textbf{1},\textbf{2M}) + 2(\textbf{1},\textbf{1},\mathbf{ M(2M-1) -1 }) + 2(\textbf{1},\textbf{N},\textbf{2M}) $
\end{enumerate}

\item $\textrm{E}_{7} \times \textrm{Sp}(N) \times \textrm{Sp}(M) $
\begin{enumerate}
\item $N,M = 5,6,7$:

\item[] $4(\textbf{56},\textbf{1},\textbf{1}) + 2(24-N-2M)(\textbf{1},\textbf{N},\textbf{1}) + 2(\textbf{1},\mathbf{ \tfrac{1}{2}N(N-1)-1 },\textbf{1}) + 2(24-N-2M)(\textbf{1},\textbf{1},\textbf{2M}) + 2(\textbf{1},\textbf{1},\mathbf{M(2M-1) -1 }) + 2(\textbf{1},\textbf{N},\textbf{2M}) $
\end{enumerate}

\item $\textrm{E}_{6} \times \textrm{Sp}(N) \times \textrm{Sp}(M)$

\begin{enumerate}
\item $N,M = 5,6,7$:
\item[] $2(\textbf{27},\textbf{1},\textbf{1}) + 2(24-N-2M)(\textbf{1},\textbf{N},\textbf{1}) + 2(\textbf{1},\mathbf{ \tfrac{1}{2}N(N-1)-1 },\textbf{1}) + 2(24-N-2M)(\textbf{1},\textbf{1},\textbf{2M}) + 2(\textbf{1},\textbf{1},\mathbf{M(2M-1) -1 }) + 2(\textbf{1},\textbf{N},\textbf{2M}) $
\end{enumerate}
\end{enumerate}

\subsection{$C \times C\times C$ models} 

\begin{enumerate}
\item $\boxed{SU(N) \times SU(M)\times SU(P)} $
\begin{enumerate}
\item $10\leq N\leq M\leq P \leq 16$:

\item[]  $(16-P)(\textbf{N},\textbf{1},\textbf{1}) + 2(\mathbf{\tfrac{1}{2}N(N-1)},\textbf{1},\textbf{1}) + (16-P)(\textbf{1},\textbf{M},\textbf{1}) + 2(\textbf{1},\mathbf{\tfrac{1}{2}M(M-1)},\textbf{1}) + (2P-N-M)(\textbf{1},\textbf{1},\textbf{P}) + (\textbf{N},\textbf{1},\textbf{P}) + (\textbf{1},\textbf{M},\textbf{P})$

\item $10 \leq N < M <P$, \quad $M\leq 16$, \quad $2M\leq N+P$:

\item[] $(16-M)(\textbf{N},\textbf{1},\textbf{1}) + 2(\mathbf{\tfrac{1}{2}N(N-1)},\textbf{1},\textbf{1}) + (2M-N-P)(\textbf{1},\textbf{M},\textbf{1}) + (16-M)(\textbf{1},\textbf{1},\textbf{P}) + 2(\textbf{1},\textbf{1},\mathbf{\tfrac{1}{2}P(P-1)}) + (\textbf{N},\textbf{M},\textbf{1}) + (\textbf{1},\textbf{M},\textbf{P})$

\item $10\leq N\leq M$, \quad $N+M\leq 24$, \quad $N+M-8\leq P\leq 16$:

\item[] $(16-P)(\textbf{N},\textbf{1},\textbf{1}) + 2(\mathbf{\tfrac{1}{2}N(N-1)},\textbf{1},\textbf{1}) + (16-P)(\textbf{1},\textbf{M},\textbf{1}) + 2(\textbf{1},\mathbf{\tfrac{1}{2}M(M-1)},\textbf{1}) + (P+8-N-M)(\textbf{1},\textbf{1},\textbf{P}) + (\textbf{1},\textbf{1},\mathbf{\tfrac{1}{2}P(P-1)}) + (\textbf{N},\textbf{1},\textbf{P}) + (\textbf{1},\textbf{M},\textbf{P})$

\item $10\leq N\leq M\leq P$, \quad $N+P \leq 24$:

\item[] $(24-N-P)(\textbf{N},\textbf{1},\textbf{1}) + 3(\mathbf{\tfrac{1}{2}N(N-1)},\textbf{1},\textbf{1}) + (M-8)(\textbf{1},\textbf{M},\textbf{1}) + (\textbf{1},\mathbf{\tfrac{1}{2}M(M+1)},\textbf{1}) + (24-N-P)(\textbf{1},\textbf{1},\textbf{P}) + 3(\textbf{1},\textbf{1},\mathbf{\tfrac{1}{2}P(P-1)}) + (\textbf{N},\textbf{1},\textbf{P}) $

\item $10\leq N\leq M < P$, \quad $P\leq 13$:

\item[] $ (24-N-M)(\textbf{N},\textbf{1},\textbf{1}) + 3(\mathbf{\tfrac{1}{2}N(N-1)},\textbf{1},\textbf{1}) + (24-N-M)(\textbf{1},\textbf{M},\textbf{1}) + 3(\textbf{1},\mathbf{\tfrac{1}{2}M(M-1)},\textbf{1}) + (P-8)(\textbf{1},\textbf{1},\textbf{P}) + (\textbf{1},\textbf{1},\mathbf{\tfrac{1}{2}P(P+1)}) + (\textbf{N},\textbf{M},\textbf{1}) $

\item $10\leq N<M\leq P$, \quad $M+P\leq 24$:

\item[] $ (N-8)(\textbf{N},\textbf{1},\textbf{1}) + (\mathbf{\tfrac{1}{2}N(N+1)},\textbf{1},\textbf{1}) + (24-M-P)(\textbf{1},\textbf{M},\textbf{1}) + 3(\textbf{1},\mathbf{\tfrac{1}{2}M(M-1)},\textbf{1}) + (24-M-P)(\textbf{1},\textbf{1},\textbf{P}) + 3(\textbf{1},\textbf{1},\mathbf{\tfrac{1}{2}P(P-1)}) + (\textbf{1},\textbf{M},\textbf{P}) $
\end{enumerate}

\item  $\boxed{SU(N) \times SU(M)\times SO(P) }$ 

\begin{enumerate}
    \item $10\leq N\leq M\leq 14\, ,$ \quad $N+M\leq 24$, \quad  $10\leq P\leq 13$:

    \item[] $(24-N-M)(\textbf{N},\textbf{1},\textbf{1}) + 3(\mathbf{ \tfrac{1}{2}N(N-1) },\textbf{1},\textbf{1}) + (24-N-M)(\textbf{1},\textbf{M},\textbf{1}) + 3(\textbf{1},\mathbf{ \tfrac{1}{2}M(M-1) },\textbf{1}) + (P-8)(\textbf{1},\textbf{1},\textbf{P}) + (\textbf{N},\textbf{M},\textbf{1})$

    \item $10\leq N\leq M\leq 11\, ,$ \quad $10\leq P\leq 13$,\quad $^*(N,M,P) = (10,11,13):$
\begin{enumerate}
    \item $ (N-8)(\textbf{N},\textbf{1},\textbf{1}) + (\mathbf{ \tfrac{1}{2}N(N+1) },\textbf{1},\textbf{1}) + (24-M-P)(\textbf{1},\textbf{M},\textbf{1}) + 3(\textbf{1},\mathbf{ \tfrac{1}{2}M(M-1) },\textbf{1}) + (P-M)(\textbf{1},\textbf{1},\textbf{P}) + k(\textbf{1},\textbf{1},\mathbf{ 2^{\lfloor \tfrac{P-1}{2}\rfloor} }) + (\textbf{1},\textbf{M},\textbf{P}) $

    \item[ii.\textsuperscript{*}] $ (24-N-P)(\textbf{N},\textbf{1},\textbf{1}) + 3(\mathbf{ \tfrac{1}{2}N(N-1) },\textbf{1},\textbf{1}) + (M-8)(\textbf{1},\textbf{M},\textbf{1}) + (\textbf{1},\mathbf{ \tfrac{1}{2}M(M+1) },\textbf{1}) + (P-N)(\textbf{1},\textbf{1},\textbf{P}) + k(\textbf{1},\textbf{1},\mathbf{ 2^{\lfloor\tfrac{P-1}{2}\rfloor} }) + (\textbf{N},\textbf{1},\textbf{P}) $\\
    $k=4$ for $P=13$, and $k=8$ for $P=10,11,12$
\end{enumerate}
    \item $10\leq N=M \leq 11$, \quad $10\leq P\leq 14:$

    \item[] $8(\textbf{1},\textbf{1},\textbf{P}) + (\textbf{1},\textbf{1},\mathbf{ \tfrac{1}{2}P(P-1) } + k(\textbf{1},\textbf{1},\mathbf{ 2^{\lfloor\tfrac{P-1}{2}\rfloor} }) + 2(\textbf{M},\textbf{N},\textbf{1})$\\
    $k=4$ for $P=13$, and $k=8$ for $P=10,11,12$

    \item $10\leq N=M \leq 11$, \quad $10\leq P\leq 14:$

    \item[] $ (P-2)(\textbf{1},\textbf{1},\textbf{P}) + \tilde{k}(\textbf{1},\textbf{1},\mathbf{ 2^{\lfloor\tfrac{P-1}{2}\rfloor} }) + 2(\textbf{N},\textbf{M},\textbf{1}) $\\
    $\tilde{k} = 3$ for $P=13$, and $\tilde{k} = 6$ for $P=10,11,12$

    \item $10\leq N=M \leq 11$, $P=10:$

    \item[] $ 4(\textbf{1},\textbf{1},\mathbf{ \tfrac{1}{2}P(P-1) }) + 6(\textbf{1},\textbf{1},\mathbf{ 2^{\lfloor\tfrac{P-1}{2}\rfloor} }) + 2(\textbf{N},\textbf{M},\textbf{1}) $
\end{enumerate}

\item $\boxed{SU(N)\times SU(M)\times Sp(P)}$ 

\begin{enumerate}
    \item $10\leq N\leq 14$, \quad $18\leq M\leq 22$, \quad $10\leq P \leq 14$, \quad $N+P\leq 24$, \quad $N+M\leq 32$, \quad $P\leq M-8$:

    \item[] $(24-N-P)(\textbf{N},\textbf{1},\textbf{1}) + 3(\mathbf{\tfrac{1}{2}N(N-1)},\textbf{1},\textbf{1}) + (M-8-P)(\textbf{1},\textbf{M},\textbf{1}) + (\textbf{1},\mathbf{\tfrac{1}{2}M(M+1)},\textbf{1}) + (32-N-M)(\textbf{1},\textbf{1},\textbf{2P}) + (\textbf{1},\textbf{1},\mathbf{P(2P-1)-1}) + (\textbf{N},\textbf{1},\textbf{2P}) + (\textbf{1},\textbf{M},\textbf{2P})$

    \item $10\leq N\leq M \leq 11$, \quad $5\leq P \leq 8$:
\begin{enumerate}
    \item $(16-2P)(\textbf{N},\textbf{1},\textbf{1}) + 2(\mathbf{ \tfrac{1}{2}N(N-1) },\textbf{1},\textbf{1}) + (16-2P)(\textbf{1},\textbf{M},\textbf{1}) + 2(\textbf{1},\mathbf{ \tfrac{1}{2}M(M-1) },\textbf{1}) + 2(2P+8-M-N)(\textbf{1},\textbf{1},\textbf{2P}) + 2(\textbf{1},\textbf{M},\textbf{2P}) + 2(\textbf{N},\textbf{1},\textbf{2P})  $

    \item $(N-8)(\textbf{N},\textbf{1},\textbf{1}) + (\mathbf{ \tfrac{1}{2}N(N+1) },\textbf{1},\textbf{1}) + (24-M-2P)(\textbf{1},\textbf{M},\textbf{1}) + 3(\textbf{1},\mathbf{ \tfrac{1}{2}M(M-1) },\textbf{1}) + 2(24-M-2P)(\textbf{1},\textbf{1},\textbf{2P}) + 2(\textbf{1},\textbf{1},\mathbf{ P(2P-1)-1 }) + 2(\textbf{1},\textbf{M},\textbf{2P}) $

    \item $(24-N-2P)(\textbf{N},\textbf{1},\textbf{1}) + 3(\mathbf{ \tfrac{1}{2}N(N-1) },\textbf{1},\textbf{1}) + (M-8)(\textbf{1},\textbf{M},\textbf{1}) + (\textbf{1},\mathbf{ \tfrac{1}{2}M(M+1) },\textbf{1}) + 2(24-N-2P)(\textbf{1},\textbf{1},\textbf{2P}) + 2(\textbf{1},\textbf{1},\mathbf{ P(2P-1)-1 }) + 2(\textbf{N},\textbf{1},\textbf{2P}) $

    \item $ (2N-M-P)(\textbf{N},\textbf{1},\textbf{1}) + (16-N-P)(\textbf{1},\textbf{M},\textbf{1}) + 2(\textbf{1},\mathbf{ \tfrac{1}{2}M(M-1) },\textbf{1}) + (48-N-M-4P)(\textbf{1},\textbf{1},\textbf{2P}) + 2(\textbf{1},\textbf{1},\mathbf{ P(2P-1)-1 }) + (\textbf{1},\textbf{M},\textbf{2P}) + (\textbf{N},\textbf{1},\textbf{2P}) + (\textbf{N},\textbf{M},\textbf{1}) $

    \item $ (16-M-P)(\textbf{N},\textbf{1},\textbf{1}) + 2(\mathbf{ \tfrac{1}{2}N(N-1) },\textbf{1},\textbf{1}) + (2M-N-P)(\textbf{1},\textbf{M},\textbf{1}) + (48-N-M-4P)(\textbf{1},\textbf{1},\textbf{P}) + 2(\textbf{1},\textbf{1},\mathbf{ P(2P-1)-1 }) + (\textbf{1},\textbf{M},\textbf{2P}) + (\textbf{N},\textbf{1},\textbf{2P}) + (\textbf{N},\textbf{M},\textbf{1}) $
\end{enumerate}
    \item $10\leq N\leq M\leq 11$, \quad $P\geq 5$, \quad $M-N-P+5 \geq 0$:

    \item[] $(16-M)(\textbf{N},\textbf{1},\textbf{1}) + 2(\mathbf{ \tfrac{1}{2}N(N-1) },\textbf{1},\textbf{1}) + (22-N-2P)(\textbf{1},\textbf{M},\textbf{1}) + 2(16-M)(\textbf{1},\textbf{1},\textbf{2P}) + (\textbf{1},\textbf{1},\mathbf{ P(2P-1)-1 }) + 2(\textbf{1},\textbf{M},\textbf{2P}) + (\textbf{N},\textbf{M},\textbf{1})$

    \item $ 10\leq N\leq M \leq 11 $, \quad $P\geq 5$,  \quad $M-P=4$:

    \item[] $ (32-2N-P)(\textbf{N},\textbf{1},\textbf{1}) + 4(\mathbf{ \tfrac{1}{2}N(N-1) } + (12-M)(\textbf{1},\textbf{M},\textbf{1}) + (M-N)(\textbf{1},\textbf{1},\textbf{2P}) + 3(\textbf{1},\textbf{M},\textbf{2P}) + (\textbf{N},\textbf{1},\textbf{2P}) $

    \item $10\leq N=M\leq 11$, \quad $5\leq P\leq 8$:

    \item[] $8(8-P)(\textbf{1},\textbf{1},\textbf{2P}) + 3(\textbf{1},\textbf{1},\mathbf{ P(2P-1)-1 }) + 2(\textbf{N},\textbf{M},\textbf{1}) $

    \item $10\leq N=M\leq 11$, \quad $P=5$:

    \item[] $45(\textbf{1},\textbf{1},\textbf{10}) + (\textbf{1},\textbf{1},\textbf{110}) + 2(\textbf{N},\textbf{M},\textbf{1}) $

    \item $10\leq N\leq 12$, \quad $15\leq M\leq 16$, \quad $10\leq P\leq 11$, \quad $N+2P\leq 2M$:

    \item[] $(16-M)(\textbf{N},\textbf{1},\textbf{1}) + 2(\mathbf{\tfrac{1}{2}N(N-1)},\textbf{1},\textbf{1}) + (2M - N - 2P)(\textbf{1},\textbf{M},\textbf{1}) + 2(16-M)(\textbf{1},\textbf{1},\textbf{P}) + (\textbf{1},\textbf{1},\mathbf{\tfrac{1}{2}N(N-1)-1}) + (\textbf{N},\textbf{M},\textbf{1}) + 2(\textbf{1},\textbf{M},\textbf{2P})$
\end{enumerate}

\bigskip
\item $ \boxed{SU(N) \times SO(M) \times SO(P)} $
\begin{enumerate}
    \item $10\leq N\leq 11$, \quad $10\leq M\leq P\leq 14$, \quad $^{*}(N,M,P) = (10,12,12),\, (10,10,14):$

    \item[] $ (24-N-P)(\textbf{N},\textbf{1},\textbf{1}) + 3(\mathbf{ \tfrac{1}{2}N(N-1) },\textbf{1},\textbf{1}) + (M-8)(\textbf{1},\textbf{M},\textbf{1}) + (P-N)(\textbf{1},\textbf{1},\textbf{P}) + k(\textbf{1},\textbf{1}, \mathbf{ 2^{ \lfloor\tfrac{P-1}{2}\rfloor } }) + (\textbf{N},\textbf{1},\textbf{P}) $\\
    $k=2$ for $P=14$, $k=4$ for $P=13$, and $k=8$ for $P=10,11,12$

    \item $10\leq N\leq 11$, \quad $N\leq M < P \leq 12: $

    \item[] $ (24-N-M)(\textbf{N},\textbf{1},\textbf{1}) + 3(\mathbf{ \tfrac{1}{2}N(N-1) },\textbf{1},\textbf{1}) + (M-N)(\textbf{1},\textbf{M},\textbf{1}) + 8(\textbf{1},\mathbf{ 2^{ \lfloor\tfrac{M-1}{2}\rfloor } },\textbf{1}) + (P-8)(\textbf{1},\textbf{1},\textbf{P}) + (\textbf{N},\textbf{M},\textbf{1}) $

    \item $N=10$, \quad $10\leq M=P\leq 12$,\quad $(N,M,P)^{*}=(10,12,12):$

    \item[] $ 2(\textbf{N},\textbf{1},\textbf{1}) + (\mathbf{ \tfrac{1}{2}N(N+1) },\textbf{1},\textbf{1}) + 8(\textbf{1},\mathbf{ 2^{ \lfloor\tfrac{M-1}{2}\rfloor } },\textbf{1}) + 8(\textbf{1},\textbf{1},\mathbf{ 2^{ \lfloor\tfrac{P-1}{2}\rfloor } }) + (\textbf{1},\textbf{M},\textbf{P}) $
\end{enumerate}

\bigskip
\item $\boxed{ SU(N)\times SO(M) \times Sp(P) }$ 
\begin{enumerate}
    \item $10\leq N\leq 13$, \quad $18\leq M\leq 22$, \quad $10\leq P \leq 14$,\\
    or $ 10\leq N \leq 11 $, \quad $ 10 \leq M \leq 21 $, \quad $5 \leq P \leq 10$:

    \item[] {$ (24-N-P)(\textbf{N},\textbf{1},\textbf{1}) + 3(\mathbf{ \tfrac{1}{2}N(N-1) },\textbf{1},\textbf{1}) + (M-P-8)(\textbf{1},\textbf{M},\textbf{1}) + (32-M-N)(\textbf{1},\textbf{1},\textbf{2P}) + (\textbf{1},\textbf{1},\mathbf{ P(2P-1)-1 }) + (\textbf{1},\textbf{N},\textbf{2P}) + (\textbf{N},\textbf{1},\textbf{2P}) $}

    \item $ 10\leq N\leq 11 $, \quad $ 10\leq M\leq 13 $, \quad $5\leq P\leq 9$:
\begin{enumerate}
    \item $ (N-8)(\textbf{(N},\textbf{1},\textbf{1}) + (\mathbf{ \tfrac{1}{2}N(N+1) },\textbf{1},\textbf{1}) + (M-P-3)(\textbf{1},\textbf{M},\textbf{1}) + 5(\textbf{1},\mathbf{ 2^{\ \lfloor\tfrac{M-1}{2}\rfloor } },\textbf{1}) + (48-M-4P)(\textbf{1},\textbf{1},\textbf{2P}) + 2(\textbf{1},\textbf{1},\mathbf{ P(2P-1)-1 }) + (\textbf{1},\textbf{M},\textbf{2P}) $

    \item $ (24-N-2P)(\textbf{N},\textbf{1},\textbf{1}) + 3(\mathbf{\tfrac{1}{2}N(N-1)},\textbf{1},\textbf{1}) + (M-8)(\textbf{1},\textbf{M},\textbf{1}) + 2(24-N-2P)(\textbf{1},\textbf{1},\textbf{2P}) + 2(\textbf{1},\textbf{1}, \mathbf{ P(2P-1)-1 }) + 2(\textbf{N},\textbf{1},\textbf{2P})  $
\end{enumerate}
    \item $N=10$, \quad $M=14$, \quad $P=5^{*}, 7$:

    \item[] $ (14-2P)(\textbf{10},\textbf{1},\textbf{1}) + 3(\textbf{45},\textbf{1},\textbf{1}) + 6(\textbf{1},\textbf{14},\textbf{1}) + 2(14-2P)(\textbf{1},\textbf{1},\textbf{2P}) + 2(\textbf{1},\textbf{1},\mathbf{ P(2P-1)-1 }) + 2(\textbf{10},\textbf{1},\textbf{2P}) $

    \item $N=11$, \quad $M=14$, \quad $P=5^{*}$:

    \item[] $3(\textbf{11},\textbf{1},\textbf{1}) + 3(\textbf{55},\textbf{1},\textbf{1}) + 6(\textbf{1},\textbf{14},\textbf{1}) + 6(\textbf{1},\textbf{1},\textbf{10}) + 2(\textbf{1},\textbf{1},\textbf{44}) + 2(\textbf{11},\textbf{1},\textbf{10}) $

    \item $10\leq N\leq 11$, \quad $10\leq M\leq14$, \quad $5\leq P \leq 8$:

    \item[] $ (16-2P)(\textbf{N},\textbf{1},\textbf{1}) + 2(\mathbf{ \tfrac{1}{2}N(N-1) },\textbf{1},\textbf{1}) + (M-P-4)(\textbf{1},\textbf{M},\textbf{1}) + \tfrac{1}{2} k (\textbf{1},\mathbf{ 2^{ \lfloor\tfrac{M-1}{2}\rfloor } },\textbf{1}) + (80-2N-M-4P)(\textbf{1},\textbf{1},\textbf{2P}) + (\textbf{1},\textbf{M},\textbf{2P}) + 2(\textbf{N},\textbf{1},\textbf{2P}) $\\
    $k=2$ for $M=14$, $k=4$ for $M=13$, and $k=8$ for $M=10,11,12$

    \item $10\leq N \leq 11$, \quad $N+M \leq 24$, \quad $P=5$:

    \item[] $ (N-8)(\textbf{N},\textbf{1},\textbf{1}) + (\mathbf{ \tfrac{1}{2}N(N+1) },\textbf{1},\textbf{1}) + (M-2P)(\textbf{1},\textbf{M},\textbf{1}) + k(\textbf{1},\mathbf{ 2^{ \lfloor\tfrac{M-1}{2}\rfloor } },\textbf{1}) + (28-2M)(\textbf{1},\textbf{1},\textbf{2P}) + 2(\textbf{1},\textbf{1},\mathbf{ P(2P-1)-1}) + 2(\textbf{1},\textbf{M},\textbf{2P}) $\\
    $k=2$ for $M=14$, $k=4$ for $M=13$, and $k=8$ for $M=10,11,12$

    \item $10\leq N\leq 11$, \quad $M=12$, \quad $P=6$:

    \item[] $ (N-8)(\textbf{N},\textbf{1},\textbf{1}) + (\mathbf{ \tfrac{1}{2}N(N+1) },\textbf{1},\textbf{1}) + 8(\textbf{1},\textbf{32},\textbf{1}) + 4(\textbf{1},\textbf{1},\textbf{12}) + 2(\textbf{1},\textbf{1},\textbf{65}) + 2(\textbf{1},\textbf{12},\textbf{12})  $

    \item $10\leq N \leq 11$, \quad $ 10\leq M \leq 12 $, \quad $5\leq P \leq 10$, \quad $2N+M\leq 32$:
\begin{enumerate}
    \item $2(N-P)(\textbf{N},\textbf{1},\textbf{1}) + (M-P-2)(\textbf{1},\textbf{M},\textbf{1}) + 6(\textbf{1},\mathbf{ 2^{ \lfloor\tfrac{M-1}{2}\rfloor } },\textbf{1}) + (32-2N-M)(\textbf{1},\textbf{1},\textbf{2P}) + (\textbf{1},\textbf{1},\mathbf{ P(2P-1)-1 }) + (\textbf{1},\textbf{M},\textbf{2P}) + 2(\textbf{N},\textbf{1},\textbf{2P})  $

    \item $ (8 - (N-8)\alpha - 2P + N)(\textbf{N},\textbf{1},\textbf{1}) + (1+\alpha)(\mathbf{ \tfrac{1}{2}N(N-1) },\textbf{1},\textbf{1}) + (M-P-3-\alpha)(\textbf{1},\textbf{M},\textbf{1}) + (5-\alpha)(\textbf{1},\mathbf{ 2^{ \lfloor\tfrac{M-1}{2}\rfloor } },\textbf{1}) + 2(\textbf{1},\textbf{1},\textbf{2P}) + (\textbf{1},\textbf{1},\mathbf{ P(2P-1)-1 }) + (\textbf{1},\textbf{M},\textbf{P}) + 2(\textbf{N},\textbf{1},\textbf{P}) $\\
    $\alpha=0,...,4$
\end{enumerate}
    \item $10\leq N \leq 11$, \quad $10\leq M\leq N$, \quad $P=N-4$:

    \item[] $ (8-P)(\textbf{N},\textbf{1},\textbf{1}) + (8-P)(\textbf{1},\textbf{M},\textbf{1}) + (\textbf{1},\mathbf{\tfrac{1}{2}M(M-1)},\textbf{1}) + 8(\textbf{1},\mathbf{ 2^{ \lfloor\tfrac{M-1}{2}\rfloor } },\textbf{1}) + (M-N)(\textbf{1},\textbf{1},\textbf{2P}) + (\textbf{1},\textbf{M},\textbf{2P}) + 3(\textbf{N},\textbf{1},\textbf{2P}) $
    
\end{enumerate}

\item $\boxed{ SU(N) \times Sp(M) \times Sp(N)} $
\begin{enumerate}
    \item $28\leq N\leq 32$, \quad $10\leq M\leq P \leq 14$, \quad $M+P\leq N-8$:

    \item[] $(N-8-M-P)(\textbf{N},\textbf{1},\textbf{1}) + (\mathbf{\tfrac{1}{2}N(N-1)},\textbf{1},\textbf{1}) ++ (32-N)(\textbf{1},\textbf{2M},\textbf{1}) + (\textbf{1},\mathbf{M(2M-1)-1},\textbf{1}) + (32-N)(\textbf{1},\textbf{1},\textbf{2P}) + (\textbf{1},\textbf{1},\mathbf{P(2P-1)-1}) + (\textbf{N},\textbf{2M},\textbf{1}) + (\textbf{N},\textbf{1},\textbf{2P})$
    
    \item $10\leq N\leq 11$, \quad $5\leq M\leq P\leq 8$
\begin{enumerate}
    \item $ 2(8-P)(\textbf{N},\textbf{1},\textbf{1}) + 2(\mathbf{\tfrac{1}{2}N(N-1)},\textbf{1},\textbf{1}) + 4(8-P)(\textbf{1},\textbf{2M},\textbf{1}) + 2(\textbf{1},\mathbf{ M(2P-1)-1 },\textbf{1}) + 2(2P-2M-N+8)(\textbf{1},\textbf{1},\textbf{2P}) + 2(\textbf{1},\textbf{2M},\textbf{2P}) + 2(\textbf{N},\textbf{1},\textbf{2P}) $

    \item $ (N-8)(\textbf{N},\textbf{1},\textbf{1}) + (\mathbf{ \tfrac{1}{2}N(N+1) },\textbf{1},\textbf{1}) + 4(12-M-P)(\textbf{1},\textbf{2M},\textbf{1}) + 2(\textbf{1},\textbf{ M(2M-1)-1 },\textbf{1}) + 4(12-M-P)(\textbf{1},\textbf{1},\textbf{2P}) + 2(\textbf{1},\textbf{1},\textbf{ P(2P-1)-1 }) + 2(\textbf{1},\textbf{2M},\textbf{2P}) $

    \item $2(N-M-P)(\textbf{N},\textbf{1},\textbf{1}) + 2(16-N)(\textbf{1},\textbf{2M},\textbf{1}) + (\textbf{1},\textbf{ M(2M-1)-1 },\textbf{1}) + 2(16-N)(\textbf{1},\textbf{1},\textbf{2P}) + (\textbf{1},\textbf{1},\textbf{ P(2P-1)-1 }) + 2(\textbf{N},\textbf{1},\textbf{2P}) + (2\textbf{N},\textbf{2M},\textbf{1}) $

    \item $ (2N-M-2P)(\textbf{N},\textbf{1},\textbf{1}) + (48-N-4M-2P)(\textbf{1},\textbf{2M},\textbf{1}) + 2(\textbf{1},\textbf{ M(2M-1)-1 },\textbf{1}) + 2(16-N-M)(\textbf{1},\textbf{1},\textbf{2P}) + (\textbf{1},\textbf{1},\textbf{ P(2P-1)-1 }) + (\textbf{1},\textbf{2M},\textbf{2P} ) + 2(\textbf{N},\textbf{1},\textbf{2P}) + (\textbf{N},\textbf{2M},\textbf{1})$

    \item $ (2N-2M-P)(\textbf{N},\textbf{1},\textbf{1}) + 2(16-N-P)(\textbf{1},\textbf{2M},\textbf{1}) + (\textbf{1},\textbf{ M(2M-1)-1 },\textbf{1}) + (48-N-2M-4P)(\textbf{1},\textbf{1},\textbf{2P}) + 2(\textbf{1},\textbf{1},\textbf{ P(2P-1)-1 }) + (\textbf{1},\textbf{2M},\textbf{2P}) + (\textbf{N},\textbf{1},\textbf{2P}) + 2(\textbf{N},\textbf{2M},\textbf{1}) $
\end{enumerate}
    \item $10\leq N\leq 11$, \quad $M=5$, \quad $5\leq P\leq 7$

    \item[] $ (2N-P-15)(\textbf{N},\textbf{1},\textbf{1}) +  3(12-N)(\textbf{1},\textbf{2M},\textbf{1}) + (48-N-4P)(\textbf{1},\textbf{1},\textbf{2P}) + (\textbf{1},\textbf{1},\textbf{ P(2P-1)-1 }) + (\textbf{N},\textbf{1},\textbf{2P}) + 3(\textbf{N},\textbf{2M},\textbf{1})$

    \item $10\leq N\leq 11$, \quad $M=5$, \quad $N-P=4$

    \item[] $ (12-N)(\textbf{N},\textbf{1},\textbf{1}) + (24-2P)(\textbf{1},\textbf{2M},\textbf{1}) + 3(\textbf{1},\textbf{ M(2M-1)-1 },\textbf{1}) + (P-6)(\textbf{1},\textbf{1},\textbf{2P}) + (\textbf{1},\textbf{2M},\textbf{2P}) + 3(\textbf{N},\textbf{1},\textbf{2P}) $
\end{enumerate}

\item $ \boxed{\textrm{SO}(N) \times \textrm{SO}(M) \times \textrm{SO}(P)}$ 
\begin{enumerate}
    \item $N=10$, \quad $10\leq M=P\leq 13^{*}$:

    \item[] $ 2(\textbf{N},\textbf{1},\textbf{1}) + k(\textbf{1},\mathbf{ 2^{\lfloor\tfrac{M-1}{2}\rfloor} },\textbf{1}) + k(\textbf{1},\textbf{1},\mathbf{ 2^{ \lfloor\tfrac{P-1}{2}\rfloor } }) + (\textbf{1},\textbf{M},\textbf{P}) $

    \item $10\leq N\leq 11^{*}$, \quad $M=P=10$:

    \item[] $ (66-6N)(\textbf{N},\textbf{1},\textbf{1}) + 4(\mathbf{ 2^{ \lfloor\tfrac{N-1}{2}\rfloor } },\textbf{1},\textbf{1}) + 6(\textbf{1},\textbf{M},\textbf{1}) + 4(\textbf{1},\mathbf{ 2^{\lfloor\tfrac{M-1}{2}\rfloor})},\textbf{1}) + 6(\textbf{1},\textbf{1},\textbf{P}) + 4(\textbf{1},\textbf{1},\mathbf{ 2^{ \lfloor\tfrac{P-1}{2}\rfloor } }) $

    \item $N=11$, \quad $M=P=10^{*}$:

    \item[] $8(\textbf{1},\textbf{16},\textbf{1}) + 8(\textbf{1},\textbf{1},\textbf{16}) + (\textbf{1},\textbf{10},\textbf{10})$
\end{enumerate}

\item $\boxed{ SO(N)\times SO(M)\times Sp(P)}$

\begin{enumerate}
    \item $10\leq N\leq 12$, \quad $N\leq M\leq 21-P$, \quad $5\leq P\leq N-3$, or $(N,M,P) = (12,13,9)$, \\ $(N,M,P)\neq (12,14,17),(12,15,6)$,\\
    $(N,M,P)^{*} = (11,14,7)$:

    \item[] $(N-P-3)(\textbf{N},\textbf{1},\textbf{1}) + 5(\mathbf{2}^{\lfloor\tfrac{N-1}{2}\rfloor},\textbf{1},\textbf{1}) + (M-8)(\textbf{1},\textbf{M},\textbf{1}) + (2(24-2P)-N)(\textbf{1},\textbf{1},\textbf{2P}) + 2(\textbf{1},\textbf{1},\mathbf{P(2P-1)-1}) +(\textbf{N},\textbf{1},\textbf{2P}) $

    \item $10\leq N<M\leq 12$,\quad $5\leq P\leq M-8$:

    \item[] $(N-9)(\textbf{N},\textbf{1},\textbf{1}) + (M-P-3)(\textbf{1},\textbf{M},\textbf{1}) + 5(\textbf{1},\mathbf{2}^{\lfloor\tfrac{M-1}{2}\rfloor},\textbf{1}) + (2(24-2P)-M)(\textbf{1},\textbf{1},\textbf{2P}) + 2(\textbf{1},\textbf{1},\mathbf{P(2P-1)-1}) + (\textbf{1},\textbf{M},\textbf{2P}) $

    \item $10\leq N\leq 14$, \quad $P+8\leq M\leq 21$, \quad $N\leq M$, \quad $N+M\leq 32$, \quad $5\leq P\leq 10$,\\
    $(N,M,P)\neq (11,21,5),(12,20,5),(13,19,5),(14,18,5),(14,18,6),(14,18,7)$,\\
    $(N,M,P)^{*}=(11,21,6),(13,19,6),(14,17,5)$:

    \item[] $(N-P)(\textbf{N},\textbf{1},\textbf{1}) + k(\mathbf{2}^{\lfloor\tfrac{N-1}{2}\rfloor},\textbf{1},\textbf{1}) + (M-8-P)(\textbf{1},\textbf{M},\textbf{1}) + (32-M-N)(\textbf{1},\textbf{1},\textbf{2P}) + (\textbf{1},\textbf{1},\mathbf{\tfrac{1}{2}P(2P-1)-1}) + (\textbf{N},\textbf{1},\textbf{2P}) + (\textbf{1},\textbf{M},\textbf{2P}) $\\
    $k=2$ for $N=14$, $k=4$ for $N=13$, $k=8$ for $N=10,11,12$

    \item $10\leq N\leq M\leq 14$, \quad $5\leq P\leq N-4$, \quad $(N,M,P)\neq (14,14,10)$:

    \item[] $(N-4-P)(\textbf{N},\textbf{1},\textbf{1}) + k_{1}(\mathbf{2}^{\lfloor\tfrac{N-1}{2}\rfloor},\textbf{1},\textbf{1}) + (M-4-P)(\textbf{1},\textbf{M},\textbf{1}) + k_{2}(\textbf{1},\mathbf{2}^{\lfloor\tfrac{M-1}{2}\rfloor},\textbf{1}) + (32-N-M)(\textbf{1},\textbf{1},\textbf{2P}) + (\textbf{1},\textbf{1},\mathbf{P(2P-1)-1}) + (\textbf{N},\textbf{1},\textbf{2P}) + (\textbf{1},\textbf{M},\textbf{2P})$ \\
    $k_1 = 1$ for $N=14$, $k_1 = 2$ for $N=13$, $k_1 = 4$ for $N=10,11,12$\\
    $k_2$ as above with $N\to M$

    \item $10\leq N\leq M\leq 14$, \quad $5\leq P\leq 7$, \quad $P\leq N-4$\\
    $(N,M,P)^{*} = (12,12,8)$:

    \item[] $(N-4-P)(\textbf{N},\textbf{1},\textbf{1}) + k_{1}(\mathbf{2}^{\lfloor\tfrac{N-1}{2}\rfloor},\textbf{1},\textbf{1}) + (M-4-P)(\textbf{1},\textbf{M},\textbf{1}) + k_{2}(\textbf{1},\mathbf{2}^{\lfloor\tfrac{M-1}{2}\rfloor},\textbf{1}) + (2(2P+8)-N-M)(\textbf{1},\textbf{1},\textbf{2P})  + (\textbf{N},\textbf{1},\textbf{2P}) + (\textbf{1},\textbf{M},\textbf{2P})$ \\
    $k_1 = 1$ for $N=14$, $k_1 = 2$ for $N=13$, $k_1 = 4$ for $N=10,11,12$\\
    $k_2$ as above with $N\to M$

    \item $10\leq N \leq 13$, \quad $M=13$, \quad $5\leq P\leq 7$:

    \item[] $(N-2-P)(\textbf{N},\textbf{1},\textbf{1}) + 3k_{1}(\mathbf{2}^{\lfloor\tfrac{N-1}{2}\rfloor},\textbf{1},\textbf{1}) + (M-6-P)(\textbf{1},\textbf{M},\textbf{1}) + (\textbf{1},\mathbf{2}^{\lfloor\tfrac{M-1}{2}\rfloor},\textbf{1}) + (32-M-N)(\textbf{1},\textbf{1},\textbf{2P}) + (\textbf{1},\textbf{1},\mathbf{\tfrac{1}{2}P(2P-1)-1})  + (\textbf{N},\textbf{1},\textbf{2P}) + (\textbf{1},\textbf{M},\textbf{2P})$ \\
    $k_1 = 1$ for $N=13$, $k_1 = 2$ for $N=10,11,12$\\

    \item $10\leq N\leq M\leq 12$, \quad $5\leq P\leq M-5$:

    \item[] $(N-P-(4-n))(\textbf{N},\textbf{1},\textbf{1}) + (4+n)(\mathbf{2}^{\lfloor\tfrac{N-1}{2}\rfloor},\textbf{1},\textbf{1}) + (M-P-(4+n))(\textbf{1},\textbf{M},\textbf{1}) + (4-n)(\textbf{1},\mathbf{2}^{\lfloor\tfrac{M-1}{2}\rfloor},\textbf{1}) + (32-M-N)(\textbf{1},\textbf{1},\textbf{2P}) + (\textbf{1},\textbf{1},\mathbf{\tfrac{1}{2}P(2P-1)-1})  + (\textbf{N},\textbf{1},\textbf{2P}) + (\textbf{1},\textbf{M},\textbf{2P})$ \\
    $n$ any integer where the coefficients are positive

    \item $10\leq N\leq 12$, \quad $N\leq M\leq 14$, \quad $P=5$,\\
    or $M=12$, \quad $P=6$,\\
    $(N,M,P)\neq (12,14,5)$, \quad $(N,M,P)^{*} = (11,14,5), (12,13,5)$:

    \item[] $(N-8)(\textbf{N},\textbf{1},\textbf{1}) + (M-2P)(\textbf{1},\textbf{M},\textbf{1}) + k(\textbf{1},\mathbf{2}^{\lfloor\tfrac{M-1}{2}\rfloor},\textbf{1}) + (2(24-2P)-2M)(\textbf{1},\textbf{1},\textbf{2P}) + 2(\textbf{1},\textbf{1},\mathbf{p(2P-1)-1}) + 2(\textbf{1},\textbf{M},\textbf{2P})$\\
    $k=8$ for $M=14$, $k=4$ for $M=13$, $k=2$ for $M=10,11,12$

    \item $10\leq N\leq 11$, \quad $N<M\leq 13$, \quad $P=5$\\
    $(N,M,P)^{*} = (11,13,5)$:

    \item[] $(N-2P)(\textbf{N},\textbf{1},\textbf{1}) +8(\mathbf{2}^{\lfloor\tfrac{N-1}{2}\rfloor},\textbf{1},\textbf{1}) + (M-8)(\textbf{1},\textbf{M},\textbf{1}) + (2(24-2P)-2N)(\textbf{1},\textbf{1},\textbf{2P}) + 2(\textbf{1},\textbf{1},\mathbf{P(2P-1)-1}) + 2(\textbf{N},\textbf{1},\textbf{2P}) $

    \item $10\leq N\leq M\leq 14$, \quad $N+M\leq 24$, \quad $5\leq P\leq 6$,\\
    $(N,M,P)\neq (10,14,5)$,\quad $(N,M,P)^{*} = (10,14,6)$:

    \item[] $(N-P-4)(\textbf{N},\textbf{1},\textbf{1}) + k_{1}(\mathbf{2}^{\lfloor\tfrac{N-1}{2}\rfloor},\textbf{1},\textbf{1}) + (M-4-P)(\textbf{1},\textbf{M},\textbf{1}) + k_{2}(\textbf{1},\mathbf{2}^{\lfloor\tfrac{M-1}{2}\rfloor},\textbf{1}) + (2(24-2P)-M-N)(\textbf{1},\textbf{1},\textbf{2P}) + 2(\textbf{1},\textbf{1},\mathbf{p(2P-1)-1})+ (\textbf{N},\textbf{1},\textbf{2P}) + (\textbf{1},\textbf{M},\textbf{2P})$\\
    $k_1 = 1$ for $N=14$, $k_1 = 2$ for $N=13$, $k_1 = 4$ for $N=10,11,12$\\
    $k_2$ as above with $N\to M$

    \item $(N,M,P) = (10,10,5),(10,11,5),(10,12,5)$:

    \item[] $8(\mathbf{2}^{\lfloor\tfrac{N-1}{2}\rfloor},\textbf{1},\textbf{1}) + (M-2P)(\textbf{1},\textbf{M},\textbf{1}) + 8(\textbf{1},\mathbf{2}^{\lfloor\tfrac{M-1}{2}\rfloor},\textbf{1}) + (32-N-2M)(\textbf{1},\textbf{1},\textbf{2P}) + (\textbf{1},\textbf{1},\mathbf{P(2P-1)-1}) + (\textbf{N},\textbf{1},\textbf{2P}) + 2(\textbf{1},\textbf{M},\textbf{2P})$

    \item $(N,M,P) = (10,11,5)$:

    \item[] $3(\mathbf{2}^{\lfloor\tfrac{N-1}{2}\rfloor},\textbf{1},\textbf{1}) + (N-P-5)(\textbf{1},\textbf{M},\textbf{1}) + 3(\textbf{1},\mathbf{2}^{\lfloor\tfrac{M-1}{2}\rfloor},\textbf{1}) + (32-2N-M)(\textbf{1},\textbf{1},\textbf{2P}) + (\textbf{1},\textbf{1},\mathbf{P(2P-1)-1}) + 2(\textbf{N},\textbf{1},\textbf{2P}) + (\textbf{1},\textbf{M},\textbf{2P})$

    \item $(N,M,P) = (11,13,5),(12,13,5),(12,13,6)$:

    \item[] $(N-6-P)(\textbf{N},\textbf{1},\textbf{1}) + 2(\mathbf{2}^{\lfloor\tfrac{N-1}{2}\rfloor},\textbf{1},\textbf{1}) + (M-2-P)(\textbf{1},\textbf{M},\textbf{1}) + 3(\textbf{1},\mathbf{2}^{\lfloor\tfrac{M-1}{2}\rfloor},\textbf{1}) + (32-M-N)(\textbf{1},\textbf{1},\textbf{2P}) + (\textbf{1},\textbf{1},\mathbf{\tfrac{1}{2}P(2P-1)-1})  + (\textbf{N},\textbf{1},\textbf{2P}) + (\textbf{1},\textbf{M},\textbf{2P})$

    \item $(N,M,P) = (13,14,5),(13,15,5)^{*}$:

    \item[] $(N-P-8)(\textbf{N},\textbf{1},\textbf{1}) + (M-P)(\textbf{1,\textbf{M},\textbf{1}}) + k(\textbf{1},\mathbf{2^{\lfloor\tfrac{N-1}{2}\rfloor}},\textbf{1}) + (32-N-M)(\textbf{1},\textbf{1},\textbf{2P}) + (\textbf{1},\textbf{1},\mathbf{P(2P-1)-1}) + (\textbf{N},\textbf{1},\textbf{2P}) + (\textbf{1},\textbf{M},\textbf{2P}) $
\end{enumerate}

\item $\boxed{ SO(N)\times Sp(M)\times Sp(P) }$

\begin{enumerate}
    \item $13\leq N\leq 21$, \quad $5\leq M\leq P \leq 10$, \quad $P\leq N-8$, \quad $N+2M\leq 32$:

    \item[] $(N-P-8)(\textbf{N},\textbf{1},\textbf{1}) + (2(24-2M)-2P)(\textbf{1},\textbf{2M},\textbf{1}) + 2(\textbf{1},\mathbf{M(2M-1)-1},\textbf{1}) + (32-2M-N)(\textbf{1},\textbf{1},\textbf{2P}) + (\textbf{1},\textbf{1},\mathbf{P(2P-1)-1}) + (\textbf{N},\textbf{1},\textbf{2P}) + (\textbf{1},\textbf{2M},\textbf{2P})$

    \item $18\leq N\leq 32$, \quad $10\leq M\leq P \leq 14$, \quad $M+P \leq N-8$,\\
    or $8\leq N\leq 21$, \quad $5\leq M\leq P\leq 6$, \quad $M+P\leq N-8 $:

    \item[] $(N-M-P-8)(\textbf{N},\textbf{1},\textbf{1}) + (32-N)(\textbf{1},\textbf{2M},\textbf{1}) + (\textbf{1},\mathbf{M(2M-1)-1},\textbf{1}) + (32-N)(\textbf{1},\textbf{1},\textbf{2P}) + (\textbf{1},\textbf{1},\mathbf{P(2P-1)-1}) + (\textbf{N},\textbf{2M},\textbf{1}) + (\textbf{N},\textbf{1},\textbf{2P})$

    \item $13 \leq N \leq 20$, \quad $5\leq M<P\leq 9$, \quad $2P+N\leq 32$, \quad $M\leq N-8$:

    \item[] $(N-M-8)(\textbf{N},\textbf{1},\textbf{1}) + (32 -2P-N)(\textbf{1},\textbf{2M},\textbf{1}) + (\textbf{1},\mathbf{M(2M-1)-1},\textbf{1}) + (2(24-2P)-2M)(\textbf{1},\textbf{1},\textbf{2P}) + 2(\textbf{1},\textbf{1},\mathbf{P(2P-1)-1}) + (\textbf{N},\textbf{2M},\textbf{1}) + (\textbf{1},\textbf{2M},\textbf{2P})$

    \item $10\leq N\leq 14$, \quad $5\leq M\leq P \leq 8$, \quad $P\leq N-4$:

    \item[] $(N-4-P)(\textbf{N},\textbf{1},\textbf{1}) + k(\mathbf{2^{\lfloor\tfrac{N-1}{2}\rfloor}},\textbf{1},\textbf{1}) + (32-4P)(\textbf{1},\textbf{2M},\textbf{1}) + (\textbf{1},\mathbf{M(2M-1)-1},\textbf{1}) + (2(2P-2M+8)-N)(\textbf{1},\textbf{1},\textbf{2P}) + (\textbf{N},\textbf{1},\textbf{2P}) + 2(\textbf{1},\textbf{2M},\textbf{2P})$\\
    $k=1$ for $N=14$, $k=2$ for $N=13$, $k=4$ for $N=10,11,12$

    \item $10\leq N\leq 14$, \quad $5\leq M\leq P\leq 7$, \quad $2M+2P \leq 24$:
    
    \item[] $(N-8)(\textbf{N},\textbf{1},\textbf{1}) + 2(24-2M-2P)(\textbf{1},\textbf{2M},\textbf{1}) + 2(\textbf{1},\mathbf{M(2M-1)-1},\textbf{1}) + 2(24-2M-2P)(\textbf{1},\textbf{1},\textbf{2P}) + 2(\textbf{1},\textbf{1},\mathbf{P(2P-1)-1}) + 2(\textbf{1},\textbf{2M},\textbf{2P})$

    \item $10\leq N\leq 12$, \quad $5\leq M\leq N-4$, \quad $P=M+1$:

    \item[] $(N-4-M)(\textbf{N},\textbf{1},\textbf{1}) + 4(\mathbf{2^{\lfloor\tfrac{N-1}{2}\rfloor}},\textbf{1},\textbf{1}) + (2(2M-2P+8)-N)(\textbf{1},\textbf{2M},\textbf{1}) + (32-4M)(\textbf{1},\textbf{1},\textbf{2P}) + (\textbf{1},\textbf{1},\mathbf{P(2P-1)-1}) + (\textbf{N},\textbf{2M},\textbf{1}) + 2(\textbf{1},\textbf{2M},\textbf{2P})$

    \item $10\leq N\leq 12$, \quad $M=5$, \quad $5\leq P\leq N-3$:

    \item[] $(N-P-3-n)(\textbf{N},\textbf{1},\textbf{1}) + (5-n)(\mathbf{2^{\lfloor\tfrac{N-1}{2}\rfloor}},\textbf{1},\textbf{1}) + 2(2M(1-n) + 8(1+n) -2P)(\textbf{1},\textbf{2M},\textbf{1}) + n(\textbf{1},\mathbf{M(2M-1)-1},\textbf{1}) + (32-N-4M)(\textbf{1},\textbf{1},\textbf{2P}) + (\textbf{1},\textbf{1},\mathbf{P(2P-1)-1}) + (\textbf{N},\textbf{1},\textbf{2P}) + 2(\textbf{1},\textbf{2M},\textbf{2P})$\\
    $n=0,...,N-P-3$

    \item $11\leq N\leq 12$, \quad $M=5\leq P$, \quad $M+P<N$:

    \item[] $(N-M-P-1-n)(\textbf{N},\textbf{1},\textbf{1}) + (7-n)(\mathbf{2^{\lfloor\tfrac{N-1}{2}\rfloor}},\textbf{1},\textbf{1}) + (2(2M(1-n) + 8(1+n) -2P)-N)(\textbf{1},\textbf{2M},\textbf{1}) + n(\textbf{1},\mathbf{M(2M-1)-1},\textbf{1}) + (32-N-4M)(\textbf{1},\textbf{1},\textbf{2P}) + (\textbf{1},\textbf{1},\mathbf{P(2P-1)-1}) + (\textbf{N},\textbf{1},\textbf{2P}) + (\textbf{N},\textbf{2M},\textbf{1})+ 2(\textbf{1},\textbf{2M},\textbf{2P})$\\
    $0\leq n<N-M-P$
\end{enumerate}

\item[] $\boxed{ Sp(N) \times Sp(M) \times Sp(P) }$

\item $ \textrm{Sp}(5) \times \textrm{Sp}(5) \times \textrm{Sp}(6) $
\begin{enumerate}
\item  { $ 8 (\textbf{10},\textbf{1},\textbf{1}) + 2 (\textbf{10},\textbf{1},\textbf{12}) + 8 (\textbf{1},\textbf{10},\textbf{1}) + 2 (\textbf{1},\textbf{10},\textbf{12}) + (\textbf{1},\textbf{44},\textbf{1}) + (\textbf{44},\textbf{1},\textbf{1}) $ } 
 
\end{enumerate}

\item $ \textrm{Sp}(5) \times \textrm{Sp}(5) \times \textrm{Sp}(7) $
\begin{enumerate}
\item  { $ 4 (\textbf{10},\textbf{1},\textbf{1}) + 2 (\textbf{10},\textbf{1},\textbf{14}) + 4 (\textbf{1},\textbf{10},\textbf{1}) + 2 (\textbf{1},\textbf{10},\textbf{14}) + 4 (\textbf{1},\textbf{1},\textbf{14}) + (\textbf{1},\textbf{44},\textbf{1}) + (\textbf{44},\textbf{1},\textbf{1}) $ } 
 
\end{enumerate}

\item $ \textrm{Sp}(5) \times \textrm{Sp}(5) \times \textrm{Sp}(8) $
\begin{enumerate}
\item  { $ 2 (\textbf{10},\textbf{1},\textbf{16}) + 2 (\textbf{1},\textbf{10},\textbf{16}) + 8 (\textbf{1},\textbf{1},\textbf{16}) + (\textbf{1},\textbf{44},\textbf{1}) + (\textbf{44},\textbf{1},\textbf{1}) $ } 
 
\end{enumerate}

\item $ \textrm{Sp}(5) \times \textrm{Sp}(6) \times \textrm{Sp}(7) $
\begin{enumerate}
\item  { $ 4 (\textbf{10},\textbf{1},\textbf{1}) + 2 (\textbf{10},\textbf{1},\textbf{14}) + 4 (\textbf{1},\textbf{12},\textbf{1}) + 2 (\textbf{1},\textbf{12},\textbf{14}) + (\textbf{1},\textbf{65},\textbf{1}) + (\textbf{44},\textbf{1},\textbf{1}) $ } 
 
\end{enumerate}

\item $ \textrm{Sp}(5) \times \textrm{Sp}(6) \times \textrm{Sp}(8) $
\begin{enumerate}
\item  { $ 2 (\textbf{10},\textbf{1},\textbf{16}) + 2 (\textbf{1},\textbf{12},\textbf{16}) + 4 (\textbf{1},\textbf{1},\textbf{16}) + (\textbf{1},\textbf{65},\textbf{1}) + (\textbf{44},\textbf{1},\textbf{1}) $ } 
 
\end{enumerate}

\item $ \textrm{Sp}(5) \times \textrm{Sp}(7) \times \textrm{Sp}(8) $
\begin{enumerate}
\item  { $ 2 (\textbf{10},\textbf{1},\textbf{16}) + 2 (\textbf{1},\textbf{14},\textbf{16}) + (\textbf{1},\textbf{90},\textbf{1}) + (\textbf{44},\textbf{1},\textbf{1}) $ } 
 
\end{enumerate}

\item $ \textrm{Sp}(6) \times \textrm{Sp}(6) \times \textrm{Sp}(8) $
\begin{enumerate}
\item  { $ 2 (\textbf{12},\textbf{1},\textbf{16}) + 2 (\textbf{1},\textbf{12},\textbf{16}) + (\textbf{1},\textbf{65},\textbf{1}) + (\textbf{65},\textbf{1},\textbf{1}) $ } 
 
\end{enumerate}

\subsection{$G_1\times G_2 \times G_3 \times G_4$ models}

\item { $\textrm{SU}(10) \times \textrm{SU}(10) \times \textrm{SU}(10) \times \textrm{SU}(15)$ 
\begin{enumerate}
\item {$ ( \textbf{10},\textbf{1},\textbf{1},\textbf{1} ) + 2 ( \textbf{45},\textbf{1},\textbf{1},\textbf{1} ) +  ( \textbf{1},\textbf{10},\textbf{1},\textbf{1} ) + 2 ( \textbf{1},\textbf{45},\textbf{1},\textbf{1} ) +  ( \textbf{1},\textbf{1},\textbf{10},\textbf{1} ) + 2 ( \textbf{1},\textbf{1},\textbf{45},\textbf{1} ) +  ( \textbf{10},\textbf{1},\textbf{1},\textbf{15} ) +  ( \textbf{1},\textbf{10},\textbf{1},\textbf{15} ) +  ( \textbf{1},\textbf{1},\textbf{10},\textbf{15} )$}
\end{enumerate}}

\item { $\textrm{SU}(10) \times \textrm{SU}(10) \times \textrm{SU}(10) \times \textrm{SU}(16)$ 
\begin{enumerate}
\item {$2 ( \textbf{45},\textbf{1},\textbf{1},\textbf{1} ) + 2 ( \textbf{1},\textbf{45},\textbf{1},\textbf{1} ) + 2 ( \textbf{1},\textbf{1},\textbf{45},\textbf{1} ) + 2 ( \textbf{1},\textbf{1},\textbf{1},\textbf{16} ) +  ( \textbf{10},\textbf{1},\textbf{1},\textbf{16} ) +  ( \textbf{1},\textbf{10},\textbf{1},\textbf{16} ) +  ( \textbf{1},\textbf{1},\textbf{10},\textbf{16} )$}
\end{enumerate}}

\item { $\textrm{SU}(10) \times \textrm{SU}(10) \times \textrm{SU}(11) \times \textrm{SU}(16)$ 
\begin{enumerate}
\item {$2 ( \textbf{45},\textbf{1},\textbf{1},\textbf{1} ) + 2 ( \textbf{1},\textbf{45},\textbf{1},\textbf{1} ) + 2 ( \textbf{1},\textbf{1},\textbf{55},\textbf{1} ) +  ( \textbf{1},\textbf{1},\textbf{1},\textbf{16} ) +  ( \textbf{10},\textbf{1},\textbf{1},\textbf{16} ) +  ( \textbf{1},\textbf{10},\textbf{1},\textbf{16} ) +  ( \textbf{1},\textbf{1},\textbf{11},\textbf{16} )$}
\end{enumerate}}

\item { $\textrm{SU}(10) \times \textrm{SU}(10) \times \textrm{SU}(12) \times \textrm{SU}(16)$ 
\begin{enumerate}
\item {$2 ( \textbf{45},\textbf{1},\textbf{1},\textbf{1} ) + 2 ( \textbf{1},\textbf{45},\textbf{1},\textbf{1} ) + 2 ( \textbf{1},\textbf{1},\textbf{66},\textbf{1} ) +  ( \textbf{10},\textbf{1},\textbf{1},\textbf{16} ) +  ( \textbf{1},\textbf{10},\textbf{1},\textbf{16} ) +  ( \textbf{1},\textbf{1},\textbf{12},\textbf{16} )$}
\end{enumerate}}

\item { $\textrm{SU}(10) \times \textrm{SU}(11) \times \textrm{SU}(11) \times \textrm{SU}(16)$ 
\begin{enumerate}
\item {$2 ( \textbf{45},\textbf{1},\textbf{1},\textbf{1} ) + 2 ( \textbf{1},\textbf{55},\textbf{1},\textbf{1} ) + 2 ( \textbf{1},\textbf{1},\textbf{55},\textbf{1} ) +  ( \textbf{10},\textbf{1},\textbf{1},\textbf{16} ) +  ( \textbf{1},\textbf{11},\textbf{1},\textbf{16} ) +  ( \textbf{1},\textbf{1},\textbf{11},\textbf{16} )$}
\end{enumerate}}

\end{enumerate}

\section{Solutions with $n_T=0$}

\begin{enumerate} 

\item \text{$G_2$}  

\begin{enumerate} 
 
\item[(a)\textsuperscript{*}]$ 1  \times  {\bf  14}+ 1  \times  {\bf273}$

\item[(b)\textsuperscript{*}]$ 1  \times  {\bf7}+1  \times  {\bf14}+1  \times  {\bf77}+1  \times  {\bf189}$

\item[(c)] $1  \times  {\bf27}+1  \times  {\bf76}+1  \times  {\bf182}$

\item[(d)\textsuperscript{*}]$ 1  \times {\bf7}+1 \times {\bf14}+7 \times {\bf27}+1 \times {\bf77}$

\item[(e)\textsuperscript{*}]$ 2 \times {\bf7}+2 \times {\bf14}+1 \times {\bf27}+1 \times {\bf64}+2 \times {\bf77}$

\item[(f)]$ (1+3n) \times {\bf7}+(13-n) \times {\bf14}+(1+n) \times {\bf27}+1 \times {\bf64}\ ,\qquad n=0,1$

\item[(g)]$ 3 \times {\bf7}+5 \times {\bf27}+2 \times {\bf64}$

\item[(h)] $1 \times {\bf7}+6 \times {\bf14}+3 \times {\bf64}$

\item[(i)]$ 3 \times {\bf14}+4 \times {\bf27}+2 \times {\bf64}$

\item[(j)] $(21+3n) \times {\bf7}+(7-3n) \times {\bf14}+ (1+n) \times {\bf27}\ ,\qquad n=0,1,2$

\item[(k)]$ (10+3n)  \times {\bf7}+(12-3n) \times {\bf14}+(1+n) \times {\bf27}\ ,\qquad n=0,...,3$

\item[(l)] $(1+3n) \times {\bf7}+(12-3n) \times {\bf14}+(3+n) \times {\bf27}\ ,\qquad n=0,...,4$

\item[(m)]$ 28 \times {\bf7}+3 \times {\bf14}+1 \times {\bf27}$

\item[(n)] $31 \times {\bf7}+2 \times {\bf27}$

\item[(p)]$ 1 \times {\bf14}+9 \times {\bf27}$

\item[(q)] $31 \times {\bf7}+1 \times {\bf27}$

\item[(r)]$ 7 \times {\bf7}+15 \times {\bf14}$

\item[(s)] $18 \times {\bf7}+10 \times {\bf14}$

\item[(t)] $25 \times {\bf7}+6 \times {\bf14}$

\item[(u)] $27 \times {\bf7}+1 \times {\bf14}$

\item[(v)] $28 \times {\bf7}+3 \times {\bf14}$

\item[(w)] $13 \times {\bf7}$

\item[(x)] $22 \times {\bf7}$

\end{enumerate}

\item $F_4$  

\begin{enumerate} 
\item $1 \times {\bf 324} $

 \item[(b)\textsuperscript{*}]$ 1 \times {\bf 52}+ \times {\bf 273} $

 \item[(c)] $6 \times {\bf 52} $

 \item[(d)]$ 6 \times {\bf 26} $

 \item[(e)] $9 \times {\bf 26} $
 
\item[(f)] $6 \times {\bf 26}+3 \times {\bf 52} $

 \item[(g)] $9 \times {\bf 26}+1 \times {\bf 52} $

\end{enumerate} 

\item $E_6$  
\begin{enumerate} 
\item[(a)\textsuperscript{*}] $1 \times {\bf 351} $

 \item[(b)\textsuperscript{*}]$ 1 \times {\bf 351'} $

 \item[(c)]$ 7 \times {\bf 27} $

 \item[(d)]$ 10 \times {\bf 27} $

 \item[(e)]$ 4 \times {\bf 27}+3 \times {\bf 78} $

 \item[(f)]$ 9 \times {\bf 27}+1 \times {\bf 78} $
\end{enumerate}

 \item $E_7$  
\begin{enumerate} 

\item $9 \times {\bf 56} $

\item $12 \times {\bf 56} $

\item $3 \times {\bf 133} $

\item$9 \times {\bf 56}+1 \times {\bf 133} $

\end{enumerate}

\item $SU(N)$
\begin{align}
&N=10:&& \textrm{(a) }  8\times{\bf 10}+6\times{\bf 45} 
\nn\\
&&& \textrm{(b) }  14\times{\bf 10}+3\times{\bf 45}
\nn\\[3pt]
&N=11:&& \textrm{(a) } 4\times{\bf 11}+6\times{\bf 55}
\nn\\
&&& \textrm{(b) } 13\times{\bf 11}+3\times{\bf 55} 
\nn
\end{align}

\item $SO(N)$
\begin{align}
&N=18:  && \textrm{(a)\textsuperscript{*} }1 \times{\bf 170}+1 \times {\bf 256}
\nn\\[3pt]
&N=17: && \textrm{(a) } 1\times{\bf 152}+1\times{\bf 256}
\nn\\[3pt]
&N=16: && \textrm{(a) } 2 \times {\bf 128}+1\times {\bf 135} 
\nn\\
&&& \textrm{(b) } 16 \times {\bf 16}+1\times {\bf 128}
\nn\\[3pt]
&N=15: && \textrm{(a) }  1\times {\bf 119}+2 \times {\bf 128}
\nn\\
&&& \textrm{(b) } 15\times {\bf 15}+1 \times {\bf 128}
 \nn\\[3pt]
&N=14: &&\textrm{(a)\textsuperscript{*} }   1 \times {\bf 364}
\nn\\
&&& \textrm{(b) }4\times {\bf 64}+1 \times {\bf 104}
\nn\\
&&& \textrm{(c) } 14\times {\bf 14}+2 \times {\bf 64}
\nn\\[3pt]
&N=13: && \textrm{(a) }8\times {\bf 64}+1 \times {\bf 90}
\nn\\
&&& \textrm{(b) } 13\times {\bf 13}+4 \times {\bf 64}
\nn\\[3pt]
&N=12:&& \textrm{(a) }16\times {\bf 32}+1 \times {\bf 77}
\nn\\
&&& \textrm{(b) } 8\times {\bf 32}+3 \times {\bf 66}
\nn\\
&&& \textrm{(c) } 9\times {\bf 12}+5 \times {\bf 32}
\nn\\
&&& \textrm{(d) }9\times {\bf 12}+9 \times {\bf 32}+1 \times {\bf 66}
\nn\\
&&& \textrm{(e) }12\times {\bf 12}+8 \times {\bf 32}
\nn\\[3pt]
&N=11:&& \textrm{(a) } 16\times{\bf 32}+1\times{\bf 65}
\nn\\
&&& \textrm{(b) }  2\times{\bf 11}+8\times{\bf 32}+3  \times  {\bf 55} 
\nn\\
&&& \textrm{(c) } 8  \times  {\bf 11}+5  \times  {\bf 32} 
\nn\\
&&& \textrm{(d) } 9  \times  {\bf 11}+9  \times  {\bf 32}+1  \times  {\bf 55} 
\nn\\
&&& \textrm{(e) } 11  \times  {\bf 11}+8  \times  {\bf 32} 
\nn\\[3pt]
& N=10:   &&\textrm{(a)\textsuperscript{*} }  2\times{\bf 54}+1\times{\bf 210}   
\nn\\
&&& \textrm{(b) } 16  \times  {\bf 16}+1  \times  {\bf 54} 
\nn\\
&&& \textrm{(c) } 4  \times  {\bf 10}+8  \times  {\bf 16}+3  \times  {\bf 45} 
\nn\\
&&& \textrm{(d) } 7  \times  {\bf 10}+5  \times  {\bf 16} 
\nn\\
&&& \textrm{(e) } 9  \times  {\bf 10}+9  \times  {\bf 16}+1  \times  {\bf 45} 
\nn\\
&&& \textrm{(f) }  10  \times  {\bf 10}+8  \times  {\bf 16} 
\nn
\end{align}

\item $Sp(N)$
\begin{align}
&   7\le N\le 10:  &&  2(24-2N)  \times  {\bf N}+ 2  \times \left( {\bf N(2N-1)-1} \right) \ ,\qquad n=0,...,3
\nn\\[3pt]
&   N=6:   &&   \textrm{(a) } 5  \times  {\bf 65} 
\nn\\
&&& \textrm{(b)\textsuperscript{*} }   1 \times  {\bf 65}+1  \times  {\bf 78}+2  \times  {\bf 208} 
\nn\\
&&& \textrm{(c) } 24  \times  {\bf 12}+2  \times  {\bf 65} 
\nn\\[3pt]
& N=5:   && \textrm{(a) } 6  \times  {\bf 10}+3  \times  {\bf 44}+1  \times  {\bf 55}+2  \times  {\bf 110} 
\nn\\
 &&& \textrm{(b)\textsuperscript{*} }   37  \times  {\bf 10}+2  \times  {\bf 44}+1  \times  {\bf 110} 
\nn\\
&&& \textrm{(c) }   16  \times  {\bf 10}+5  \times  {\bf 44} 
\nn\\
&&& \textrm{(d) }   28  \times  {\bf 10}+2  \times  {\bf 44} 
\nn
\end{align}

$\boxed{E\times E}$
\medskip

\item $\textrm{G}_{2}\times \textrm{G}_{2}$  
\begin{enumerate}
    \item[(a)\textsuperscript{*}] $({\bf 14,1}) + ({\bf 1,7}) + ({\bf 1,14}) + ({\bf 1,77}) + ({\bf 7,27})$
\end{enumerate}

\item $\textrm{G}_{2}\times \textrm{E}_{6}$   
\begin{enumerate}
    \item[(a)\textsuperscript{*}] $({\bf 14,1}) + 3({\bf 27,1}) + 3({\bf 1,27}) + ({\bf 7, 27})$

    \item[(b)\textsuperscript{*}] $({\bf 7,1}) + 2({\bf 14,1}) + ({\bf 64,1}) + ({\bf 77,1}) + ({bf 7,27})$
\end{enumerate}

\medskip
$\boxed{E\times C}$
\medskip

\item $\textrm{G}_{2}\times \textrm{Sp}(5)$  
\begin{enumerate}
    \item[(a)\textsuperscript{*}] $2({\bf 7,1}) + 9({\bf 1,10}) + 2({\bf 1,44}) + ({\bf 1,110}) + 4({\bf 7,10})$
\end{enumerate}

\item $\textrm{E}_6 \times \textrm{SO}(10)$  
\begin{enumerate}
\item[(a)\textsuperscript{*}] $(\textbf{27},\textbf{10})$
\end{enumerate}

\medskip
\item $\boxed{SU(N)\times SU(M)}$
\medskip

\begin{enumerate}
    \item $10\leq N\leq M\leq 11$:

    \item[] $(24-N-M)(\textbf{N},\textbf{1}) + 3(\mathbf{\tfrac{1}{2}N(N-1)},\textbf{1}) + (24-N-M)(\textbf{1},\textbf{M}) + (\textbf{1},\mathbf{\tfrac{1}{2}M(M-1)}) + (\textbf{N},\textbf{M})$
\end{enumerate}

\medskip
\item $\boxed{SU(N)\times SO(M)}$
\medskip

\begin{enumerate}
    \item $10\leq N\leq M$, \quad $N\leq M\leq 14 $, \quad $N+M\leq 24$:

    \item[] $(24-N-M)(\textbf{N},\textbf{1}) + 3(\mathbf{\tfrac{1}{2}N(N-1)},\textbf{1}) + (M-N)(\textbf{1},\textbf{M}) + k(\textbf{1},\mathbf{2^{\lfloor\tfrac{M-1}{2}\rfloor}})$\\
    $k=2$ for $M=14$, $k=4$ for $M=13$, and $k=8$ for $M=10,11,12$
\end{enumerate}

\item $\boxed{SU(N)\times Sp(M)}$

\begin{enumerate}
    \item $10\leq N\leq 11$, \quad $5\leq M\leq 7$, \quad $N+2M\leq 24$:

    \item[] $(24-N-2M)(\textbf{N},\textbf{1}) + 3(\mathbf{\tfrac{1}{2}N(N-1)},\textbf{1}) + 2(24-2M-N)(\textbf{1},\textbf{2M}) + 2(\textbf{1},\mathbf{M(2M-1)-1})\\ +2(\textbf{N},\textbf{2M})$
\end{enumerate}

 $\boxed{SO(N)\times SO(M)}$

\begin{enumerate}
    \item $10\leq N=M\leq 16$:

    \item[] $k(\mathbf{2^{\lfloor\tfrac{N-1}{2}\rfloor}},\textbf{1}) + k(\textbf{1},\mathbf{2^{\lfloor\tfrac{M-1}{2}\rfloor}}) + (\textbf{N},\textbf{M})$\\
    $k=1$ for $M=15,16$, $k=2$ for $M=14$, $k=4$ for $M=13$, and $k=8$ for $M=10,11,12$

    \item $(N,M)^{*} = (10,16)$:

    \item[] $ ({\bf 1,128})+({\bf 16,16})+({\bf 54,1})$
\end{enumerate}

\item $ \boxed{Sp(N) \times SO(M)} $

\begin{enumerate}
    \item $5\leq N\leq 9$, \quad $10\leq M\leq 12$, \quad $N\leq M-3$:

    \item[] $(N-3-M)(\textbf{1},\textbf{M}) + 5(\textbf{1},\mathbf{2^{\lfloor\tfrac{M-1}{2}\rfloor}}) + (2(24-2N)-M)(\textbf{2N},\textbf{1}) + 2(\mathbf{N(2N-1)-1},\textbf{1})$

    \item $N=5$, \quad $10\leq M\leq 14$,\\
    or $(N,M) = (6,12)$:

    \item[] $2(24-2N-M)(\textbf{2N},\textbf{1}) + 2(\mathbf{N(2N-1)-1},\textbf{1}) + (M-2N)(\textbf{1},\textbf{M}) + k(\textbf{1},\mathbf{2^{\lfloor\tfrac{M-1}{2}\rfloor}})$\\
    $k=2$ for $M=14$, $k=4$ for $M=13$, and $k=8$ for $M=10,11,12$

    \item $(N,M) = (5,10)$:

    \item[] $5 ({\bf 44,1}) + 7 ({\bf 1,10}) + ({\bf 10,16})$
\end{enumerate}

\item $\boxed{Sp(N)\times Sp(M)}$

\begin{enumerate}
    \item $5\leq N\leq M\leq 7$, \quad $N+M\leq 12$:

    \item[] $2(24-2N-2M)(\textbf{2N},\textbf{1}) + 2(\mathbf{N(2N-1)-1},\textbf{1}) + 2(24-2N-2M)(\textbf{1},\textbf{2M}) + 2(\textbf{1},\mathbf{M(2M-1)-1}) + 2(\textbf{2N},\textbf{2M})$
\end{enumerate}
 
 \end{enumerate}

\end{appendix}
\bibliographystyle{jhep}
\bibliography{bibliography}
\end{document}